\pdfoutput=1
\documentclass[paper=a4, fontsize=10pt, numbers=noenddot]{scrartcl}
\usepackage[margin=2.299cm]{geometry}

\usepackage[english]{babel}
\usepackage[T1]{fontenc}
\usepackage[scaled=1.04]{biolinum}

\usepackage{fourier}
\usepackage[scaled=0.83]{beramono}
\usepackage{microtype,xspace,soul}
\soulregister{\peight}{7}

\usepackage{amsmath,amssymb}
\usepackage{mathtools,relsize}
\DeclareMathAlphabet{\mathcal}{OMS}{cmsy}{m}{n}

\usepackage[dvipsnames]{xcolor}
\usepackage{graphicx}
\usepackage{booktabs,tabularx,caption}
\usepackage{enumitem,afterpage}
\usepackage{hyperref,doi}
\usepackage{siunitx}
\usepackage[nomessages]{fp}

\usepackage[sort&compress,numbers]{natbib}

\def\epjc{Eur.\ Phys.\ J.\ C}

\def\aj{AJ}%
\def\apj{ApJ}%
\def\apjl{ApJ}%
\def\apjs{ApJS}%
\def\jcap{JCAP}%
\def\mnras{MNRAS}%
\def\prd{Phys.\ Rev.\ D}%
\def\physrep{Phys.\ Rep.}%
\def\epjc{Eur.~Phys.~J.~C}%
\definecolor{Blue}{RGB}{0,0,122}
\definecolor{Red}{RGB}{173,42,26}
\hypersetup{colorlinks=true, linktoc=all, linkcolor=Blue, citecolor=Red, urlcolor=Blue}
\newcommand{\updated}[1]{#1}

\newcommand{\parsecode}[1]{\textsf{#1}\xspace}
\newcommand{\fermist}{\parsecode{Fermitools}}
\newcommand{\mn}{\parsecode{MultiNest}}
\newcommand{\tw}{\parsecode{T-Walk}}
\newcommand{\diver}{\parsecode{Diver}}
\newcommand{\dmfit}{\parsecode{DMFIT}}
\newcommand{\pythia}{\parsecode{Pythia}}
\newcommand{\ndwarfs}{27}
\FPeval{\nparams}{clip(2+2*\ndwarfs)}
\newcommand{\peight}{\textsc{Pass~8}\xspace}
\newcommand{\peightrtwo}{\textsc{Pass~8\ (R2)}\xspace}
\newcommand{\peightrthree}{\textsc{Pass~8\ (R3)}\xspace}
\newcommand{\pseven}{\textsc{Pass~7}\xspace}
\newcommand{\rettwo}{\textsc{Ret~II}\xspace}
\newcommand{\sigv}{\left\langle\sigma\! v\right\rangle}
\newcommand{\lgJ}{\log_{10}J}
\newcommand{\bbb}{b\bar{b}}
\newcommand{\tptm}{\tau^+\!\tau^-}
\newcommand{\mrm}[2]{#1_\text{#2}}
\newcommand{\prrange}[2]{$\left[\num{#1}, \, \num{#2}\right]$}
\newcommand{\p}{p}
\newcommand{\data}{d}
\newcommand{\obs}{D}
\newcommand{\prob}[2]{\p\left(#1 \, \middle| \, #2 \right)}
\newcommand{\kde}[2]{\hat{\p}\left(#1 \, \middle| \, #2 \right)}
\newcommand{\altprob}[2]{\p(#1 \, | \, #2 )}
\newcommand{\prior}[1]{\p\left(#1\right)}
\newcommand{\dd}{\mathrm{d}}
\newcommand{\ee}{\mathrm{e}}
\newcommand{\vc}[1]{\mathbf{#1}}
\newcommand{\reffig}[1]{Fig.~\ref{#1}}
\def\lsim{\mathrel{\rlap{\lower4pt\hbox{$\sim$}}\raise1pt\hbox{$<$}}}
\def\gsim{\mathrel{\rlap{\lower4pt\hbox{$\sim$}}\raise1pt\hbox{$>$}}}

\title{\Large A Global Analysis of Dark Matter Signals from \ndwarfs~Dwarf Spheroidal Galaxies \updated{using 11 Years} of Fermi-LAT Observations}
\author{
	\large \href{mailto:hoof@uni-goettingen.de}{Sebastian Hoof},\textsuperscript{1,2} \href{mailto:a.geringer-sameth@imperial.ac.uk}{Alex Geringer-Sameth},\textsuperscript{2} and \href{mailto:r.trotta@imperial.ac.uk}{Roberto Trotta}\textsuperscript{2,3}\\
	\footnotesize\textsuperscript{1}Institut f\"ur Astrophysik, Georg-August Universit\"at G\"ottingen, Friedrich-Hund-Platz 1, 37077 G\"ottingen, Germany\\[-0.25em]
	\footnotesize\textsuperscript{2}Department of Physics, Imperial Centre for Inference and Cosmology, Imperial College London, London SW7 2AZ, UK\\[-0.25em]
	\footnotesize\textsuperscript{3}Data Science Institute, William Penney Laboratory, Imperial College London, London SW7 2AZ, UK
}
\date{}

\begin{document}

\sffamily
\maketitle

\rmfamily
\thispagestyle{empty}
\renewcommand{\baselinestretch}{1.15}\normalsize

\vspace{-1cm}

\begin{center}
\large\bfseries\sffamily Abstract\\[0.5ex]
\end{center}
\begin{abstract}
\noindent We search for a dark matter signal in 11~years of Fermi-LAT gamma-ray data from \ndwarfs~Milky Way dwarf spheroidal galaxies with spectroscopically measured $J$-factors. Our analysis includes uncertainties in $J$-factors and background normalisations and compares results from a Bayesian and a frequentist perspective. We revisit the dwarf spheroidal galaxy Reticulum~II, confirming that the purported gamma-ray excess seen in Pass~7 data is much weaker in Pass~8, independently of the statistical approach adopted. We introduce for the first time posterior predictive distributions to quantify the probability of a dark matter detection from another dwarf galaxy given a tentative excess. A global analysis including all \ndwarfs~dwarfs shows no indication for a signal in nine annihilation channels. We present stringent new Bayesian and frequentist upper limits on the dark matter cross~section as a function of dark matter mass. The best-fit dark matter parameters associated with the Galactic Centre excess are excluded by at least 95\% confidence level/posterior probability in the frequentist/Bayesian framework in all cases. However, from a Bayesian model comparison perspective, dark matter annihilation within the dwarfs is not strongly disfavoured compared to a background-only model. These results constitute the highest exposure analysis on the most complete sample of dwarfs to date. Posterior samples and likelihood maps from this study are publicly available.
\end{abstract}

\vspace{0.5cm}
\renewcommand{\baselinestretch}{1.0}\normalsize
\tableofcontents
\renewcommand{\baselinestretch}{1.25}\normalsize
\newpage

\section{Introduction}\label{sec:introduction}
Cold dark matter~(DM) makes up about 84\% of all the matter in the Universe today~\cite{1807.06209}, but the identity of the DM~particles is still unknown. If DM has some coupling to the Standard Model then we may expect DM to self-annihilate in astrophysical environments of sufficient density. Weakly-interacting DM particles with masses $\gsim \si{\GeV}$ are both theoretically and observationally well-motivated~\cite[e.g.][]{hep-ph/9506380} and the final state of their annihilation generally includes the emission of gamma~rays. The search for this emission is called indirect detection (see e.g. Refs~\cite{1604.00014,1705.11165} for reviews) and ideally focuses on objects containing a large number of DM particles in a suitably small region of space. From that perspective, dwarf spheroidal galaxies~(dSphs) are promising targets for such searches since they mostly consist of DM and neither contain many stars, nor much gas. They therefore present an environment with comparably low astrophysical backgrounds~\cite{0709.1510}. Their relative proximity to us and large separations from poorly understood photon sources are additional benefits for indirect searches in dSphs. Gamma-ray observations towards dSphs from instruments such as the Large Area Telescope~(LAT)~\cite{0902.1089} onboard the Fermi Gamma-ray Space Telescope have been shown to be powerful for detecting~--~or at least constraining~--~DM~models~\cite{0909.3300,1001.4531,1007.4199,1108.2914,1108.3546,1111.2604,1203.6731,1205.3620,1310.0828,1503.02320,1503.02632,1503.02641,1503.06209,1510.00389,1511.09252,1601.06590,1611.03184,1702.05266,1705.00426,1711.04696,1803.05508,1804.07542}.

However, an excess in gamma rays can only be established with high fidelity if the background contributions and systematics are well understood or otherwise accounted for. Assuming that all of the DM consists of the same type of particles in every dSph, we also have to demand consistency between data from different dSphs. We address both these issues using a global Bayesian analysis in addition to the usual, purely frequentist treatment.

\textsc{Reticulum~II}~(\rettwo)~\cite{1503.02079,1503.02584} is one of the more recently discovered dSphs and has been investigated by several authors using data from the~LAT. Two studies claimed a gamma-ray excess above background with a significance of $3.7\sigma$~\cite{1503.02320} and $3.2\sigma$~\cite{1503.06209}. The study in Ref.~\cite{1503.02320} used an event weighting technique~\cite{1410.2242}, while Ref.~\cite{1503.06209} adopted a maximum-likelihood approach. Follow-up studies, however, did not find an indication for a DM origin of the observed signal~\cite{1503.02632,1611.03184}. It was argued that this discrepancy is due to the differences in the data sets used~\cite{1503.02632,1807.08740}: the Fermi-LAT \pseven data release in Refs~\cite{1503.02320,1503.06209} and \peightrtwo in Refs~\cite{1503.02632,1611.03184}.

In light of the diverging conclusions from previous studies, it is important to understand why ostensibly similar (although not identical) analyses obtain different results. For \rettwo, this is particularly interesting since the significance of a DM~signal seems to be increasing with time, even in the lower-significance \peight data~\cite{1805.06612}.

More generally, any gamma-ray excess from dSphs is likely to be initially of a very marginal significance. In this case, one can expect that different statistical approaches, when applied to the same data, will yield different conclusions as to the statistical significance and origin of such an excess (see Refs~\cite{Cousins:1994yw,1405.5010} for examples of how the choice of method affects parameter estimation). One of the aims of this study is therefore to investigate the dependency of the conclusions on the viability of a DM signal from \rettwo on the methodology (Bayesian vs frequentist) as well as on the data set used (\pseven vs \peight). This study is also the first fully Bayesian analysis of the \rettwo gamma-ray observations. We demonstrate the power and usefulness of our approach and perform a Bayesian model comparison to assess in a quantitative way the viability of the DM signal hypothesis. We do this for \rettwo alone as well as for a combined global fit of all dwarfs with constrained DM halos. Given the number of dwarfs analysed in this way along with an 11-year exposure, this study is the most complete of its kind to date.

In the next section, we introduce our methodology and discuss the various inputs for our analysis. In Sec.~\ref{sec:results}, we first apply our method to \rettwo and introduce posterior predictive distributions as a diagnostic tool before performing a global analysis of \ndwarfs~dSphs. We compare our findings with previous work and discuss the results before concluding in Sec.~\ref{sec:conclusions}.

\section{Methodology}\label{sec:methods}
We use the publicly available \fermist\footnote{We use \texttt{v9r33p0} for analysing \pseven data \updated{and version~\texttt{1.0.9} of the new \textsf{Python} interface for analysing \peightrthree data.} More information and download links for these tools are available at \url{https://fermi.gsfc.nasa.gov/ssc/data/analysis/software/}.} for LAT data extraction and preparation. We consider a DM matter candidate~$\chi$ (a weakly interacting particle, or WIMP) with thermally-averaged annihilation cross~section~$\sigv$ and mass~$m_\chi$ inside a dSph with DM density distribution~$\rho_\chi$. The differential photon flux (in units of photons per time and area) per energy~$E$ and solid angle~$\Omega$, coming from sky direction~$\vc{n}$, is given by
\begin{equation}
	\frac{\dd^2 \Phi(E,\vc{n})}{\dd E \, \dd \Omega} = \underbrace{\frac{\sigv}{8\otherpi \, m_\chi^2} \; \left(\sum_{f}^{} \frac{\dd N_f}{\dd E} \; b_f\right)}_{\text{``particle physics''}}\; \underbrace{\int_{}^{} \! \dd \ell \, \rho_\chi^2(\ell;\, \vc{n} ) \vphantom{\sum_{f}^{}}}_{\equiv \tfrac{\dd J(\vc{n})}{\dd \Omega}} \, , \label{eq:flux}
\end{equation}
where $b_f$ is the branching fraction for annihilation into various final states~$f$ and $\tfrac{\dd N_f}{\dd E}$ is the spectrum of gamma~rays emitted from annihilation into final state~$f$. The so-called ``$J$-profile'' $\dd J(\vc{n})/\dd\Omega$ is determined by the astrophysics of the system, i.e. the macroscopic distribution of DM in the dSph~\cite{astro-ph/9712318, astro-ph/0305075, 0808.2641}. It is the square of the dark matter density integrated along the line of sight in the direction~$\vc{n}$. We separately consider the $\bbb$ and $\tptm$ channels as our benchmark final states. While this choice allows us to compare our results to almost all of the existing literature, the DM interpretation of the Galactic~Centre excess in Ref.~\cite{1411.4647} for the $\tptm$~channel presents an interesting model that has not been decisively excluded yet. Choosing the heaviest fermionic final states as benchmark cases has been advocated in the literature using helicity arguments~\cite{1983_Goldberg}~(see also Ref.~\cite{hep-ph/9506380} for a review). This favours fermions with larger masses~$m_f$ due to a~$(m_f/m_\chi)^2$ suppression of the cross~section, making~$\bbb$ and~$\tptm$ the most interesting channels since the top~quark is very heavy compared to the full mass range of interest.

\updated{In addition to our main results, we provide limits for seven other channels~($e^+e^-$, $\mu^+\mu^-$,$c\bar{c}$, $t{\bar{t}}$, $gg$, $W^+W^-$, and $ZZ$) in Appendix~A to show how they compare to the benchmark channels.}

We bin the data spatially on the sky and in energy (the details can be found in Sec.~\ref{sec:data}). The DM signal in each bin can be calculated using the~\fermist, which convolve~(\ref{eq:flux}) with the instrument response (effective area, point-spread function~(PSF), and energy dispersion). As discussed in Sec.~\ref{sec:jfactors}, we treat each dSph as a point source of gamma~rays and so the convolution with the PSF yields a model prediction that depends only on the scalar quantity~$J$ (the so-called ``$J$-factor''), the integral of $\dd J/\dd\Omega$ over the solid angle. For reference values of~$J$ and~$\sigv$, we pre-compute and tabulate the DM~signal for 125~mass values from \SIrange{2}{e4}{\GeV} for each energy bin (100~log-spaced values from \SIrange{2}{e2}{\GeV} and 25~log-spaced values from \SIrange{e2}{e4}{\GeV}). This was done by generating source maps using \texttt{gtsrcmaps} for given fixed WIMP and background model parameters and subsequently obtaining the binned counts cube files (auxiliary output from \texttt{gtlike}). We interpolate to obtain the DM signal at arbitrary mass and have verified that the interpolation is accurate to within~5\% for any energy bin in both channels. Since the gamma-ray signal is proportional to $J \times \sigv$, we can linearly rescale the pre-computed reference signal counts with the appropriate values of~$J$ and $\sigv$ and speed up the likelihood evaluations.

The LAT detects individual photons such that the resulting data $\data$ can be described by a Poisson process. Events are independent, so the binned likelihood function is a product of Poisson distributions for the number of observed counts $n_{i,j}$ in each energy bin~$i$ and spatial bin~$j$,
\begin{equation}
	\prob{\data}{m_\chi,\, \sigv;\,\beta,\, \lgJ}= \prod_{i,j} \frac{\lambda_{i,j}^{n_{i,j}}}{n_{i,j}!} \; \ee^{-\lambda_{i,j}} \, , \label{eq:lnL:poisson}
\end{equation}
where $\lambda_{i,j}$ is the combined background and signal count expectation value for bin $i,j$. The latter is given by
\begin{equation}
	\lambda_{i,j} = b^{\text{iso}}_{i,j} + b^{\text{src}}_{i,j} + \beta \, b^{\text{gal}}_{i,j} + s^{\text{DM}}_{i,j}\left(m_\chi, \sigv,\, J\right) \, , \label{eq:backgrounds}
\end{equation}
with $b^{\text{iso}}_{i,j}$ and $b^{\text{gal}}_{i,j}$ being the isotropic and Galactic diffuse background contributions in the $i$th energy bin and $j$th spatial bin, respectively, while $b^{\text{src}}_{i,j}$ is the contribution from nearby point sources. We introduce a scaling parameter $\beta$ for the $b^{\text{gal}}_{i,j}$ component~(see Sec.~\ref{sec:bkgmodel} for more details about the background model). The scaling parameter accounts for some systematic uncertainties in the diffuse background model. The signal contribution from DM is given by $s^{\text{DM}}_{i,j}$.

When considering multiple dSphs, all of the above parameters -- except $m_\chi$ and $\sigv$ -- are specific to each dwarf~$k$. We ensure with our data selection procedure in Sec.~\ref{sec:data} that the data obtained in the dwarfs' vicinities are independent, such that the total likelihood is given by the product of all individual Poisson likelihoods,
\begin{equation}
	\prob{\data}{m_\chi,\, \sigv; \, \boldsymbol{\beta},\, \log_{10} \mathbf{J}}= \prod_{k} \; \prob{\data_{k}}{m_\chi,\, \sigv; \,\beta_{k},\,\lgJ_k} \, , \label{eq:likelihood}
\end{equation}
where $\boldsymbol{\beta}=\{\beta_{k}\}$ and $\log_{10} \mathbf{J}=\{\lgJ_k\}$ are the collection of $J$-factors and background normalisations for each dSph, respectively.

Finally, we also multiply~(\ref{eq:likelihood}) with an additional likelihood (or prior, in the Bayesian approach) for the nuisance parameters~$J$, as we discuss in Sec.~\ref{sec:jfactors}. To the degree that the $J$-factors are well constrained, we can break the degeneracy between~$J$ and~$\sigv$ in~(\ref{eq:flux}) and place direct limits on the cross~section. The advantage of fully incorporating background and $J$-factor uncertainties in the Bayesian analysis is to propagate them through to the posterior distributions and the resulting dark matter constraints.

\subsection{Data selection}\label{sec:data}
To perform our analysis, we only include dSphs with kinematically determined $J$-factors. The largest uniform analysis of dSph $J$-factors is currently that of Ref.~\cite{1802.06811}, who provide $J$-factor estimates for 37~out of the 41~dSphs in Table~A2 (\textit{ibid.}). Out of those~37, we focus on the 28~Milky~Way dSphs and, for the three~dSphs where two $J$-factors are given (\textsc{Horologium~I}, \rettwo, and \textsc{Tucana~II}), we use the values based on data from Ref.~\cite{1503.02584}.

To guarantee the independence of the LAT events, we require that our spatial regions-of-interest~(ROIs) for any two dSphs do not overlap. Since we choose a $\ang{1}\times\ang{1}$ square ROI around each target, the minimal permissible separation is $\sqrt{2}\text{\textdegree}\approx \ang{1.4}$. All 28~Milky Way dSphs meet this requirement.

Finally, we omit \textsc{Willman~1} from our analysis as it shows strong evidence for tidal disruption and/or non-equilibrium kinematics~\cite{astro-ph/9612025,1007.3499,1408.0002}. Tidal effects and other kinematic disturbances generally inflate measured velocity dispersions, which propagates into overestimates of~$J$. The size of such systematics have not been quantified and the $J$~determinations of dSphs such as \textsc{Willman~1} are therefore unreliable in an uncontrolled way.

Removing \textsc{Willman~1} reduces the total number of dSphs that we consider to~\ndwarfs: \textsc{Aquarius~II}, \textsc{Bo\"otes~I}, \textsc{Canes Venatici~I}, \textsc{Canes Venatici~II}, \textsc{Carina}, \textsc{Carina~II}, \textsc{Coma Berenices}, \textsc{Draco}, \textsc{Draco~II}, \textsc{Fornax}, \textsc{Grus~I}, \textsc{Hercules}, \textsc{Horologium~I}, \textsc{Leo~I}, \textsc{Leo~II}, \textsc{Leo~IV}, \textsc{Leo V}, \textsc{Pegasus~III}, \textsc{Pisces~II}, \textsc{Reticulum~II}, \textsc{Sculptor}, \textsc{Segue~1}, \textsc{Sextans}, \textsc{Tucana~II}, \textsc{Ursa Major~I}, \textsc{Ursa Major~II}, and \textsc{Ursa Minor}. This represents the most complete sample with measured $J$-factors used for DM~searches.

For each dSph in our global fit, we use 579\,weeks ($\approx$~11\,years) of \peightrthree \texttt{SOURCE} class data, using \texttt{gtselect}, \texttt{gtktime}, and \texttt{gtltcube} to extract the data, determine good time intervals, and calculate the livetime and instrument response.\footnote{\updated{We use weeks~9--511 and 512--579 and follow the Fermi Collaboration's recommendations for \peightrthree data selection} (available at \url{https://fermi.gsfc.nasa.gov/ssc/data/analysis/documentation/Cicerone/Cicerone_Data_Exploration/Data_preparation.html}) as well as their procedure for performing a binned analysis (available at \url{https://fermi.gsfc.nasa.gov/ssc/data/analysis/scitools/binned_likelihood_tutorial.html}). The non-default options for the \fermist are \texttt{evclass=128}, \texttt{evtype=3}, \texttt{zmax=90} (in \texttt{gtselect}); \texttt{roicut=no}, \texttt{filter=(DATA\_QUAL>0)\&\&(LAT\_CONFIG)==1} (in \texttt{gtktime}); and \texttt{zmax=90} (in \texttt{gtltcube}). We enable energy dispersion in our analysis following the instructions on \url{https://fermi.gsfc.nasa.gov/ssc/data/analysis/documentation/Pass8_edisp_usage.html}. As recommended, we set the global variable \texttt{USE\_BL\_EDISP=true} and apply the energy dispersion correction to all model components except the isotropic component, for which we specify \texttt{apply\_edisp=false} in the model file.}

We then bin the data in 15~log-spaced energy bins from \SIrange{0.5}{500}{\GeV} and 100~spatial bins using \texttt{gtbin} ($10\times 10$ square bins of $\ang{0.1}\times\ang{0.1}$ each). The DM signal for a given value of the DM parameters can be calculated using \texttt{gtmodel} from the~\fermist (which uses the \dmfit package~\cite{0808.2641} based on \pythia~\cite{hep-ph/0603175,0710.3820}).

The only difference for our dedicated study of \rettwo is that we only select 339\,weeks ($\approx$~6.5\,years) of \peightrthree data~(using the 3FGL catalogue to match the \pseven setup) to compare with the same amount of \pseven data.\footnote{We follow the Fermi Collaboration's recommendations for \pseven~\textsc{Reprocessed} data selection (available at \url{https://fermi.gsfc.nasa.gov/ssc/data/p7rep/analysis/documentation/Cicerone/Cicerone_Data_Exploration/Data_preparation.html}). The non-default settings applied to the \fermist are \texttt{evclass=2}, \texttt{zmax=100} (in \texttt{gtselect}) and \texttt{roicut=yes}, \texttt{filter=(DATA\_QUAL>0)\&\&(LAT\_CONFIG)==1} (in \texttt{gtktime}).} These correspond to weeks~9--347, as used in Ref.~\cite{1503.02320}. Selecting significantly more data for this comparison was not possible, since \pseven was discontinued after week~368 and we want to use the same observation time for both \pseven and \peight to enable a meaningful comparison.

\subsection{Background model}\label{sec:bkgmodel}
The three components of the background model that contribute to the signal-plus-background counts in~(\ref{eq:backgrounds}) are the isotropic, Galactic diffuse, and point source components. The isotropic background\footnote{Available as \texttt{iso\_source\_v05.txt} for \pseven and \texttt{iso\_P8R3\_SOURCE\_V2\_v1.txt} for \peightrthree in the \fermist or at \url{https://fermi.gsfc.nasa.gov/ssc/data/access/lat/BackgroundModels.html}.} was determined by the Fermi Collaboration via a full-sky fit. It is rather well constrained: the uncertainties on the energy spectrum amount to no more than 1.3\% for the most important energies below about~$\SI{30}{\giga\electronvolt}$ and less than about 9\% in the remaining energy range we consider.\footnote{These numbers are based on \peightrtwo; for \peightrthree, no error estimates are available anymore in the isotropic background files.} For this reason, we do not introduce nuisance parameters for the isotropic background and instead fix its contribution to the value given by the model.

The contribution of Galactic diffuse emission is captured by the Fermi Collaboration's Galactic interstellar emission model\footnote{Available as \texttt{gll\_iem\_v05\_rev1.fit} for \pseven and \texttt{gll\_iem\_v07.fits} for \peightrthree in the \fermist or at \url{https://fermi.gsfc.nasa.gov/ssc/data/access/lat/BackgroundModels.html}.}~\cite{1602.07246}, created from a full-sky fit to physically-motivated templates derived with the \textsf{GALPROP}~\cite{1008.3642} cosmic ray propagation code\footnote{The model also incorporates residuals in the data above a particular spatial scale. See \url{https://fermi.gsfc.nasa.gov/ssc/data/analysis/software/aux/4fgl/Galactic_Diffuse_Emission_Model_for_the_4FGL_Catalog_Analysis.pdf} for the details of the \texttt{gll\_iem\_v07.fits} model.}.
The uncertainties in this model are not easily quantified and hence we introduce energy-independent normalisation factors~$\beta_{k}$ for each dwarf~$k$ to account for possible local deviations from the reference value. An analogous approach was taken in previous studies of the background~\cite[e.g.][]{1612.03175}. The introduction of such dwarf-dependent scaling factors is important since the empirically derived background surrounding the dSphs has been shown to deviate from the Fermi Galactic interstellar emission model in ways beyond what is expected from Poisson fluctuations~\cite{1310.0828,1409.1572,1410.2242}. \updated{We note that using an energy-independent rescaling factor for the background may not fully account for an additional effect due to the background residual's spectral shape varying from location to location.}

Finally, nearby point sources could contribute to the photon counts inside our ROI due to the size of the PSF. To account for this effect, we include all nearby sources in the 3FGL~\cite{1501.02003}~(for the \rettwo analysis in Sec.~\ref{sec:results_ret2}; for consistency with Refs~\cite{1503.02320,1503.06209}) \updated{and 4FGL~\cite{1902.10045}~(for the global analysis in Sec.~\ref{sec:globalfit}) catalogues} that are up to \ang{10} away from the ROI centre.\footnote{We use the \texttt{make3FGLxml.py} and \texttt{make4FGLxml.py} scripts by T.~Johnson, available at \url{https://fermi.gsfc.nasa.gov/ssc/data/analysis/user/}.} The photon flux from all point sources is fixed to their best-fit values. This contribution is always small, and it is never larger than the two other background contributions combined. In fact, we found for the data used in our global analysis that point sources amount to less than 10\% of the combined isotropic and Galactic diffuse background in 91\% of all energy bins in all dwarfs. Although point sources are a very sub-dominant background, we nevertheless include them for completeness.

\subsection{J-factors}\label{sec:jfactors}
There is an extensive literature on determining $J$-factors of dSphs~\cite{1104.0412,1309.2641,1408.0002,1504.02048,1504.02889,1504.03309,1603.07721,1603.08046,1604.05599,1608.07111,1702.00408,1712.03188,1802.06811,1810.09917,1909.13197}. Typically, studies constrain the dark matter distributions within dSphs by using their member stars as tracers of the gravitational potential. When in statistical equilibrium, these tracers obey the Jeans equation~\cite{astro-ph/0611925}. Applying this equation to spectroscopically-determined line of sight velocities of individual stars yields constraints on the $J$-factor.

The studies that have carried out systematic analyses of large numbers of dSphs have taken a Bayesian approach~\cite{1309.2641,1408.0002,1504.02048,1603.08046,1802.06811} and their results are presented in the form of marginal posterior distributions for individual dSph $J$-factors. We adopt these posteriors as priors in our Bayesian analysis. In a frequentist context, however, it is not straightforward to include these constraints on $J$-factors: there is no simple way to incorporate a prior. Previous studies, e.g. Refs~\cite{1108.3546,1503.02641}, have re-interpreted the $J$-factor posterior as a likelihood and multiplied it with the gamma-ray likelihood. This re-interpretation poses a conceptual difficulty as discussed in Ref.~\cite[][Sec.~IX.C]{1410.2242} and recent progress has been made in creating a frequentist likelihood function for the spectroscopic observations~\cite{1608.07111,1712.03188,1810.09917}. However, due to the small number of stars in most of the known dSphs, it is not feasible to treat the majority of dSphs in the frequentist framework. Therefore, when adopting a frequentist approach, we implicitly make the conceptual leap of re-interpreting the posteriors on $J$-factors as likelihoods, and multiply them with~(\ref{eq:likelihood}) to obtain a total likelihood which is assumed to describe the gamma-ray and spectroscopic data. This follows previous practice, but we point out that it is not self-consistent from a statistical point of view. There is no such difficulty in the Bayesian approach, where it is straightforward to reinterpret the posterior from one analysis (in this case, of the spectroscopic data) as a prior for the next (the gamma-ray data analysis).

The posterior distributions for the $J$-factors are generally well-approximated by log-normal distributions, which have been used in previous work~\cite{1108.3546,1503.02641}. However, while this approximation provides mostly a good fit to the dSphs without long tails in their posteriors,\footnote{Except, perhaps, for some dSphs such as \textsc{Leo~II} or \textsc{Sculptor}, whose posteriors are noticeably skewed.} it does not in the case of dSphs with such tails~\cite{1802.06811}. Including the tails of the distribution is important for two reasons: First, the value of the 15.87th percentile of the $J$-factor mode alone, as quoted by the authors in Ref.~\cite[Table~A2]{1802.06811}, should be close to the 15.87th percentile of the full distribution if a log-normal about the mode is a good approximation. However, in \textsc{Draco~II}, \textsc{Grus~1}, and \textsc{Leo~IV}, the 15.87th percentile of mode actually corresponds to about the 70th percentile of the full distribution, thus demonstrating that a log-normal approximation is poor. The situation is less problematic for \textsc{Leo~V} (36th percentile), \textsc{Pegasus~III} (37th percentile), or \textsc{Pisces~II} (30th percentile), but the log-normal approximation still fails to capture a sizeable part of the distribution.

The second reason is that the tails extend towards lower values of~$\lgJ$ compared to the mode of the distributions in all of the cases listed above. A~log-normal approximation around the mode is therefore \emph{not} conservative; it tends to overestimate the probability of a large $J$-factor for dSphs with long tails, thus systematically increasing the DM signal contribution to the gamma-ray data for a given annihilation cross~section. For the first three systems listed in the previous paragraph, the difference is quite severe: in the log-normal approximation of the distribution, 84\% of the probability lies above the 16th percentile value, but using the full probability distribution (without a log-normal approximation), only 30\% of probability is actually above that value in each case.

To obtain a better description of the $J$-factor constraints, we approximate the posteriors using a Gaussian kernel density estimator~(KDE)~\cite{Parzen1962,Rosenblatt1956}, based on the posterior samples for an integration angle of~\ang{0.5} provided by the authors of Ref.~\cite{1802.06811}.\footnote{The posterior samples are part of the auxiliary material for Ref.~\cite{1802.06811}, available at \url{https://github.com/apace7/J-Factor-Scaling}.} The KDE~approximation to the posterior for the $J$-factor from kinematic data, $d_\text{kin}$, is then given by
\begin{equation}
	\kde{\lgJ}{d_\text{kin}} = \mathlarger{\sum}\limits_{i=1}^{N} \; \frac{w_i}{\sqrt{2\otherpi\sigma_B^2}} \, \exp \left[- \frac{\left(\lgJ - \log_{10} J_i\right)^2}{2\sigma_{B}^2}\right] \label{eq:kde} \, ,
\end{equation}
with $w_i$ and $J_i$ being the weight and $J$-factor value of the $i$th posterior sample, respectively. Since $\sum_{i} \! w_i = 1$, the KDE is normalised in the variable~$\lgJ$. The quantity~$\sigma_{B}$ is the bandwidth of the KDE and its optimal value can be estimated via e.g.\ ``Scott's rule''~\cite{ScottMultiDE}, according to which $\hat{\sigma}_{B} = \sigma N^{-1/5}$ for $N$~samples in one-dimensional data, where $\sigma$ is the standard deviation of the $\lgJ_i$ samples. We find that $\hat{\sigma}_B \approx 0.15\,\text{dex}$ for all dSph samples provided in the auxiliary~material~of~Ref.~\cite{1802.06811}. We inspect the resulting KDEs and adjust the value of the bandwidth for each dSph to ensure that the KDEs approximate well the shape of each posterior.\footnote{The resulting bandwidths for all dSphs are (in alphabetical order): 0.1, 0.075, 0.025, 0.05, 0.025, 0.1, 0.075, 0.025, 0.25, 0.01, 0.4, 0.1, 0.1, 0.025, 0.025, 0.3, 0.3, 0.25, 0.25, 0.075, 0.01, 0.2, 0.01, 0.1, 0.05, 0.1, and 0.02.} We tabulate the log of $\kde{\lgJ}{d_\text{kin}}$ for each dSph with a spacing of $0.005\,\text{dex}$ in~$\lgJ$ and use linear interpolation to calculate it for intermediate values of~$\lgJ$.

In the Bayesian framework we simply adopt this posterior from the kinematic tracer data analysis as a prior on~$J$ for our work; in a frequentist context, we must re-interpret this posterior as a likelihood function for~$J$. In either case, it is appropriate to multiply $\kde{\lgJ}{d_\text{kin}}$ with the likelihood for the Fermi-LAT data~(\ref{eq:likelihood}), while noting the different meaning in each statistical context. This results replacing the likelihood with the following expression:\footnote{Technically, this expression is not a likelihood in the Bayesian framework, but we will ignore this rather minor semantic point for the sake of simplicity}
\begin{equation}
	\prob{\data}{m_\chi,\, \sigv; \, \boldsymbol{\beta},\, \mathbf{J}}= \prod_{k} \; \prob{\data_{k}}{m_\chi,\, \sigv; \,\beta_{k},\,\lgJ_k} \; \kde{\lgJ_k}{d_\text{kin}} \, . \label{eq:totallikelihood}
\end{equation}

Besides the problematic case of \textsc{Willman~1}, which may suffer from tidal disruption or non-equilibrium kinematics as mentioned in Sec.~\ref{sec:data}, we point out that some caveats apply with respect to possible systematics in the determination of the $J$-factors. These arise due to the dependence on the halo model, the possibility of non-sphericity~\cite{1603.08046}, or a possible influence of the adopted priors~\cite{1608.07111}. Regarding the effect of triaxiality, for example, it has been shown that the arising systematic uncertainties for the classical dSphs can be about twice as large as the statistical ones~\cite{1407.7822,1504.03309}. Nonetheless, our use of $J$-factors determined by Ref.~\cite{1802.06811} allows us to treat all the dSphs in a uniform way, which is essential to test the consistency of DM signals amongst them.

We also note that our analysis treats dark matter annihilation as a point source of emission from each dwarf. This is a good approximation if $\dd J/\dd \Omega$ in~(\ref{eq:flux}) is more concentrated than the gamma-ray PSF, which is approximately \ang{0.8} at 1~GeV. Treating the DM signal as a point source is corroborated by Ref.~\cite{1503.06209}, which finds no evidence of extended gamma-ray emission in the 35~dSphs they searched. In any case, the possible contribution to~$J$ from dark matter annihilating beyond the PSF scale is typically negligible compared to the uncertainties in the overall $J$-factors. For the example of \rettwo, increasing the integration angle from~\ang{0.5} to~\ang{1}, far beyond our ROI, only increases the $J$-factor by $0.2\,\text{dex}$ while the uncertainty in~$J$ itself is around $1\,\text{dex}$~\cite{1504.03309}. DSph dark matter halos are seldom constrained beyond~\ang{0.5} because of the lack of spectroscopically observed member stars at such large radii. For those classical dwarfs that do allow such measurements, we use the results of Ref.~\cite{1408.0002} to estimate the increase in~$J$ when integrating from~\ang{0.5} to~\ang{1.0}. Only for \textsc{Draco} and \textsc{Sextans} do we find this increase to be potentially significant, though even for these two the median estimate for the increase in~$J$ is smaller than the uncertainty in~$J$ itself. The authors of Ref.~\cite[][Sec.~IV.~F]{1410.2242} quantify the reduction in sensitivity in treating an extended dSph as a point source and find the effect to be small. We therefore proceed by treating each dSph as a point source of gamma~rays.

\subsection{Statistical framework \label{sec:statframework}}
One of the aims of this work is to carry out a detailed comparison of the conclusions that can be obtained by analysing the same data from a Bayesian and a frequentist point of view. We briefly summarize how to perform parameter inference and model comparison/hypothesis testing in the two frameworks.

The Bayesian posterior distribution, for some model parameters~$\boldsymbol{\theta}$, given data~$\data$, is obtained as a normalised product of the prior probability density function~(PDF), $\prior{\boldsymbol{\theta}}$, and the likelihood, $\prob{\data}{\boldsymbol{\theta}}$, via Bayes' theorem~\cite{1763_Bayes}:
\begin{equation}
	\prob{\boldsymbol{\theta}}{\data} = \frac{\prior{\boldsymbol{\theta}}\prob{\data}{\boldsymbol{\theta}}}{\prior{\data}} \, .
\end{equation}

In a frequentist framework, the prior is not defined, and the likelihood is the quantity on which parameter inference is based -- albeit with a different interpretation from the Bayesian posterior (see e.g. Ref.~\cite{1701.01467}).

Since we are usually only interested in summarising inferences on one or two parameters at a time, one needs to eliminate in a suitable manner the parameters that are not the focus of attention. Let $\prob{\boldsymbol{\theta}}{\data}$ be the $n$-dimensional posterior of some model parameters~$\boldsymbol{\theta}=\left(\theta_1, \theta_2, \, \theta_3, \ldots, \theta_n\right)$. Then the Bayesian approach is to \textit{marginalise} over the nuisance parameters. If, e.g., parameters $\theta_1$ and $\theta_2$ are of interest, then the two-dimensional marginalised posterior is given by
\begin{equation}
	\prob{\theta_1, \, \theta_2}{\data} = \int \! \prob{\boldsymbol{\theta}}{\data} \; \dd\theta_3 \ldots \dd\theta_n \, .
\end{equation}
Credible regions~(CRs) for the parameters can be derived by finding regions over which the posterior integrates to a specified probability, using some scheme for determining the integration regions (here, we use highest posterior density regions).

In contrast, the frequentist approach is to \textit{profile} over the nuisance parameters, i.e. to eliminate them from the likelihood by replacing them with their most likely values, $\hat{\theta}_3, \ldots, \hat{\theta}_n$, that maximise the likelihood~$\prob{\data}{\boldsymbol{\theta}}$ given \emph{specific} values of~$\theta_1$ and~$\theta_2$:
\begin{equation}
	L_\text{p}(\theta_1, \, \theta_2) \equiv \prob{\data}{\theta_1, \, \theta_2, \, \hat{\theta}_3(\theta_1, \, \theta_2), \ldots, \hat{\theta}_n(\theta_1, \, \theta_2)} \equiv \sup_{\theta_3, \ldots, \theta_n} \prob{\data}{\boldsymbol{\theta}} \, . \label{eq:proflike}
\end{equation}
The construct~$L_\text{p}$ is called \textit{profile likelihood}, and it maps out the best-fitting solutions for the problem at hand.

Regarding selection of the ``best'' models, the Bayesian answer can again be given using only the posterior probability to determine the degree of belief in a given hypothesis, which will depend on one's prior belief in the hypothesis. Since this is a general statement, it is possible to consider two hypotheses, $H_0$ and $H_1$, consisting of different models and sets of parameters $\boldsymbol{\theta_0}$ and $\boldsymbol{\theta_1}$. The ratio of posterior probabilities, or posterior odds, is then given by
\begin{align}
	\frac{\prob{H_1}{\data}}{\prob{H_0}{\data}} &= \underbrace{\frac{\int \! \prob{\data}{\boldsymbol{\theta_1}, \, H_1}\prob{\boldsymbol{\theta_1}}{H_1} \; \dd \boldsymbol{\theta_1}}{\int \! \prob{\data}{\boldsymbol{\theta_0}, \, H_0}\prob{\boldsymbol{\theta_0}}{H_0} \; \dd \boldsymbol{\theta_0}}}_{\equiv \mathcal{B}_{10}} \, \frac{\prior{H_1}}{\prior{H_0}} \, , \label{eq:probratio}
\end{align}
where $\mathcal{B}_{10}$ is the so-called \textit{Bayes factor} between the two hypotheses ($\mathcal{B}_{10} > 1$ favours hypothesis~$H_1$) and $\prob{\boldsymbol{\theta_0}}{H_0}$ and $\prob{\boldsymbol{\theta_1}}{H_1}$ are the priors for $\boldsymbol{\theta_0}$ and $\boldsymbol{\theta_1}$ under hypotheses $H_0$ and $H_1$, respectively. In the case where we assign equal prior probability to both hypotheses, $\prior{H_1}=\prior{H_0}$, the Bayes factor, $\mathcal{B}_{10}$, is equal to the posterior odds~(\ref{eq:probratio}). The Bayes factor can thus favour either $H_0$ or $H_1$, and it includes an automatic ``Occam's razor'' effect, disfavouring models that have large numbers of parameters that are not required to fit the data~(see Ref.~\cite{0803.4089} for details). From a Bayesian perspective, the best model is the one that balances quality-of-fit (measured by the maximum likelihood value) and predictivity (measured by the inverse of the Occam's factor).

Hypothesis testing from a frequentist perspective is concerned with rejecting the null hypothesis~$H_0$, which states that the effect one is looking for is absent. The null hypothesis is rejected when the probability of obtaining data as ``extreme'' or ``more extreme'' than what has been observed is small under the null hypothesis. This is usually achieved by defining a test statistic (a function of data) and prescription for which values of the test statistic would result in rejecting~$H_0$. An often-used test statistic is the likelihood-ratio,
\begin{equation}
	\Lambda = \frac{\prob{\data}{H_1}}{\prob{\data}{H_0}} \, . \label{eq:likeratio}
\end{equation}

The Neyman-Pearson lemma states that~(\ref{eq:likeratio}) is the optimal test statistic for testing two simple hypotheses, i.e. without nuisance parameters~\cite{NeymanPearson}. If we want to only consider, e.g., two parameters of interest, we define the profile likelihood ratio
\begin{equation}
	\mrm{\Lambda}{p}(\theta_1, \, \theta_2) = \frac{L_\text{p}(\theta_1, \, \theta_2)}{\altprob{\data}{\boldsymbol{\hat{\theta}}}} \, , \label{eq:profilelikeratio}
\end{equation}
where $\boldsymbol{\hat{\theta}}$ denotes the global maximum-likelihood estimator and $\mrm{L}{p}$ is the profile likelihood~(\ref{eq:proflike}). Wilks' theorem now states that, under some regularity conditions, the distribution of $-2\ln \, \mrm{\Lambda}{p}(\theta_1, \, \theta_2)$ is asymptotically $\chi^2$-distributed (with two degrees of freedom)~\cite{Wilks} and~(\ref{eq:profilelikeratio}) can easily be turned into a statistical test to obtain a $p$-value, the probability of the test statistic being more extreme than the observed data (under the null hypothesis). The boundary of the confidence region at confidence level~$\alpha$ is then found by setting the $p$-value equal to~$\alpha$ and determining the corresponding parameter values that bound the region (inside which the $p$-value is larger than $\alpha$). This leads to the familiar prescription that, for example, the 68.27\% or ``$1\sigma$'' confidence interval for one parameter $\theta$ is bounded by values where $2\ln \Lambda_\text{p}(\theta)$ has dropped by one unit from its maximum value.

One of the regularity conditions of Wilks' theorem is that the hypothesis being tested must not lie on the boundary of the parameter space. For considering the null hypothesis of no DM signal, which is equivalent to setting~$\sigv$ to its boundary value~$\sigv = 0$, the regularity condition is not met and Wilks' theorem does not apply.\footnote{In this case, Chernoff's theorem can be used for hypothesis testing instead~\cite{Chernoff}.} There is no guarantee that the ensuing distribution for the test statistic is anywhere near the $\chi^2$~distribution (see e.g. Ref.~\cite{1509.01010}). We demonstrate in Sec.~\ref{sec:coverage} that for small cross sections the distribution of $-2\ln \Lambda_\text{p}$ does indeed deviate from a $\chi^2$. However, the deviation is such that $p$-values (upper tail probabilities) are smaller than what would be expected from a $\chi^2$ distribution. Using Wilks' theorem to perform null hypothesis tests and construct confidence intervals is therefore {\em conservative}. In other words, assuming a $\chi^2$ distribution of the test statistic will yield rejections of the null hypothesis that are less significant, and confidence regions that are larger, than would be obtained with the true distribution of $-2\ln \Lambda_\text{p}$. Section~\ref{sec:coverage} also establishes that at larger cross sections the $\chi^2$ approximation holds to high accuracy. We therefore safely proceed to use Wilks' theorem to construct frequentist confidence regions.

\subsection{Priors}
\begin{table}
	\renewcommand{\arraystretch}{1.05}
	\caption{Prior distributions used in this study. We use two priors on~$\sigv$ and adjust the lower end of the prior on~$m_\chi$ as appropriate for the given annihilation channel. The priors in the bottom part of the table apply to all dwarfs~$k = 1,\ldots, \ndwarfs$ and $\sigv_{-26} \equiv \sigv/\SI{e-26}{\centi\metre^3\second^{-1}}$. Priors on $J$ are kernel density estimates of the posteriors found in Ref.~\cite{1802.06811}.\label{tab:priors}}
	\centering
	\begin{tabularx}{0.63\textwidth}{cccX}
		\toprule
		\textbf{Parameter} & \textbf{Prior type} & \textbf{Channel} & \textbf{Range} \\
		\midrule
		$\left. m_\chi\right/\si{\giga\electronvolt}$ & log-uniform & $\bbb$ & \prrange{4.2}{e4} \\
		& log-uniform & $\tptm$ & \prrange{2}{e4} \\
		$\sigv_{-26}$ & log-uniform & both & \prrange{0.01}{100} \\
		& uniform & both & \prrange{0}{100} \\
		\midrule
		$\lgJ_k$ & KDE & both & \\
		$\beta_{k}$ & uniform & both & \prrange{0}{2} \\
		\bottomrule
	\end{tabularx}
\end{table}

The choices of priors on the model parameters are listed in Table~\ref{tab:priors}. With a log-uniform prior on the WIMP~mass, $m_\chi$, we encode our ignorance of the scale of new physics. Due to energy-momentum conservation, the mass of the outgoing particle sets a natural scale for the lower limit on the $m_\chi$~prior for any given WIMP~annihilation channel.

Note that it is possible to use a more informative prior on~$m_\chi$ by incorporating previous results and theoretical considerations. For example, the absence of evidence for supersymmetry at the Large Hadron Collider could be seen as prior information that favours higher and disfavours lower values of~$m_\chi$. It would also be possible to take into account naturalness of supersymmetric scenarios~\cite[e.g.][]{hep-ph/0601089,0812.0536,1302.6587}. However, additional information is generally only available after specifying an underlying theoretical model for dark matter. We want to avoid this in our phenomenological approach. In section \ref{sec:globalfit} (lower panels of \reffig{fig:globalresults}), we also present Bayesian limits on~$\sigv$, conditioned on~$m_\chi$ (as dashed lines), which are independent of the prior on the mass.

For $\sigv$, a similar rationale could be applied, but there are other choices of prior which have been used in the literature -- such as a prior that is uniform in $\sigv$ itself or one that is proportional to $\sigv^{-1/2}$~\cite{1203.6731,1503.02641}.

Choosing a prior proportional to $\sigv^{-1/2}$ can be reasoned for in this context since the Jeffreys~prior\footnote{This is the unique choice of prior (for a given likelihood) that leads to a posterior that is invariant under an arbitrary parameter transformation.} for the rate~$\lambda$ in a Poisson likelihood is proportional to $\lambda^{-1/2}$~\cite{Jeffreys1946}. This is however the Jeffreys~prior for the background-free case, and is also the so-called \textit{reference prior} for this case.

A prior that is uniform in the log of $\sigv$ requires both a lower and an upper cut-off to be proper. This choice gives equal \textit{a~priori} weight to all orders of magnitude in $\sigv$, which reflects indifference as to the scale of the cross~section. It has however the disadvantage that Bayesian upper limits on the cross~section (in the absence of a detection) and the model selection outcome both depend explicitly (if weakly) on the chosen lower cut-off, which is somewhat arbitrary (we justify our choice below, using an argument based on the expected observational sensitivity).

Finally, one can choose a prior that is uniform in $\sigv$ itself, which is bounded from below by~0 but still requires an upper cut-off to be proper. When the quantity being constrained may \textit{a~priori} be compatible with~0, i.e. when searching for a signal that could be absent, this prior has the advantage of including that possibility~\cite{1992_Loredo}. The disadvantage, however, is that the upper cut-off effectively sets a scale with higher \textit{a~priori} weight for the parameter in question.

One can argue that the natural scale for $\sigv$, under the DM hypothesis, is of the order of the thermal cross~section, which is a few times \SI{e-26}{\centi\metre^3\second^{-1}} for masses above \SI{10}{\GeV} and only mildly depends on the WIMP mass~\cite{1204.3622}. If~$\sigv$ is expressed in those units, then a choice of prior that is uniform in $\sigv$ correctly expresses our theoretical expectation that its value should be close to that order of magnitude (if non-zero) and reproduces the observed DM~density, $\Omega_\text{DM} h^2 \approx 0.12$~\cite{1807.06209}.

Comparing the results for different choices of priors is essential in a Bayesian framework. Since Ref.~\cite{1503.02641} found that the limits derived from a prior uniform in~$\sigv$ and the $\sigv^{-1/2}$ prior are similar to within a factor of~1.5 (in what they called a ``hybrid Bayesian analysis''), we will adopt two priors, which are expected to bracket possible reasonable prior choices, namely a prior uniform in the log of $\sigv$ and one that is uniform in $\sigv$ itself (both with appropriately chosen cut-off values).

The choice of lower cut-off is trivial for the prior uniform in $\sigv$, as the lower cut-off is naturally $\sigv = 0$. It is far more subtle for the log-uniform prior: below a certain value for $\sigv$, the likelihood becomes flat, since the WIMP signal falls to zero, and hence the posterior follows the shape of the prior in this region. This is in contrast with the region of larger and larger~$\sigv$, where the likelihood drops rapidly towards zero and the posterior is driven by the data. This means that both the upper limit on the cross~section from the posterior distribution and the model selection result depend on the chosen lower bound for the $\sigv$ prior. We therefore need a physical argument to set it, lest the result becomes arbitrary.

In principle, one could use theoretical constraints on models with a DM candidate (e.g. supersymmetry) to inform the lower cut-off on~$\sigv$. Unfortunately, the details depend on the spectrum of the supersymmetric particles. It has been shown for concrete realisations of supersymmetric models that possible and experimentally allowed cross~sections can be as low as $\sigv \sim \SI{e-30}{\cm^3/s}$~\cite[e.g.][]{1705.07917,1705.07935}. As a consequence, the corresponding values of~$\sigv$ are several orders of magnitude below the thermal cross~section as well as the sensitivity of existing and even planned future experiments~\cite{1705.07917}.

Instead, we adopt a variation of the argument presented in Ref.~\cite{astro-ph/0504022}, using the expected signal to define a criterion by which the model with a DM signal becomes indistinguishable from the background-only model. Specifically, we compute the value of $\sigv$ for which the DM signal -- in all energy bins for every dwarf and channel -- is less than one photon. To obtain this estimate, we fix the $J$-factors to the values at the modes of their distributions. For the \rettwo only analysis, the minimum value (for \pseven and \peight; both channels) is $\sigv_{-26} \approx \num{0.04}$, while for all dwarfs (\peight and global fit data; both channels) the minimum value is $\sigv_{-26} \approx \num{0.008}$ (for \textsc{Ursa Major~II}). We therefore deem all points in parameter space for $\sigv_{-26} < \num{0.01}$ to be empirically indistinguishable from a background-only model. We could have even used a larger threshold, since in practice uncertainty in the background model means that even signal models with a larger number of photons become effectively unidentifiable. Our choice is however conservative, in that it gives a slightly larger prior parameter space to the signal model, therefore disfavouring it via the Occam's razor effect against the background-only alternative.

For the upper cut-off in both priors, a prior uniform in the log of~$\sigv$ and uniform in~$\sigv$ itself, we make use of the argument for the thermal cross~section presented before: if the DM in the dSphs is expected to be mostly constituted by WIMPs, the natural scale for the cross~section is of the order of a few times \SI{e-26}{\centi\metre^3\second^{-1}}~\cite{1204.3622}. Since $\Omega_\text{DM} h^2 \propto \sigv^{-1}$, a WIMP with $\sigv_{-26} > 100$ is expected to contribute less than a few percent to the DM in the dSph, thus making it unviable as the~DM and as a source of gamma~rays. We therefore use $\sigv_{-26} = 100$ as an upper cut-off.

Regarding the background normalisation~$\beta_k$ for dwarf~$k$, the reference scale is~$\beta_k = 1$, since this corresponds to the value obtained in an all-sky fit. The natural lower cut-offs are at~$\beta_k=0$. We therefore choose a uniform prior around the reference value, allowing for a rather conservative upwards deviation up to $\beta_k = 2$, which we adopt as the upper cut-off value. For the $J$-factor of dwarf~$k$, the prior on $\lgJ_k$ is the KDE~approximation to the kinematic data analysis result, as explained in Sec.~\ref{sec:jfactors}.

The choice of priors is important for the outcome of the Bayesian model comparison, which always depends on it (differently from parameter inference, where the posterior is asymptotically independent of the prior). This is because the strength of the Occam's razor effect is controlled by the relative volume enclosed by the support of the likelihood vs that of the prior. Therefore, particular attention must be paid to the prior selection in order to obtain interpretable results with Bayesian model comparison.

Firstly, the Savage-Dickey density ratio shows that priors on parameters that are common between the background only model and the background-plus-signal model do not influence the outcome of model selection between them~\cite{astro-ph/0504022}. Therefore, the only priors we need to be concerned about are those on the WIMP mass and cross~section.

Secondly, the WIMP mass is unconstrained by the dSph data when all other parameters have been marginalised out. Therefore, the prior and posterior volume on the mass are almost identical (with any reasonable choice of prior) and the model selection outcome does not depend on the prior choice on the mass. We are thus left with only having to worry about plausible choices for the prior on the cross~section.

Regarding~$\sigv$, we have argued above that both priors, uniform in the log of~$\sigv$ and uniform in~$\sigv$ itself, are plausible choices. However, such priors must be proper, and the scale of the cut-offs will impact the model selection result. Fortunately, the dependency of the Bayes factor is only logarithmic in the chosen cut-off scale, hence relatively weak, given that the Jeffreys' scale for interpretation of the model comparison result is also logarithmic.

\subsection{Coverage study \label{sec:coverage}}
We introduced the background normalisation parameters, $\beta_k$, to allow for deviations in the background model. However, a possible concern in allowing such increased flexibility is that a dark matter signal in the data might be erroneously absorbed by these rescaling parameters, yielding no detection when there should be one. Conversely, our cross section upper limits might be too stringent if the background model is so flexible that it absorbs any upward statistical fluctuation in the data, thus leaving less room for a dark matter signal.

We address both questions by directly checking the coverage of our frequentist confidence intervals for a single dwarf \textsc{Ursa Major~II}, which has the largest $J$-factor. We argue that if a single dwarf passes the coverage test, then there is no reason to believe that the combined likelihood from all 27 of them, which uses more data, should be any different.

For each possible set of true values for $\beta_\textsc{UrMaII}$ and $\sigv$, we generate $10^4$~mock data sets by drawing data from the gamma-ray likelihood in Eq.~\eqref{eq:likelihood}. Then, for each data set we apply our frequentist procedure to identify whether the resulting $1\sigma$ and $2\sigma$ confidence intervals on $\sigv$ contain the true cross section. The coverage is defined as the fraction of mock data sets where the true value of $\sigv$ is contained within the confidence interval. Ideally, the coverage of $1\sigma$ and $2\sigma$ intervals should be 68.3\% and 95.4\%.

Specifically, for each mock data set we find the maximum likelihood value, $\hat{L}_1$, with both $\sigv$ and $\beta_\textsc{UrMaII}$ free to vary. We also determine the maximum likelihood value, $\hat{L}_2$, when $\sigv$ is set to its true value and $\beta_\textsc{UrMaII}$ is left free~(in all cases restricting $\beta_\textsc{UrMaII}$ to the range $[0,2]$). The true value of $\sigv$ is contained in the $1\sigma$ (68\%) or $2\sigma$ (95\%) confidence interval if $\ln(\hat{L}_2/\hat{L}_1) < 1$ or $\ln(\hat{L}_2/\hat{L}_1) < 4$, respectively. In all cases we fix $m_\chi$ and $J$ to fiducial values.

\begin{figure}[t]
	\centering
		\includegraphics[width=0.48\textwidth]{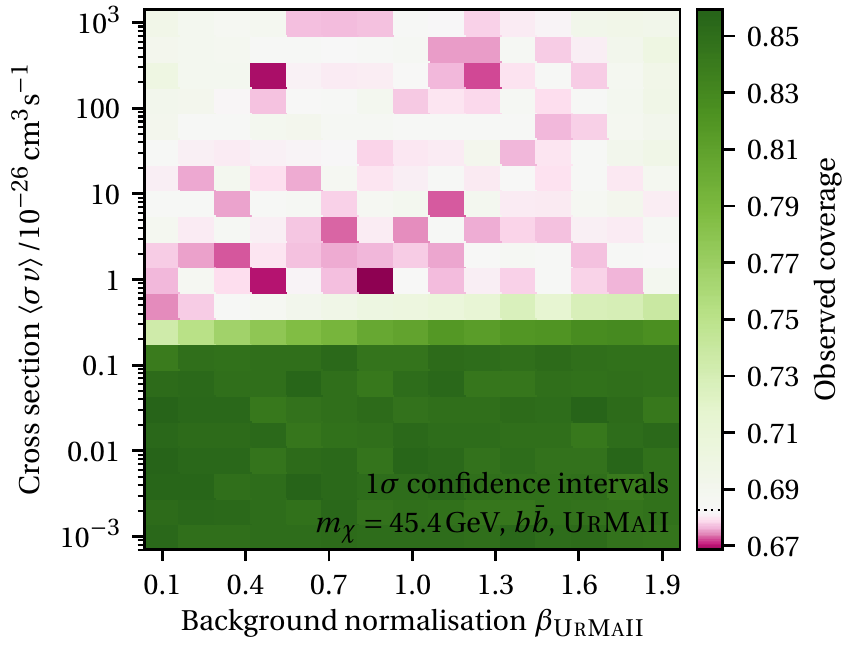}
		\hfil
		\includegraphics[width=0.48\textwidth]{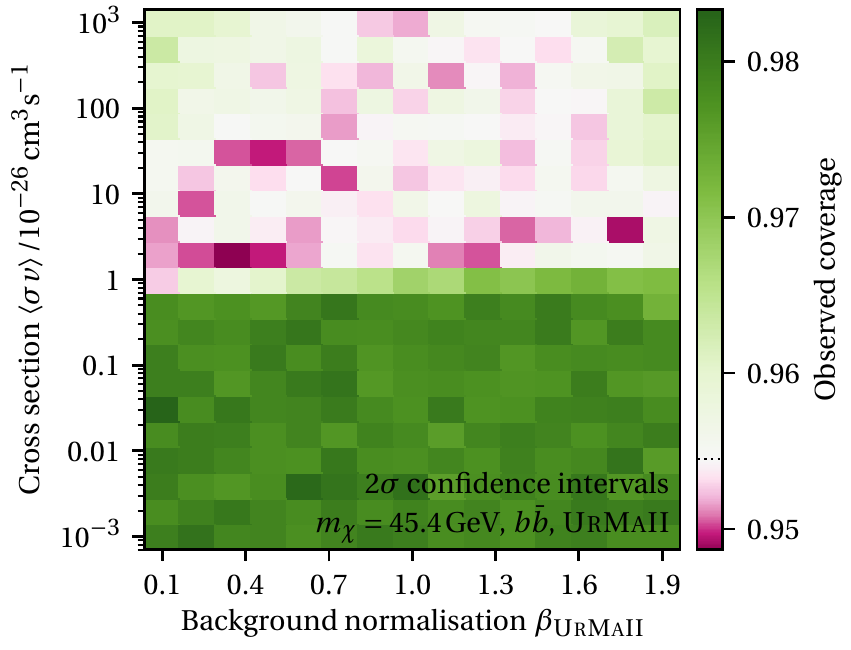}
	\caption{Simulated data for \textsc{Ursa Major~II}, demonstrating the coverage of our frequentist intervals for $\sigv$ as a function of the \emph{true} cross section $\sigv$ and \emph{true} background normalisation parameter $\beta_\textsc{UrMaII}$. \textit{Left:} Coverage of nominal ``$1\sigma$'' confidence intervals, which ideally should contain the true $\sigv$ 68.3\% of the time~(white colour, value denoted by a dotted line in the colour bar). \textit{Right:} Coverage of nominal ``$2\sigma$'' intervals, which ideally should contain the true $\sigv$ 95.4\% of the time. Green parameter values overcover (i.e., confidence intervals are conservative), while pink parameter values undercover, albeit very slightly (less than 1\% under-coverage everywhere). \label{fig:coverage2D}}
\end{figure}

We choose 21~cross sections between $10^{-29}$ and \SI{e-23}{\cm^3/\s}, 15 values of $\beta_\textsc{UrMaII}$ between 0.1 and 1.9, and 4 dark masses ($m_\chi$= 10.1, 45.4, 209, and 1097 GeV) and perform the coverage test for all 1,260 possible combinations of these parameters (all in all, 12 million mock data sets). Figure~\ref{fig:coverage2D} shows the resulting coverage for a dark matter mass $m_\chi=45.4$ GeV (other choices of mass give a similar picture). Each rectangle in the figure corresponds to $10^4$ data sets where the true values of $\sigv$ and $\beta_\textsc{UrMaII}$ are those at the centre of the rectangle. We observe over-coverage for small values of $\sigv$ (i.e.\ conservative limits), and essentially exact coverage once $\sigv$ increases beyond a threshold. Coverage holds for every possible value of $\beta_\textsc{UrMaII}$ between 0.1 and 1.9. This test not only establishes coverage for different true values of the cross section, but also for different possibilities for the background normalisation.

\begin{figure}[t]
	\centering
		\includegraphics[width=1.0\textwidth]{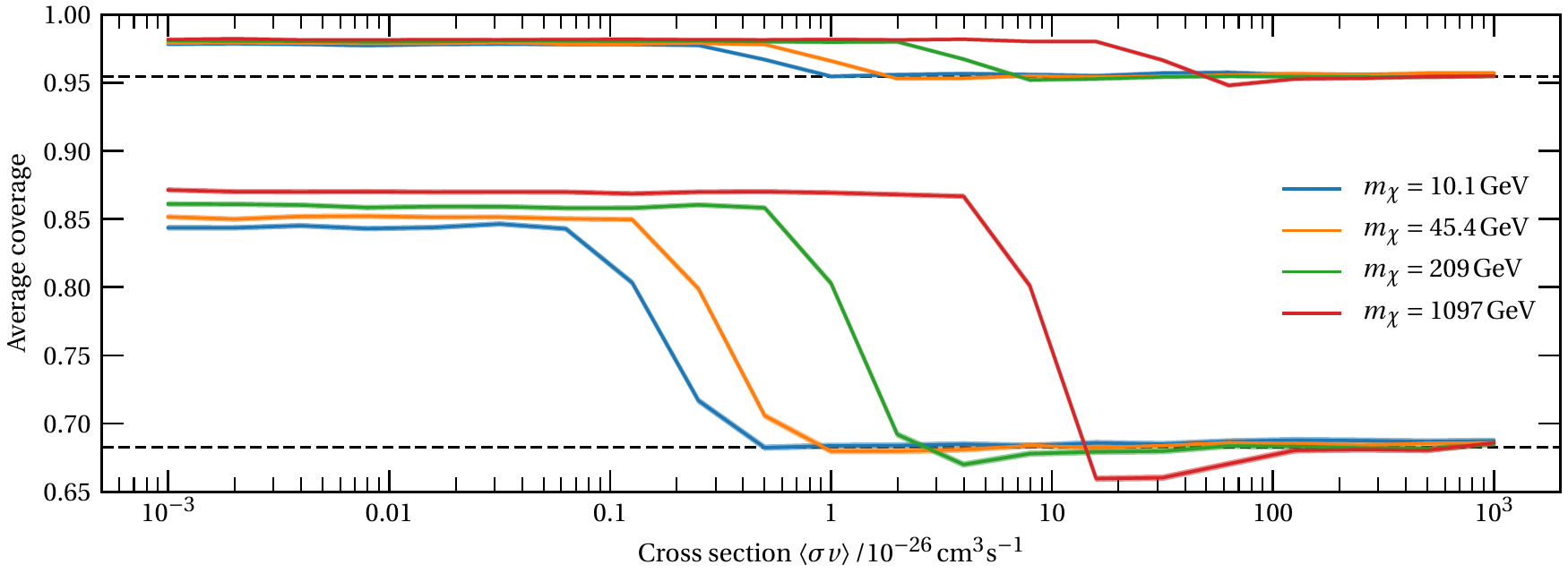}
	\caption{Coverage of confidence intervals for $\sigv$, averaged over the true value of the background normalisation parameter $\beta_\textsc{UrMaII}$, as a function of true cross section for several values of dark matter mass, $m_\chi$. The upper and lower sets of lines show $1\sigma$ and $2\sigma$ confdience intervals. Shaded bands around the lines (barely visible) show the standard deviation of our estimate of the coverage. Horizontal dashed lines are at 0.683 and 0.954, the values of exact coverage. \label{fig:coverage1D}}
\end{figure}

This is made clearer in Fig.~\ref{fig:coverage1D}, which shows the observed coverage as a function of $\sigv$, averaged over values of the true~(simulated) $\beta_\textsc{UrMaII}$, for each choice of mass we tested (narrow shaded bands around each line give the standard deviation of our estimate of the coverage). The 15 values of $\beta_\textsc{UrMaII} \in [0.1, 1.9]$ are weighted equally in the average. Horizontal dashed lines are at 0.683 and 0.954, the values of exact coverage. We can see that the nominal $1\sigma$~limits overcover (i.e., are conservative) for $\sigv$ values smaller than \SI{3e-27}{\cm^3/\s} (for $m_\chi=\SI{10}{\GeV}$) or smaller than \SI{e-25}{\cm^3/\s} (for $m_\chi=\SI{1096}{\GeV}$) (depending on the mass), which is the range of cross~section of interest to us. For larger cross sections (which are irrelevant for our purposes), coverage is close to exact, with a possible small amount of undercoverage for the largest mass value we tested.

The picture for the $2\sigma$~limits is similar, exhibiting overcoverage (i.e., conservatism) for $\sigv$ values smaller than \SI{e-26}{\cm^3/\s} (for $m_\chi=\SI{10}{\GeV}$) and smaller than \SI{e-24}{\cm^3/\s} (for $m_\chi=\SI{1096}{\GeV}$), with exact coverage above these values.

In summary, we conclude that the coverage properties of our frequentist intervals are excellent.

\section{Results and discussion}\label{sec:results}
In this section, we present and discuss the results of the~\rettwo analysis as well as the global analysis involving all \ndwarfs~dSphs. We use various algorithms\footnote{We make use of the \textsf{ScannerBit}~\cite{1705.07959} interface for those software packages, which is a part of \textsf{GAMBIT}~\cite{gambit}.} for the different computational tasks (integration, sampling, maximization). We use \mn~\cite{0704.3704,0809.3437,1306.2144} for calculating the Bayesian evidence. \mn is a nested sampling algorithm~\cite{2004_Skilling}, closing in on the regions of highest likelihood in nested shells. In \mn, these shells are approximated with an ellipsoidal decomposition, and contain sets of live points that are updated in each iteration step by replacing the point with the lowest likelihood by a new one from the prior distribution under the constraint that is has a larger likelihood value.

We use \tw~\cite{1705.07959} for sampling posterior distributions, which is an ensemble Markov Chain Monte Carlo algorithm based on Ref.~\cite{2010_Christen}. It consists of a fixed number of chains, one of which is advanced at each iteration. This selection is random and the proposal distribution for the advancement, which is selected from a pool of different ``moves'', depends on the remaining chains.

Finally, \diver~\cite{1705.07959} is used to map out the profile likelihood (which requires dedicated tools, since the typical Bayesian sampling of the posterior offers insufficient resolution for profile likelihood mapping~\cite{1101.3296}). It is a differential evolution algorithm~\cite{1995_Storn} and consists of a ``population'' of parameter points, whose parameter values are its ``genes''. The population evolves over time via mutation and crossover (of ``genes''), and selection (of the ``fittest individuals'', where fitness is measured by the log-likelihood value), hence mimicking the process of natural selection. This heuristically aims at achieving the highest possible ``fitness'' amongst some of the parameter points, i.e. the maximum likelihood value.

\subsection{Analysis of Reticulum~II}\label{sec:results_ret2}
First, we investigate the WIMP~parameter space of mass, $m_\chi$, and thermally-averaged annihilation cross~section, $\sigv$, using only the \rettwo data described in Sec.~\ref{sec:data}. The additional nuisance parameters for \rettwo are therefore the background scaling, $\mrm{\beta}{\rettwo}$, and the $J$-factor, $\log_{10}\mrm{J}{\rettwo}$. Since this part of the study is mainly to illustrate our methodology and since qualitative conclusions from the $\tptm$ and $\bbb$~channels are similar, we restrict our \rettwo analysis to the $\tptm$ channel only.

The non-default settings for \diver (\texttt{NP}: \num{5e4}, \texttt{convthresh}: \num{e-5}, \texttt{jDE}: true, \texttt{lambdajDE}: false), \mn (\texttt{nlive}: \num{2e4}, \texttt{tol}: \num{e-4}), and \tw (\texttt{sqrtR}: \num{1.001} with 528~\textsf{MPI} processes) were informed by a previous study using these algorithms~\cite{1705.07959}.

\begin{figure}[t]
	\centering
	{
		\includegraphics[width=0.49\textwidth]{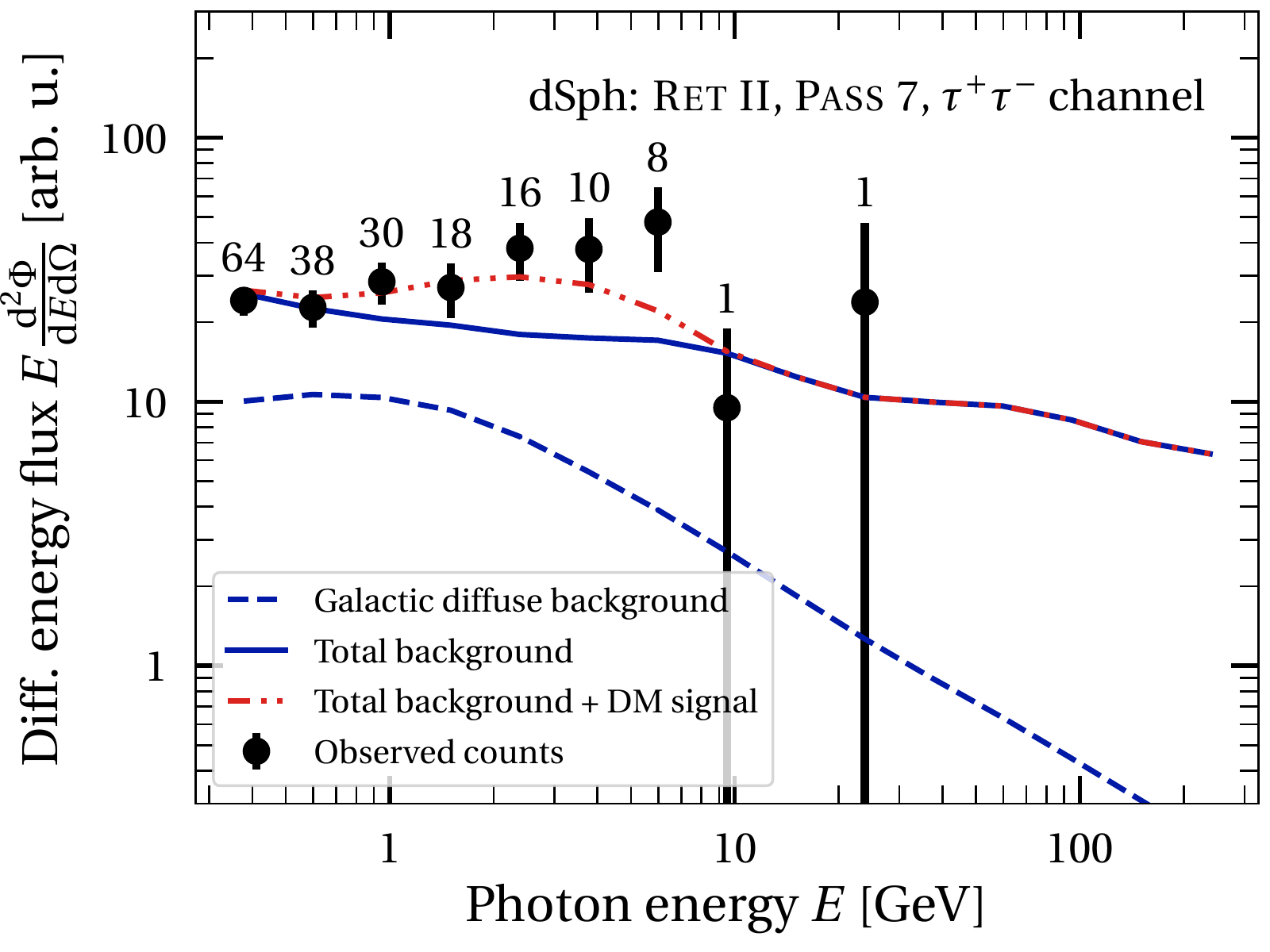}
		\hfill
		\includegraphics[width=0.49\textwidth]{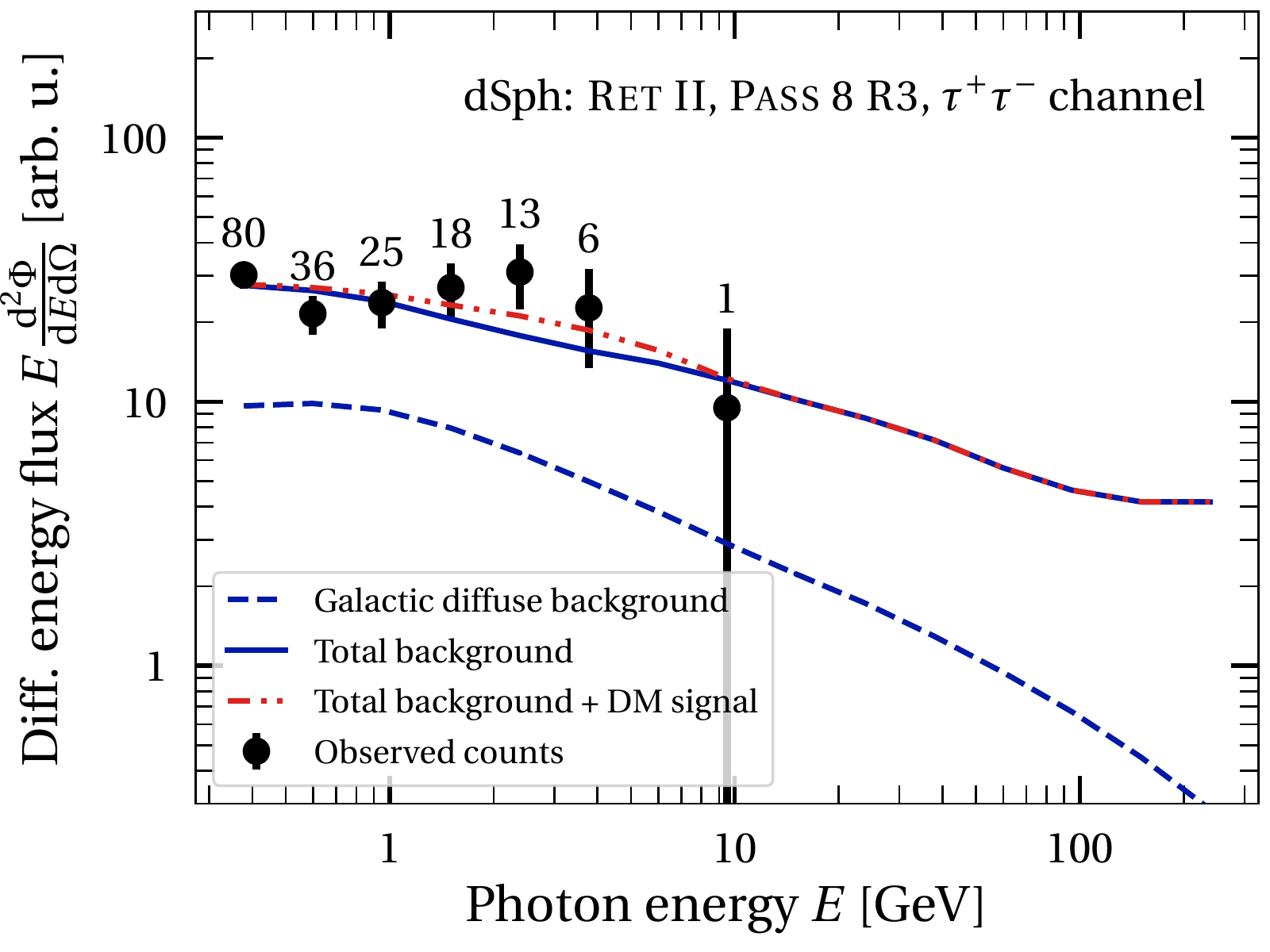}
	}
	\caption{Spectra for 6.5\,years of \pseven (\textit{left}) and \peight (\textit{right}) data for \rettwo. We show the observed counts~(black circles and numbers) with Poisson error bars, as well as the backgrounds~(blue lines) and background plus DM signal~(red lines), according to the best-fit parameters in the $\tptm$ channel. The best-fit DM mass is $m_\chi = \SI{13.3}{\GeV}$ (left panel) and $m_\chi = \SI{14.2}{\GeV}$ (right panel). \label{fig:spectra:p7vsp8}}
\end{figure}
Since the number of weeks for our data selection is the same as in Ref.~\cite{1503.02320}, we expect to find an indication for a signal using \pseven data. This indeed is confirmed by \reffig{fig:spectra:p7vsp8}, which shows \pseven and \peightrthree data for \rettwo together with the best-fit spectra from the fit that will be discussed later in this section (red lines). In the energy region around a few~GeV, there appears to be an excess above the fitted background (blue lines) for \pseven, which is less prominent for \peight. This energy region is therefore able to accommodate an additional signal contribution from DM annihilation. The number of ``excess photons'' amounts to about 26~photons in \pseven\ \updated{and 7~photons in \peight across the whole energy range considered.}

The lowest energy bins, on the other hand, place the strongest constraints on the background normalisation, due to the high number of observed counts in them. This is an important consideration for any analysis that simultaneously fits background and signal contributions. The lowest energy bin, i.e. photons with energies around~\SI{0.4}{\GeV}, is more important in \peight than in \pseven, \updated{given that there are 16~additional photons in \peight.} However, the energy bins in the right half of the bump contain fewer photons in \peight compared to \pseven.

\begin{figure}[t]
	\centering
	{
		\includegraphics[width=0.45\textwidth]{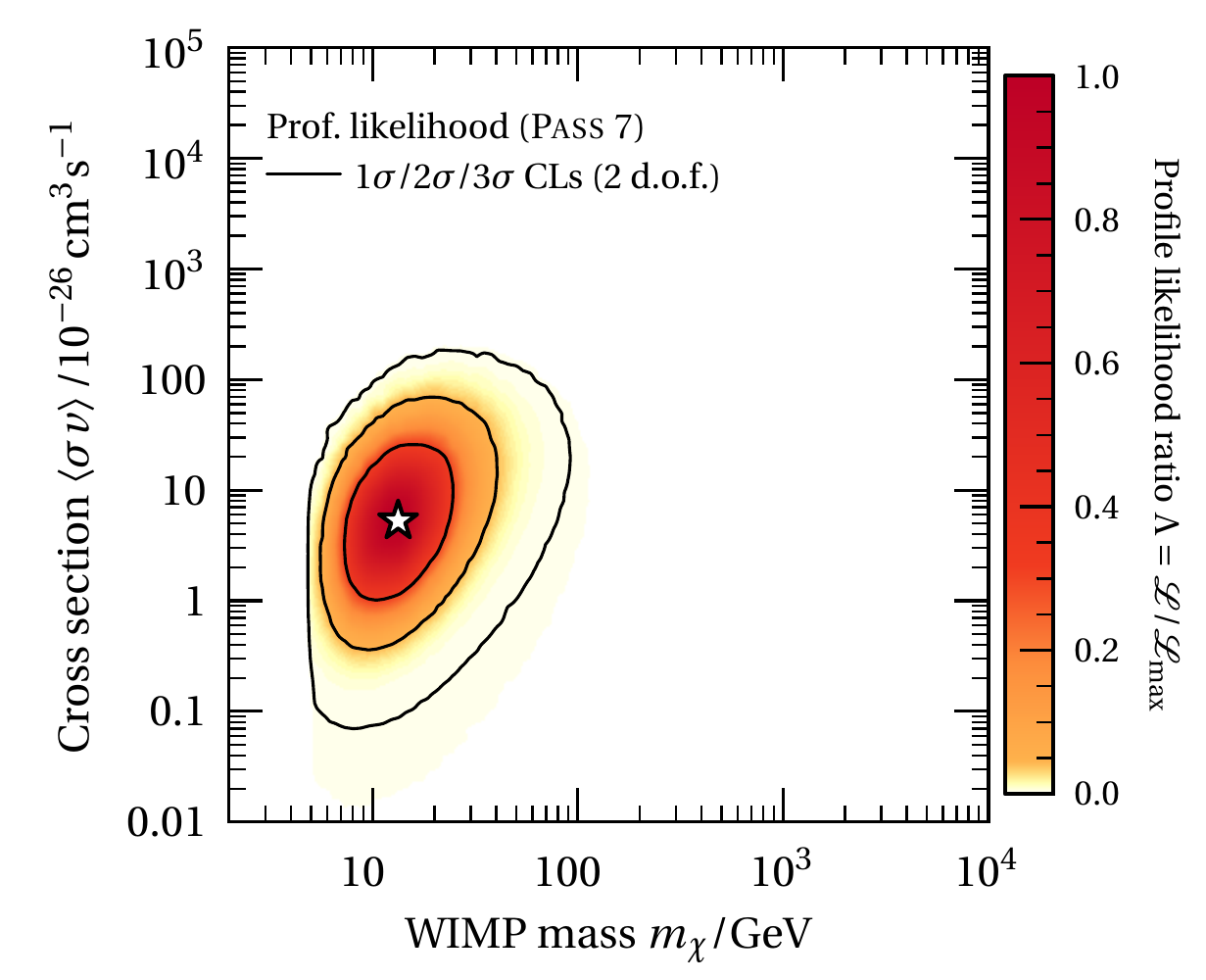}
		\hfil
		\includegraphics[width=0.45\textwidth]{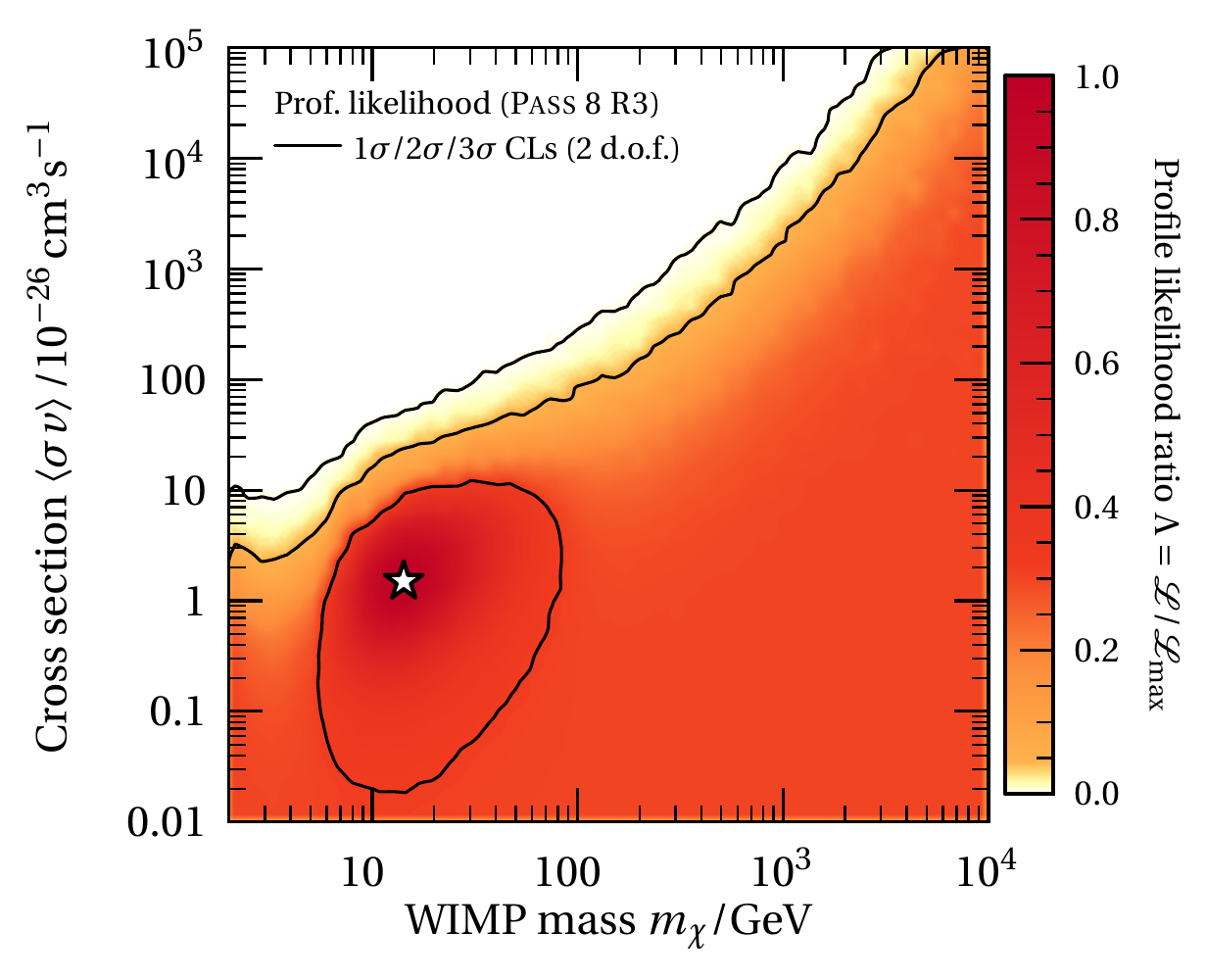}
	}
	{
		\includegraphics[width=0.45\textwidth]{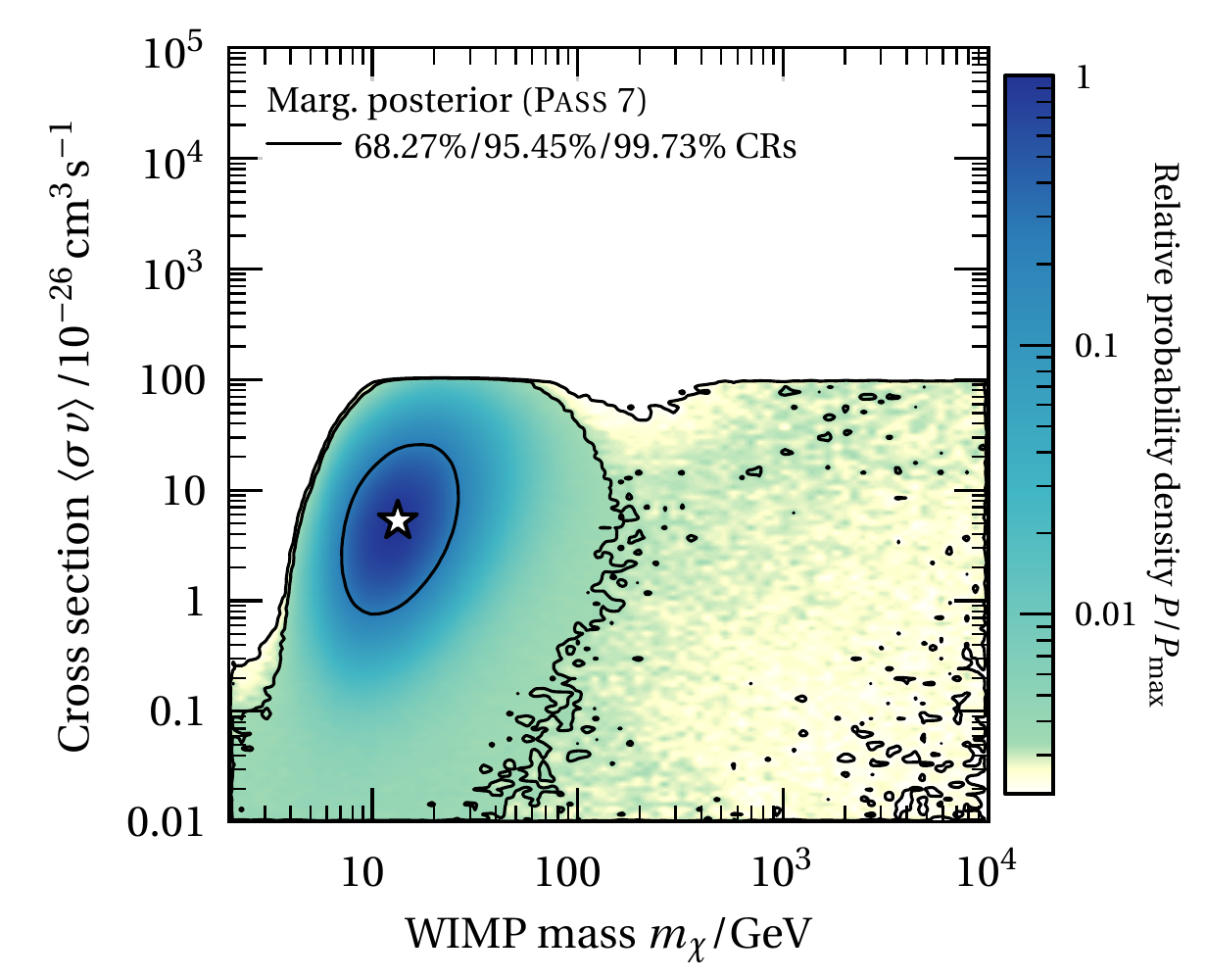}
		\hfil
		\includegraphics[width=0.45\textwidth]{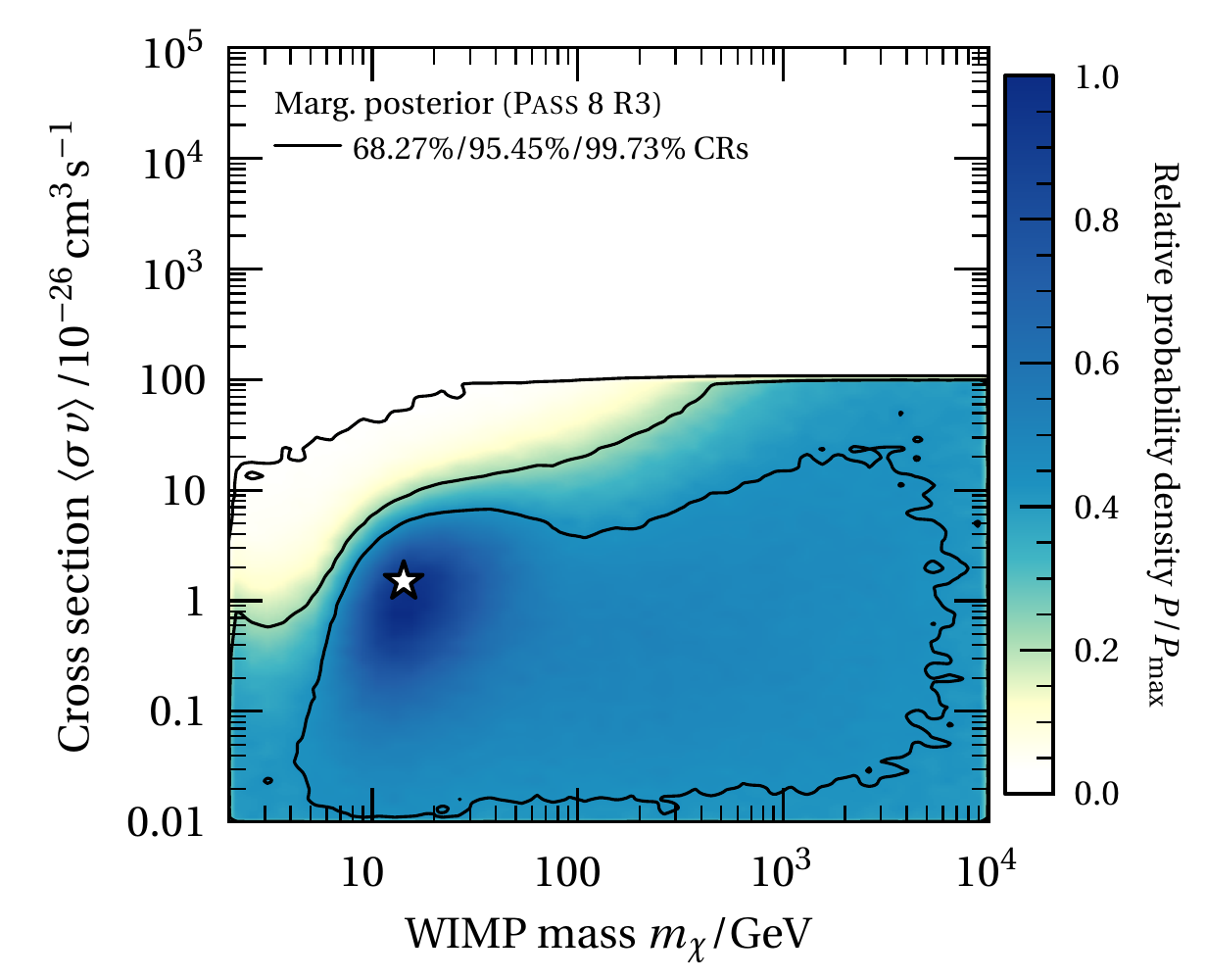}
	}
	\caption{Constraints on WIMP parameters from 6.5\,years of \rettwo data. We show profile likelihoods~(\textit{top}) and marginal posteriors~(\textit{bottom}), using \pseven~(\textit{left}) and \peight~(\textit{right}) data for the $\tptm$ channel and a log-uniform prior on~$\sigv$. The star denotes the best-fit point. We show two-sided CLs and highest posterior density CRs for profile likelihoods and posteriors, respectively. Note that the posteriors are restricted to the region $\sigv_{-26} < 100$ due to the prior range on~$\sigv$.\label{fig:ret2:p7vsp8}}
\end{figure}
In \reffig{fig:ret2:p7vsp8}, we show profile likelihoods (top panels) and the marginal posteriors (bottom panels) for the WIMP~mass and cross~section, where the nuisance parameters $\mrm{\beta}{\rettwo}$ and $\log_{10}\mrm{J}{\rettwo}$ have been profiled out and marginalised over, respectively.\footnote{We make use the software \textsf{pippi}~\cite{1206.2245} for plotting the marginalised posterior distributions and profile likelihoods.} The profile likelihood for \pseven (top left panel) shows a stronger than $3\sigma$ preference for a DM~signal. Such a preference is reduced to $1\sigma$ when using $\peight$ data (top right panel).

As we saw in \reffig{fig:spectra:p7vsp8}, the energy region of the putative excess results in a preferred value for the WIMP mass~$m_\chi$. The inclusion of the likelihood for~$\mrm{J}{\rettwo}$, on the other hand, allows for a direct inference on~$\sigv$, which includes both the uncertainty in the $J$-factor measurement and that from the Poisson likelihood.

The best-fit parameters of the WIMP~properties for the $\tptm$~channel in \pseven (\peight) data are $\widehat{m}_\chi = \SI{13.3}{\GeV}$ and $\widehat{\sigv}_{-26} = \num{5.48}$ \updated{($\widehat{m}_\chi = \SI{14.2}{\GeV}$ and $\widehat{\sigv}_{-26} = \num{1.55}$)}. The best-fit mass values are similar for \pseven and \peight because the excess is around the same energy region for both data sets. Since the number of ``excess photons'' in this region is higher for \pseven, it is also not surprising that the best-fit parameter for~$\sigv$ reflects this.

While the statistical interpretation of the posterior credible regions~(CRs) in a Bayesian analysis is different from the confidence levels~(CLs) in a frequentist understanding, the Bayesian posteriors in the bottom panels of \reffig{fig:ret2:p7vsp8} paint a qualitatively similar picture in terms of the statistical conclusions. The \pseven marginal posterior distribution shows a very slight presence of an additional signal to the background; the 68.27\%~CR (corresponding to $1\sigma$ in the Gaussian case) is a closed contour, but the 95.45\%~CR (corresponding to $2\sigma$ in the Gaussian case) does not exclude the lower prior cut-off. For \peight data (bottom right panel), the 68.27\%~CR in the same plot extends to most of the parameter space. This region becomes difficult to sample because the posterior distribution flattens out and thus the sampled posterior looks ``patchy'' in the plot. Comparing the top (frequentist) and bottom (Bayesian) panels, we notice that the Bayesian inference tends to be more conservative, as the marginalisation over the nuisance parameters typically produces wider CRs when compared to profiled likelihood CLs in cases where there is significant ``volume effect'' in the hidden dimensions~(see e.g. Ref.~\cite{0809.3792}). In light of this result for the Bayesian parameter inference, we do not expect to find Bayesian evidence for the DM~signal hypothesis from either data set when we perform a Bayesian model comparison later in this section.

\begin{figure}[t]
	\centering
	\setlength\tabcolsep{1pt}
	\begin{tabular}{llll}
		\includegraphics[width=0.21062\textwidth]{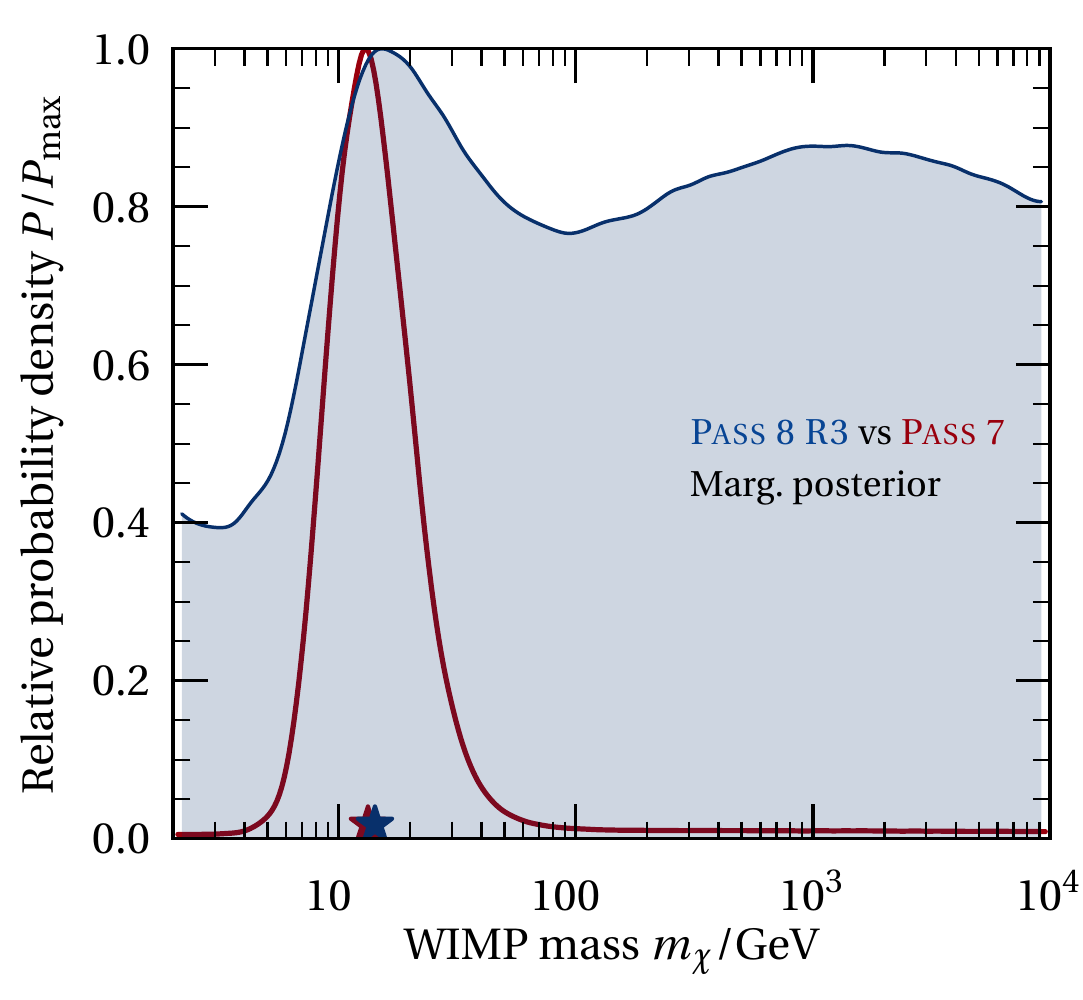}
		& & & \\
		\includegraphics[width=0.23934\textwidth]{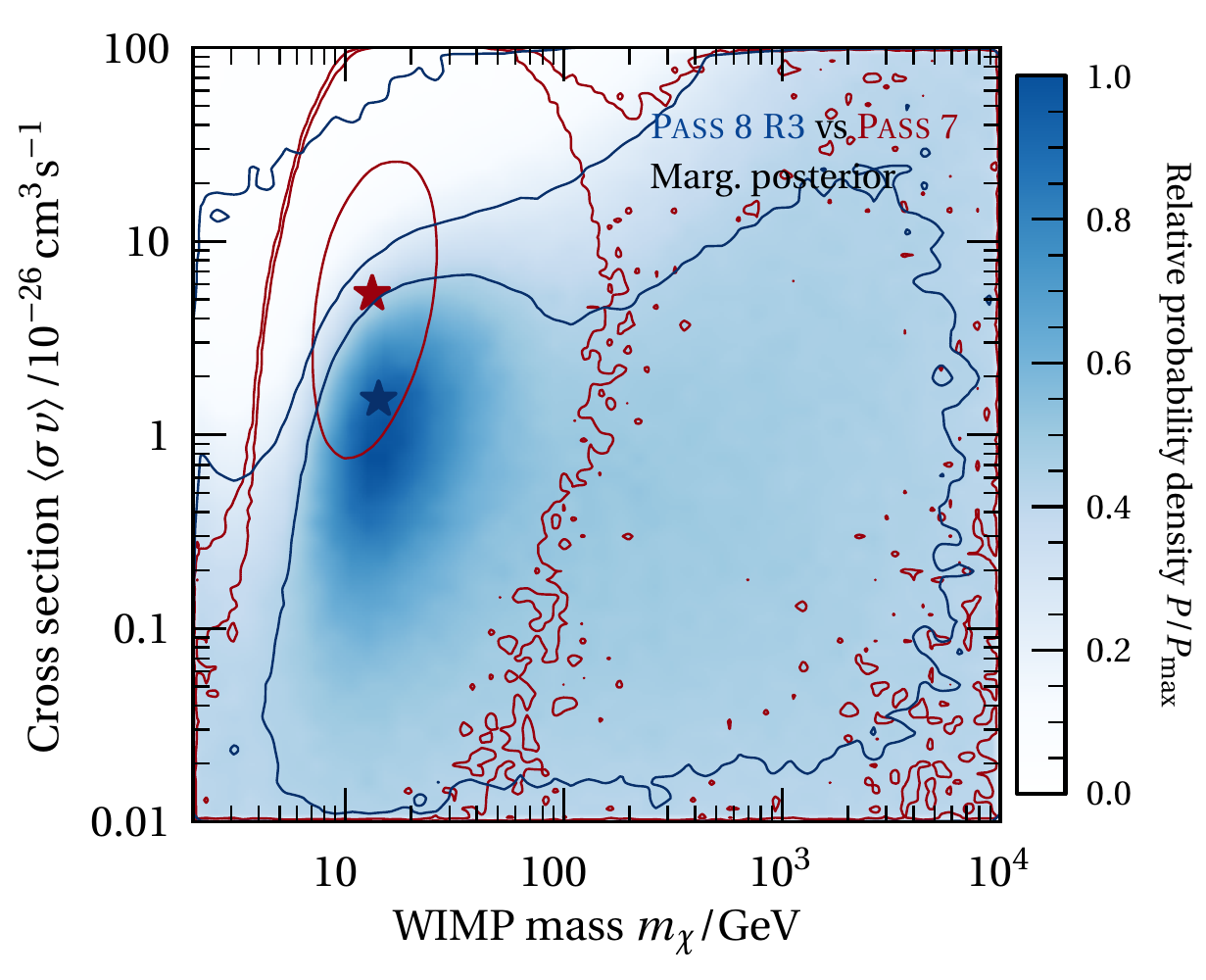}
		& \includegraphics[width=0.21062\textwidth]{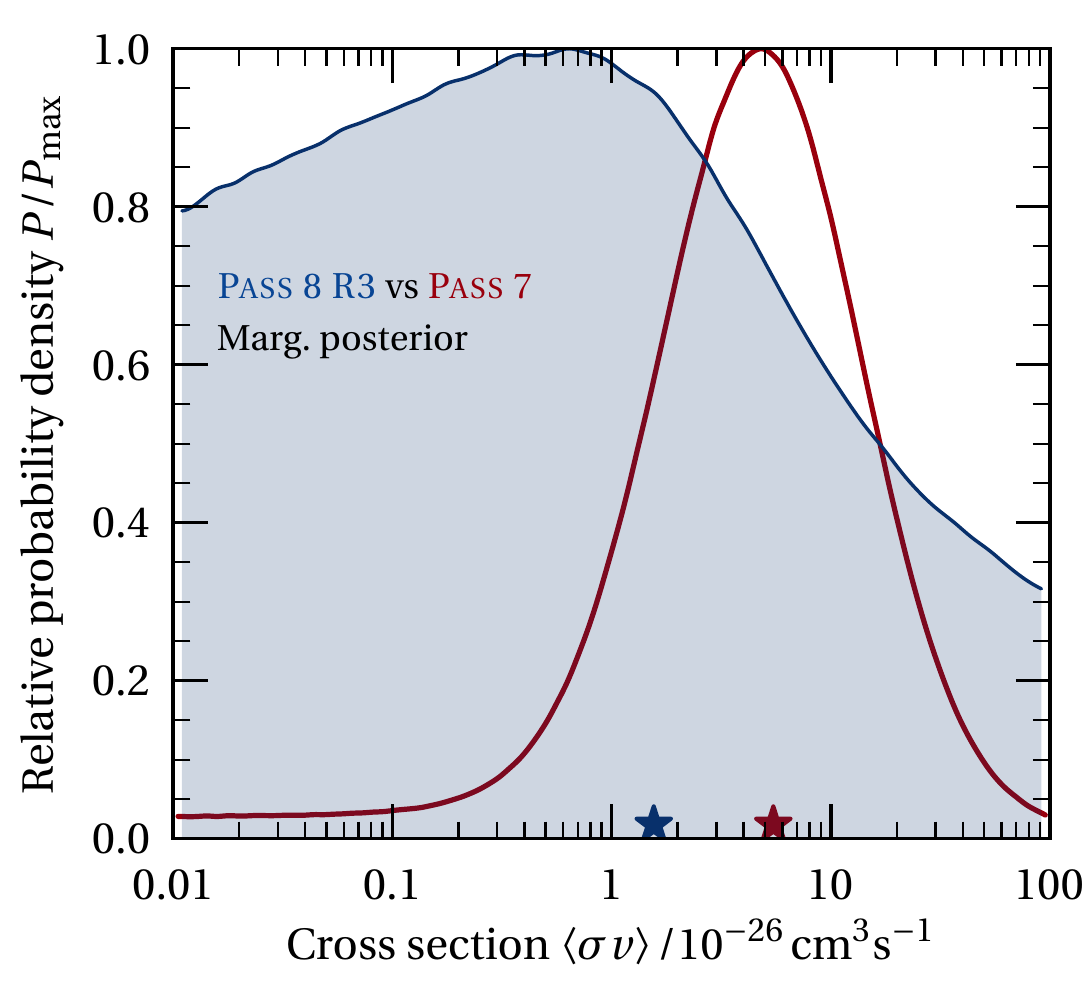}
		& & \\
		\includegraphics[width=0.23934\textwidth]{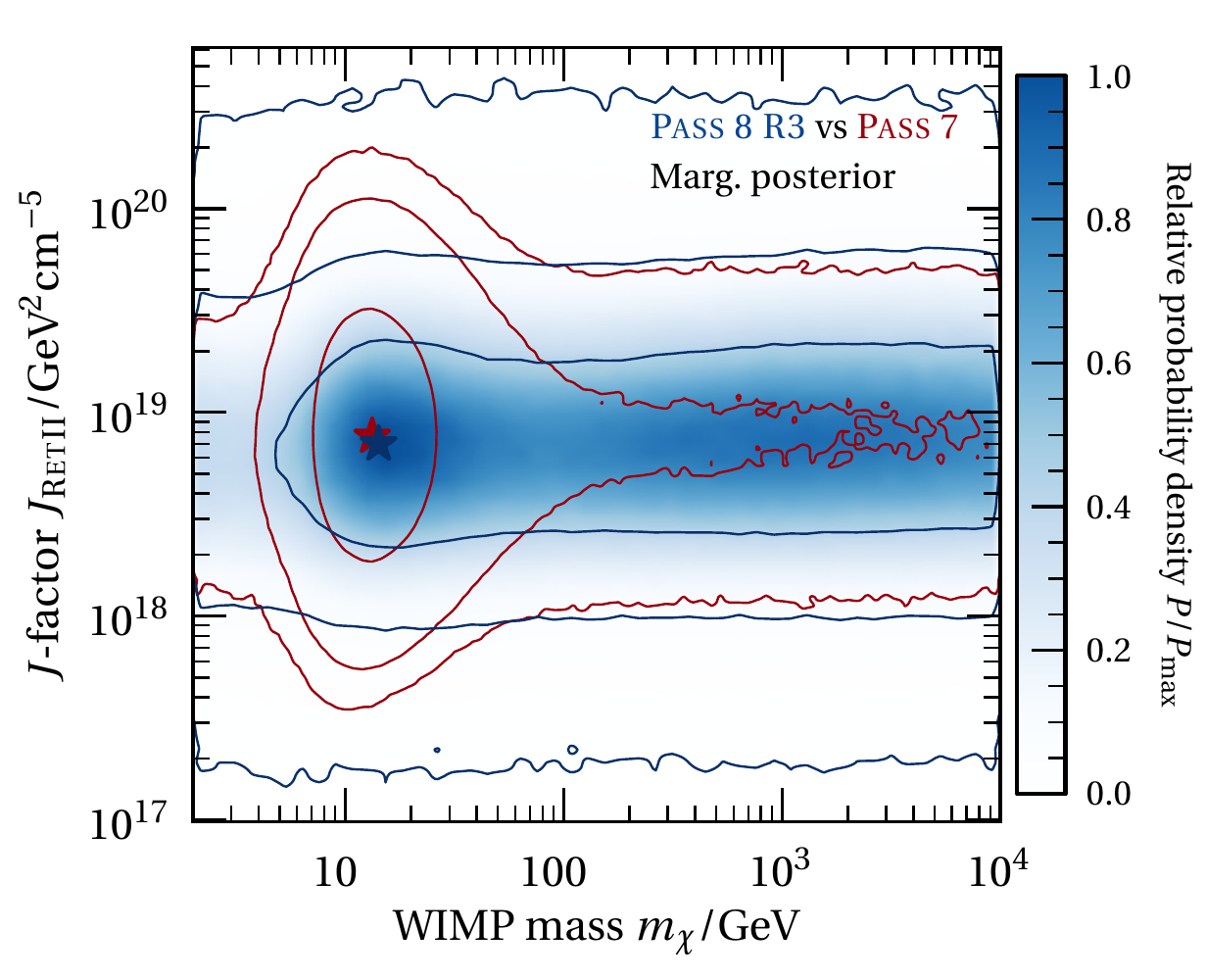}
		& \includegraphics[width=0.23934\textwidth]{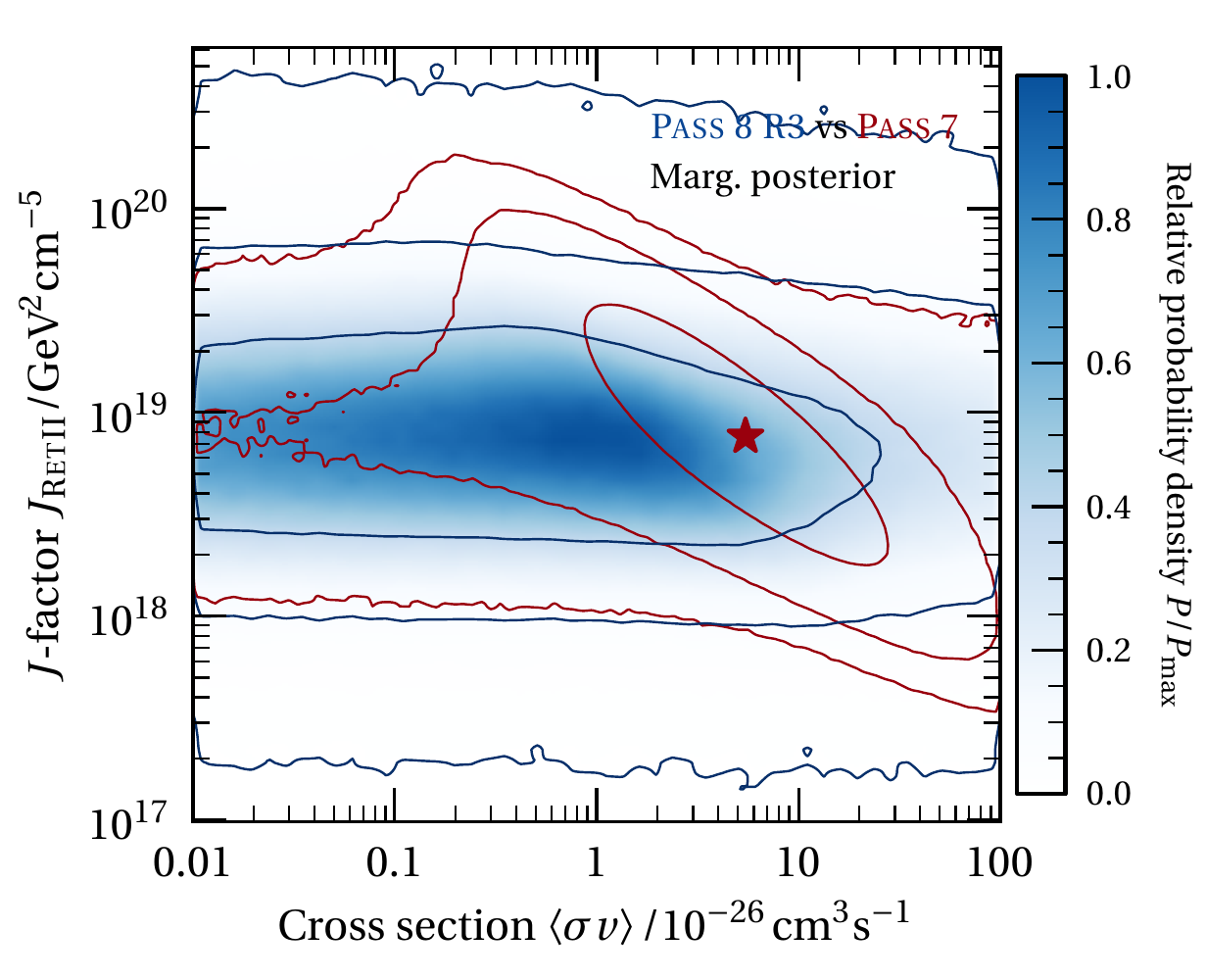}
		& \includegraphics[width=0.21062\textwidth]{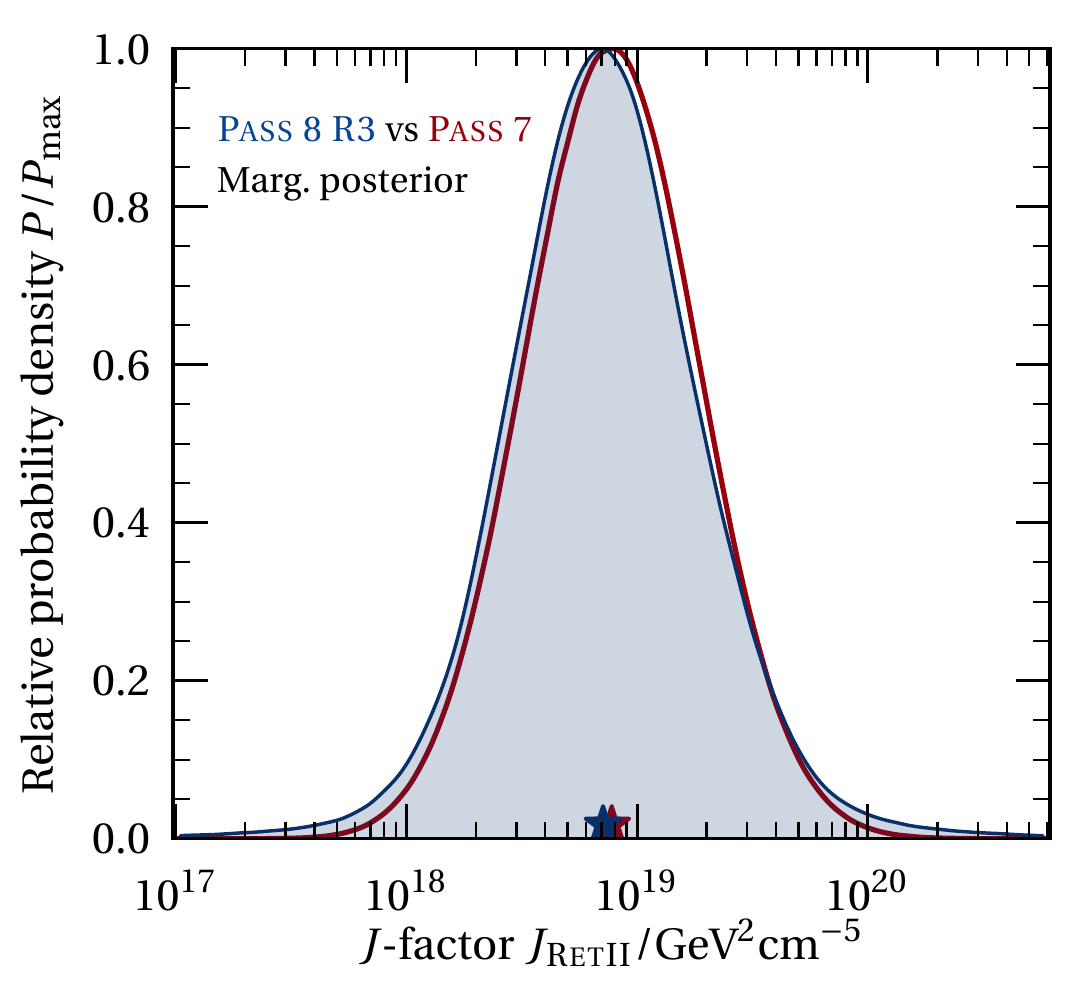}
		& \\
		\includegraphics[width=0.23934\textwidth]{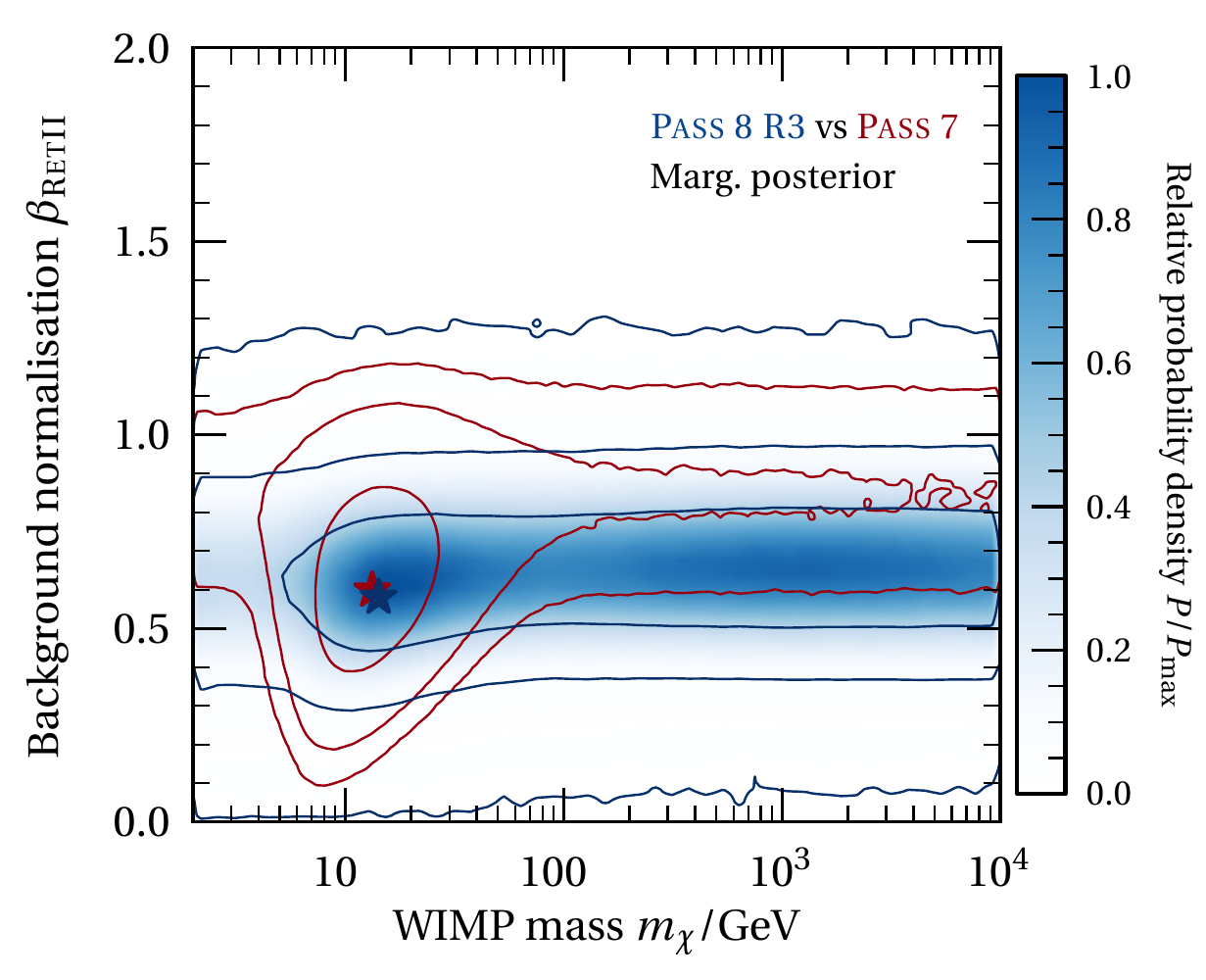}
		& \includegraphics[width=0.23934\textwidth]{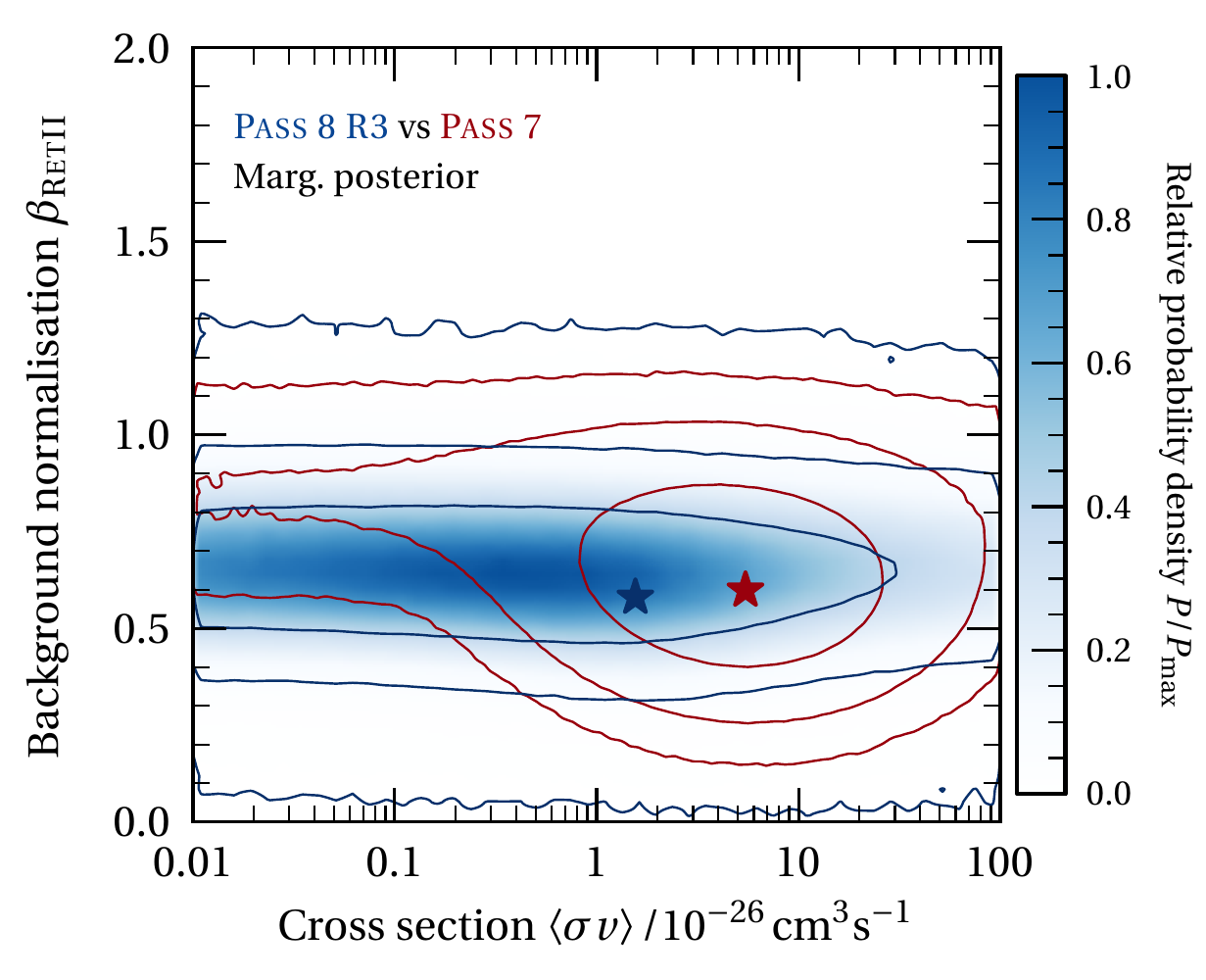}
		& \includegraphics[width=0.23934\textwidth]{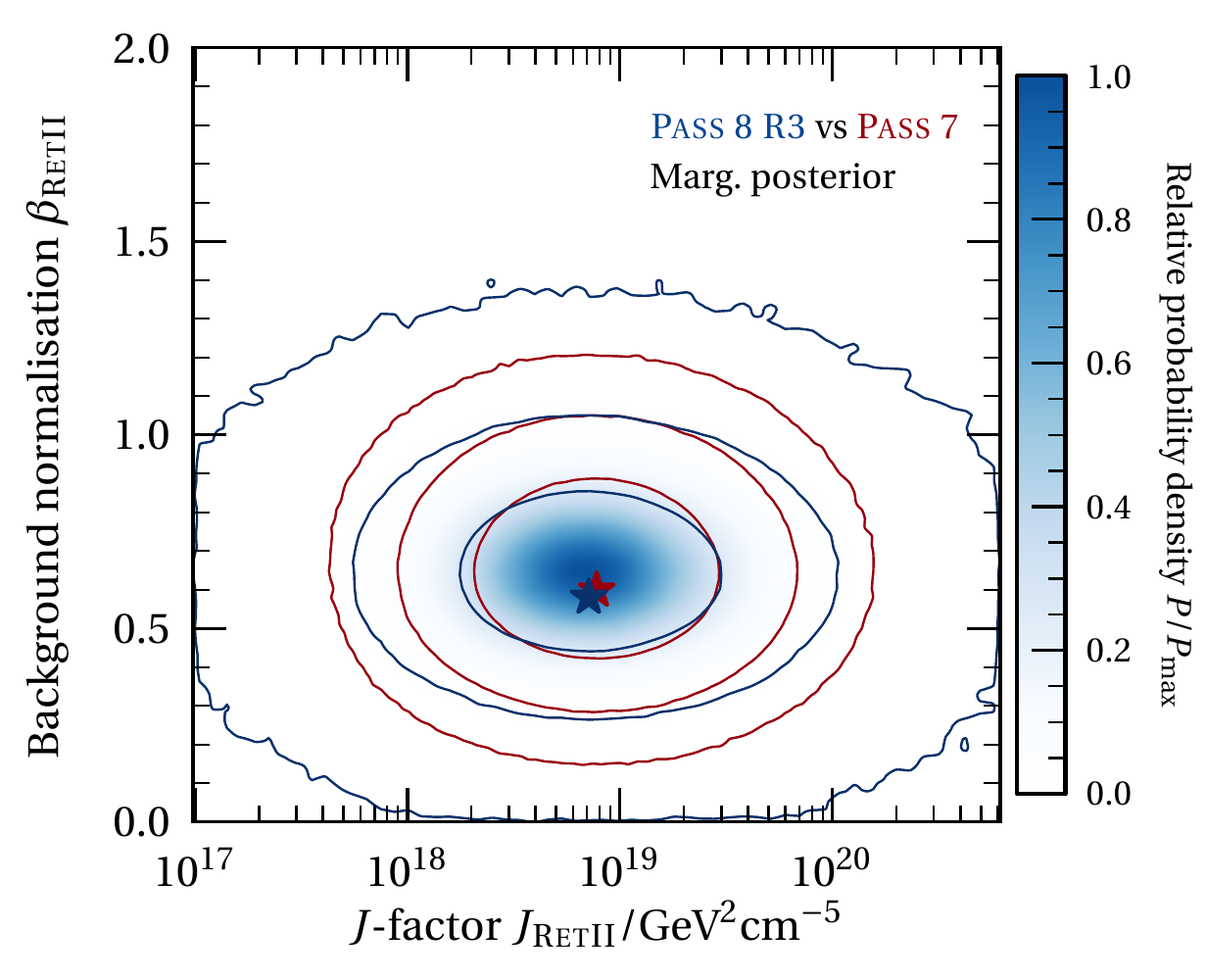}
		& \includegraphics[width=0.21062\textwidth]{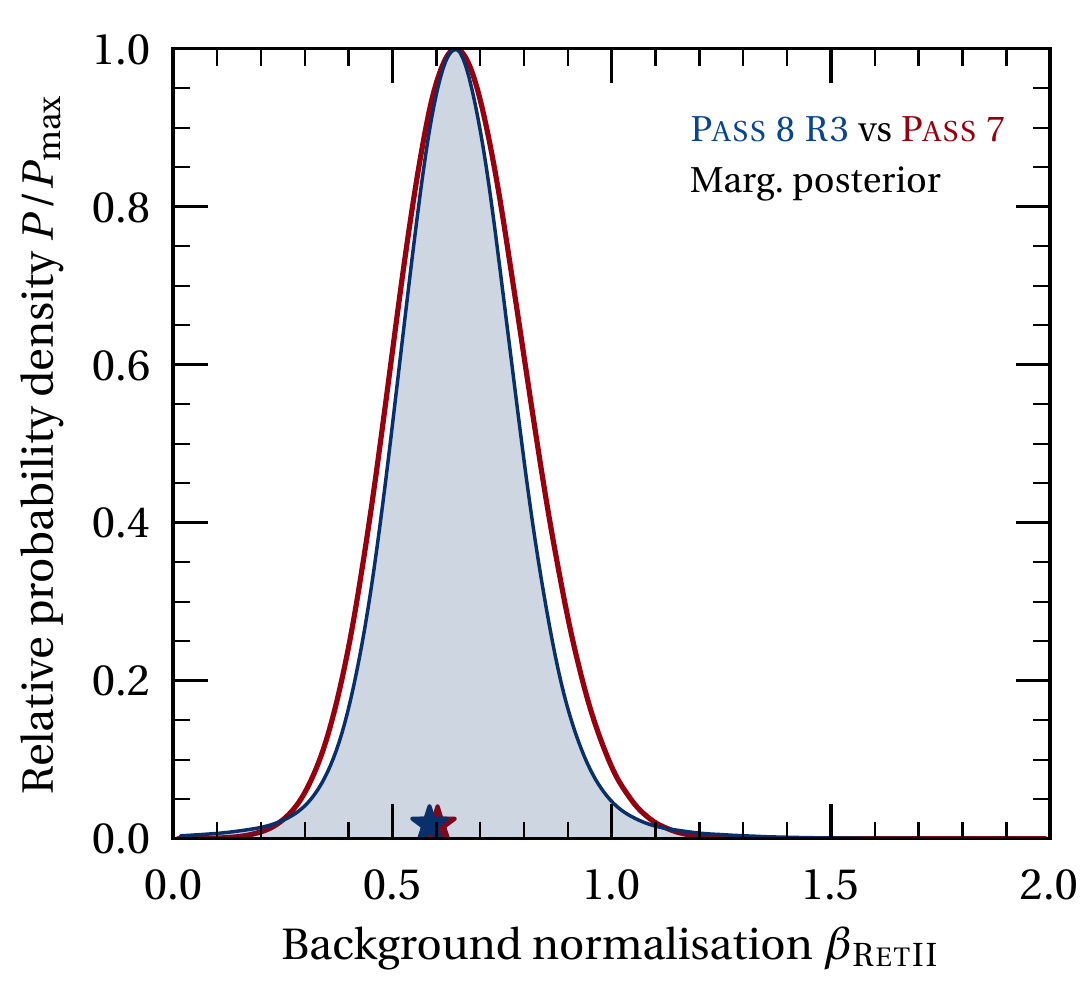}
	\end{tabular}
	\caption{One- and two-dimensional marginal posteriors for \rettwo (6.5\,years, $\tptm$ channel, log-uniform prior on~$\sigv$), showing the 68.27\%/95.45\%/99.73\% credible regions from \rettwo data. The red colour maps, contours and shaded regions are for \pseven, while the blue contours are for \peight. Red and blue stars indicate the best-fit points for the respective data sets.\label{fig:ret2:triangle}}
\end{figure}
In \reffig{fig:ret2:triangle}, we show all one- and two-dimensional marginal posteriors for \pseven vs \peight data and a~log-uniform prior on~$\sigv$. This illustrates again that, in \pseven, there is a clearly preferred range for~$\sigv$, while in \peight there is not. However, the regions of highest posterior density~(HPD) in both data sets are roughly in the same regions in parameter space as each other and the highest profile likelihood regions in the frequentist analysis. This means that the priors do not have a large impact on the analysis and the parameter regions singled out by the analysis are mostly data-driven.

Figure~\ref{fig:ret2:triangle} is also useful for visualising correlations between parameters, such as the expected anti-correlation between~$\mrm{J}{\rettwo}$ and~$\sigv$ from \pseven data, where a signal is preferred. This is because the DM signal is proportional to the product of both. Furthermore, the one-dimensional marginalised posteriors summarise the constraints that can be put on the individual parameters. Regarding~$m_\chi$ and~$\sigv$, they follow the expected behaviours in case of a strong~(\pseven) or weak~(\peight) preference for a signal, while $\mrm{J}{\rettwo}$ behaves as expected for a constrained nuisance parameter.

We note an interesting result for the background normalisations: In \pseven (\peight), we have a best-fit value of $\mrm{\hat{\beta}}{\rettwo}\approx 0.60$ \updated{($\mrm{\hat{\beta}}{\rettwo}\approx 0.58$)}. These values are considerably lower than the all-sky value of~$1$, which lies just outside the 95\%~CRs for the resulting posterior distributions. \updated{This implies that the background in the immediate vicinity around \rettwo is quite a bit lower than the average in a larger surrounding area. Since we introduced the scaling parameters~$\beta_k$ to account for the possibility of a local discrepancy of the background with respect to the reference global background model, it will be interesting to compare results for a larger number of dSphs in the next section. The shapes of the marginal posteriors for~$\mrm{\beta}{\rettwo}$ are also similar, with the marginalised posterior using \pseven data being slightly broader than in \peight.

While the background components' spectral shapes are slightly different between \pseven and \peight, Fig.~\ref{fig:spectra:p7vsp8} shows that the best-fitting models of the total background are nearly identical in the region of the excess (between about \num{1}~and \SI{10}{\GeV}). This points to the conclusion that the \pseven--\peight difference arises not because of a different background fit, but rather because of a genuine reduction in the number of excess photons in going from \pseven to \peight.}

\begin{figure}[t]
	\centering
	\setlength\tabcolsep{1pt}
	\begin{tabular}{ll}
		\includegraphics[width=0.43120\textwidth]{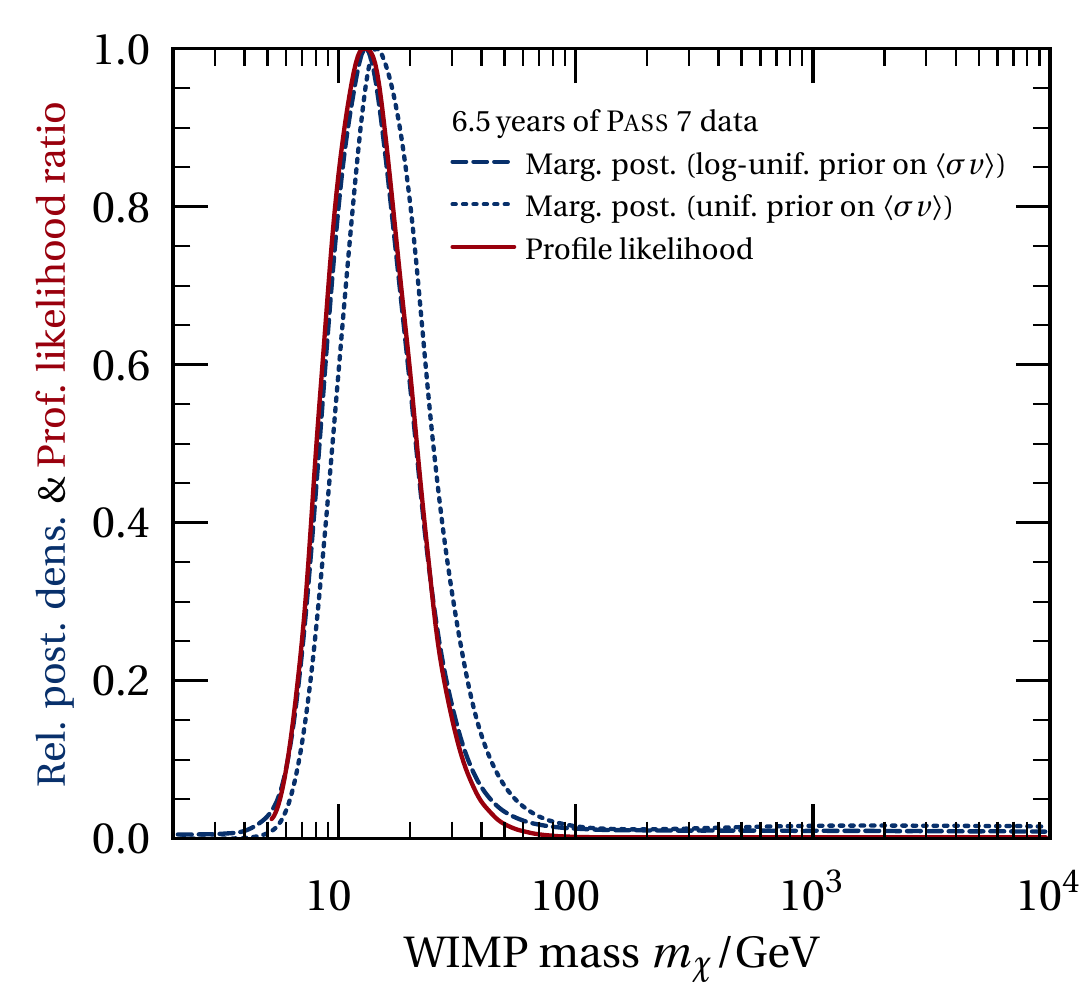} &
		\includegraphics[width=0.43120\textwidth]{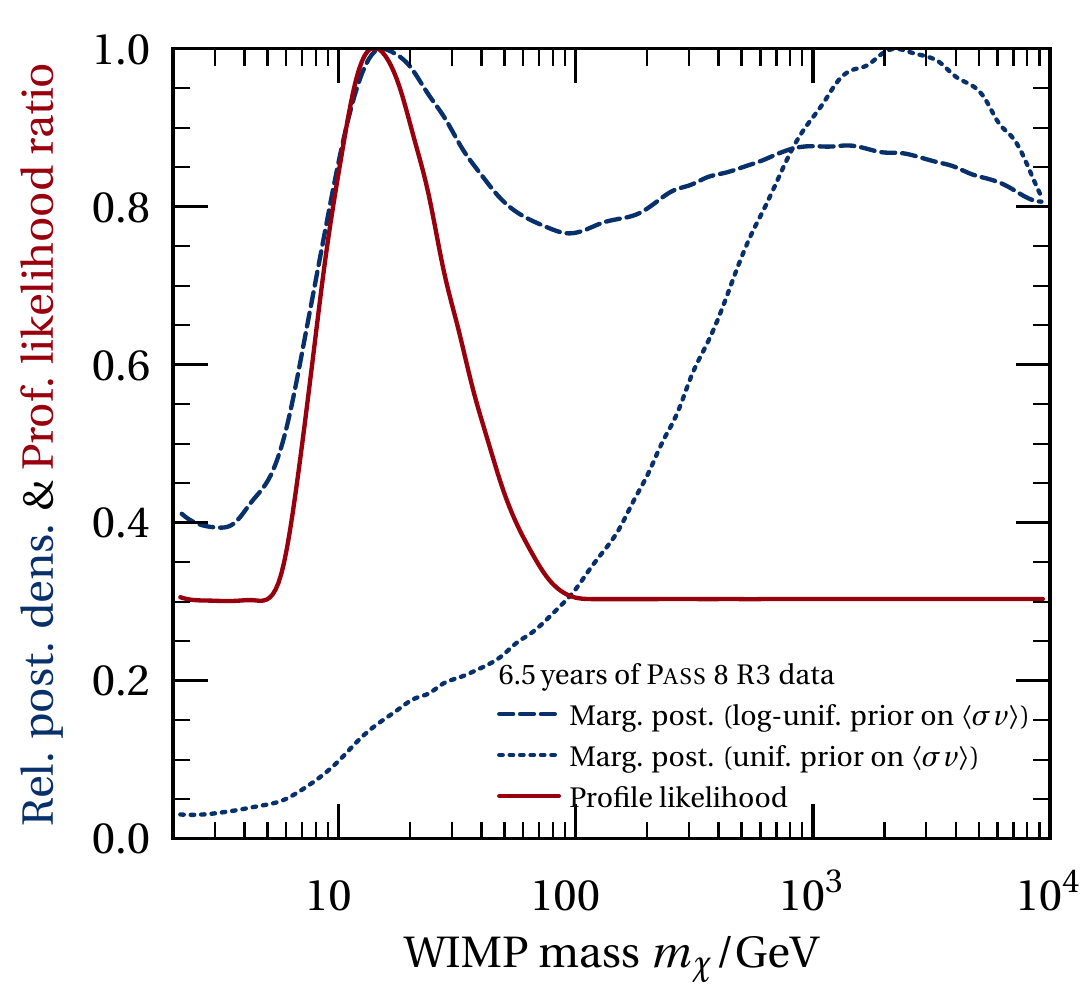} \\
		\includegraphics[width=0.43120\textwidth]{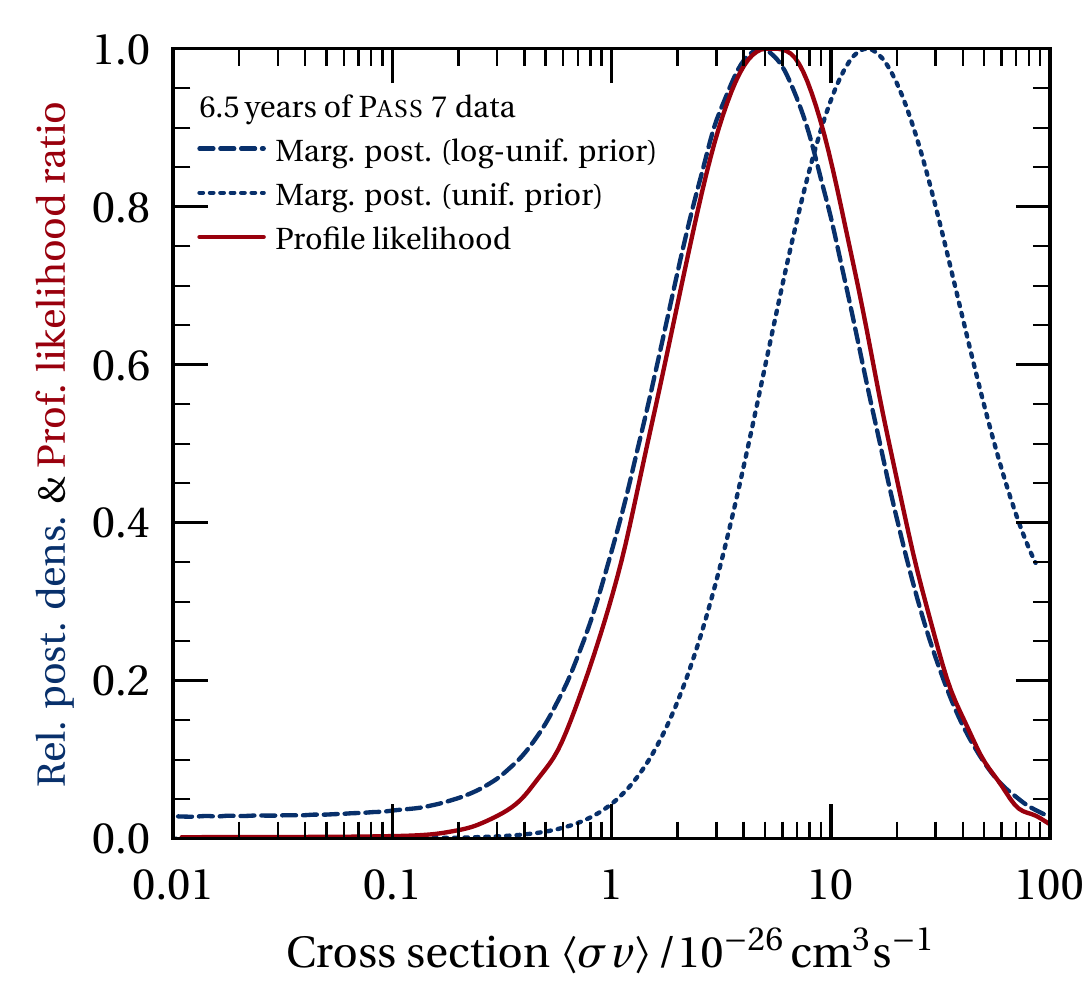} &
		\includegraphics[width=0.43120\textwidth]{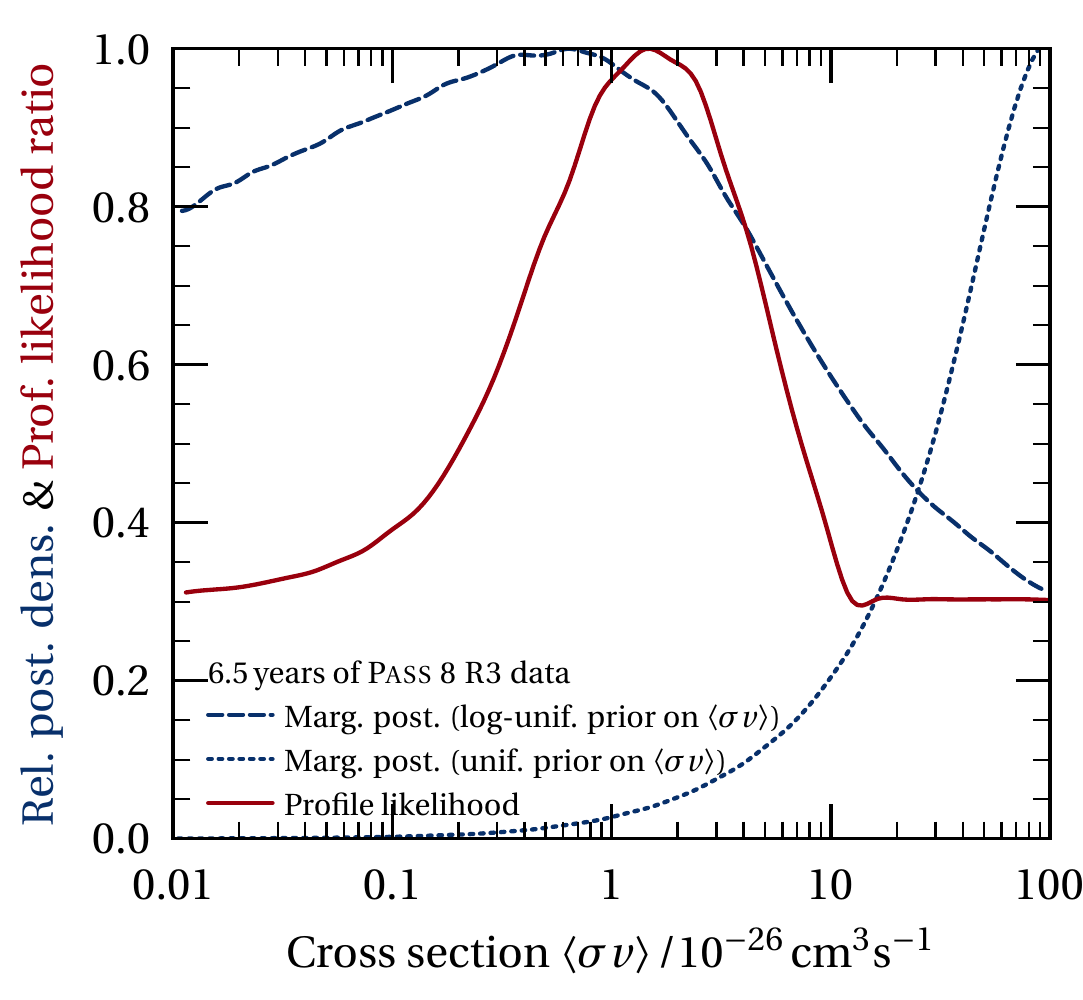}
	\end{tabular}
	\caption{One-dimensional marginal posteriors~(blues lines) and profile likelihoods~(red lines) for $m_\chi$~(\textit{top}) and $\sigv$~(\textit{bottom}). We show these for 6.5\,years of \pseven~(\textit{left}) and \peight~(\textit{right}) data (\rettwo, $\tptm$~channel).\label{fig:ret2:p7vsp8:flat}}
\end{figure}
Figure~\ref{fig:ret2:p7vsp8:flat} shows the one-dimensional marginal posterior distributions and profile likelihoods for~$m_\chi$ and~$\sigv$ and the two priors on~$\sigv$ adopted in this study. \updated{For a log-uniform prior on~$\sigv$, the resulting CRs peak at a similar location in both \pseven and \peight.} However, this is not the case when using a uniform prior on~$\sigv$. In the left panel of \reffig{fig:ret2:p7vsp8:flat} we can see that, for \pseven, the modes of the profile likelihood and posteriors agree rather well, even though the mode for~$\sigv$ using a uniform prior on~$\sigv$ is found at a higher value compared to the other two, reflecting the higher prior density for larger $\sigv$ values under the uniform prior on this quantity. For \peight (right panel), on the other hand, the main modes of the marginal posteriors for~$m_\chi$ and~$\sigv$ (using a uniform prior on~$\sigv$) get both shifted to higher values compared to the profile likelihoods and marginal posteriors using a log-uniform prior on~$\sigv$. Also, the marginal posteriors with the two different priors are quite different from each other, indicating strong prior dependence as a consequence of less constraining data. This is due to the absence of a strong preference for a signal in \peight and the higher prior weight on larger values for the cross~section affects the posterior on the WIMP~parameters. While a significant part of the posterior density is concentrated around the best-fit points for a log-uniform prior on~$\sigv$, a uniform prior on~$\sigv$ dominates the posterior density in \peight and causes the marginal posterior of the cross~section (and hence also the mass) to move to higher values.

\subsubsection{Model comparison for Reticulum~II}
\begin{table}
	\renewcommand{\arraystretch}{1.05}
	\caption{Bayes factors for comparing a background-only model with a model including an additional DM signal from $\tptm$ channel annihilation (using 6.5\,years of \rettwo data). A positive (negative) value of $\ln\left(\mathcal{B}_{10}\right)$ indicates evidence in favour of (against) the model with an additional DM~signal. The value of $\mathcal{B}_{10}$ gives the posterior odds between the DM model and the background only model if each model has equal prior probability.\label{tab:ret2:modelcomp}}
	\vspace{0.25em}
	\centering
	\begin{tabular}{lcccc}
		\toprule
		& \multicolumn{2}{c}{\pseven} & \multicolumn{2}{c}{\peight} \\
		\cmidrule(lr){2-3}\cmidrule(lr){4-5}
		Prior on $\sigv$ & uniform & log-uniform & uniform & log-uniform \\
		\midrule
		Bayes factor~$\ln\left(\mathcal{B}_{10}\right)$ & $1.61 \pm 0.02$ & $1.92 \pm 0.02$ & \updated{$-0.65 \pm 0.02$} & \updated{$-0.12 \pm 0.02$} \\
		\phantom{Bayes factor~$\ln ($}$\mathcal{B}_{10}$ & 5:1 & 7:1 & \phantom{$-$}1:2 & \phantom{$-$}1:1 \\
		\bottomrule
	\end{tabular}
\end{table}
To quantify the preference (or lack thereof) for a DM~signal contribution to the gamma-ray spectrum of \rettwo, we compute the Bayesian evidence for the background-only and the background-plus-signal model. The conceptual difference between Bayesian model comparison and frequentist hypothesis testing is that the latter can only reject the null hypothesis, while the former can show a preference for a simpler model whenever the added complexity of the more complicated model is not warranted by the data. In other words, the Bayesian model comparison framework includes an automatic Occam's razor effect.

We define the background-only model via setting $\sigv = 0$, meaning that the DM mass parameter becomes non-identifiable. The resulting Bayes factor for both data sets, \pseven and \peight, as well as the two adopted choices of prior on $\sigv$ are given in Table~\ref{tab:ret2:modelcomp}.

We use a commonly applied scale for categorising how strongly one model is favoured over the other, dating back to Jeffreys~\cite{Book_Jeffreys,KassRaftery}, with the nomenclature adopted from Ref.~\cite{1701.01467}. This ``Jeffreys' scale'' has thresholds at $\left| \ln\left(\mathcal{B}_{10}\right) \right| = 1.0$, $2.5$, and $5.0$, which we call respectively \textit{weak}, \textit{moderate}, and \textit{strong evidence}. From Table~\ref{tab:ret2:modelcomp}, we can see that only \pseven data gives \textit{weak evidence} (i.e. a Bayes factor of more than~3:1) for the DM hypothesis regardless of the adopted prior. On the other hand, the model comparisons using \peight neither favour a DM~signal, as already anticipated in the previous section, nor do they favour the background-only model. The \peight data are simply insufficiently informative to reach a conclusion either way.

We notice that, as expected, the outcome of the Bayesian model comparison is much more conservative than what would be obtained using a $p$-value frequentist approach~\cite{2001_Selke}: even in the case of \pseven data, which gives a more than $3\sigma$ significance for a non-zero $\sigv$, the Bayes factor for a log-uniform prior is only~7:1, shy of the threshold for even \textit{moderate evidence} at~12:1. This is an example of a well known statistical phenomenon called the Lindley paradox: the outcome of hypothesis testing and Bayesian model comparison differs (even asymptotically for large amount of data), because the two approaches ask fundamentally different questions~(see Refs~\cite{astro-ph/0504022,0803.4089} for a detailed discussion and further references).

In any case, the model comparison further quantifies the mostly qualitative findings that emerged so far in this section. Confirming previously proposed explanations for the difference in significance for a potential excess in \rettwo~\cite{1503.02632,1807.08740}, we find that this is due to the differences in \pseven and \peight since the results of the Bayesian and the frequentist analysis agree when using \rettwo data based on the same selection criteria.

\subsubsection{Posterior predictive distributions for signal strengths in other dwarfs}\label{sec:ppds}
After having re-visited the purported gamma-ray excess in \rettwo, we investigate how our conclusions might change when considering the other dSphs, both in terms of WIMP parameter constraints and model selection outcome. Our starting point is to note that physical consistency requires that the WIMP parameters must be the same across \emph{all} dSphs. Therefore, given a possible excess from \rettwo, it is helpful to quantify the probability that an excess due to the \emph{same} dark matter candidate will be seen in other dwarfs. One possibility is to use the best-fit WIMP~parameters from \rettwo to establish the strength of the DM signal in other dSphs. However, this approach neglects the uncertainties in the WIMP~parameters as well as the uncertainties arising from the $J$-factor and background rescaling for the dwarf for which the prediction is being made. It also does not quantify the probability of achieving a statistical significant measurement in another dwarf.

In this section, we introduce the posterior predictive distribution~(PPD) as a tool to precisely quantify the probability of seeing a DM-related signal in one dSph, conditional on the observation in another one~(in this case, \rettwo). The same approach can also be used to make predictions for future observations of the same dwarf, i.e. over longer integration times.

The advantage of this approach is three-fold: firstly, the ensuing predicted distribution is a \emph{probability distribution} for the yet unobserved data, which fully accounts for all relevant sources of uncertainty. Secondly, this approach clarifies that not seeing a DM~signal from a dwarf where one would not expect it (e.g. because the $J$-factor is too low) is not an indication against the DM~model. Indeed, the contribution to the Bayes factor from such a dwarf is null: if the distribution of data under both the background-only model and the background-plus-signal model are observationally indistinguishable, then making the observation is not going to teach us anything about the relative viability of each model. Finally, PPDs indicate the most promising targets to improve the model comparison result, for example to test the DM model further. A natural extension of PPD is Bayesian decision theory and experimental design, which we do not however pursue further in this work (see Ref.~\cite{2009_Trotta} for an example and discussion). More generally, it is important to note that the PPD can be based on posterior samples from \emph{any} experimental search, not just dSphs.

The PPD for any observable $\obs$ (which might be future or not-yet-analysed data), given previously analysed data $\data$, is
\begin{equation}
	\prob{\obs}{\data} =\int_{\Theta}^{} \! p(\obs, \boldsymbol{\theta} | \data) {\dd} \boldsymbol{\theta} = \int_{\Theta}^{} \! \prob{\obs}{\boldsymbol{\theta},\,\data} \, \prob{\boldsymbol{\theta}}{\data} \, \dd \boldsymbol{\theta} = \int_{\Theta}^{} \! \prob{\obs}{\boldsymbol{\theta}} \, \prob{\boldsymbol{\theta}}{\data} \, \dd \boldsymbol{\theta} \, , \label{eq:PPD}
\end{equation}
where $\prob{\obs}{\boldsymbol{\theta}}$ is the likelihood for data $\obs$ given parameters $\boldsymbol{\theta}$, weighted by the posterior distribution from current data, $\prob{\boldsymbol{\theta}}{\data}$, and integrated over all values for the parameters, $\boldsymbol{\theta} \in \Theta$. One can easily see that this generalises the ``best fit prediction'', obtained directly from the best-fit estimate for~$\boldsymbol{\theta}$, which is in particular appropriate if the uncertainty on~$\boldsymbol{\theta}$ is relevant. The best-fit prediction is recovered from~\eqref{eq:PPD} by setting $\prob{\boldsymbol{\theta}}{\data} = \delta(\boldsymbol{\theta} - \boldsymbol{\hat{\theta}})$, where~$\boldsymbol{\hat{\theta}}$ is the maximum-likelihood estimator of the parameters.

To obtain the PPDs, we select \num{5e5}~random samples out of the equally weighted posterior samples from the analysis of 6.5\,years of \pseven data using \rettwo only. We then generate a realisation of background and DM~signal counts for each dSph by drawing them from Poisson distributions with rates given by~\eqref{eq:backgrounds}. In order to do so, we require values for the $J$-factor and background normalisation entering the Poisson rate for the dwarf for which the prediction is made. To obtain these, we draw random samples from informative prior distributions: for the $J$-factors, we draw $J$-factor samples directly from the posterior samples supplied in the auxiliary material of Ref.~\cite{1802.06811}.

For the background normalisations, we draw samples from a normal distribution with mean~1, expecting that the background should not need any rescaling on average. \updated{The value of the standard deviation, 0.27, was obtained from the posterior obtained by combining \emph{all} background normalisation parameters $\beta_k$ in the global analysis in Sec.~\ref{sec:globalfit}. This represents the observed spread of the background scaling parameter across the \ndwarfs~dwarfs in the \peightrthree data set for 11\,years of Fermi-LAT observations. As this procedure averages over all dwarfs, it is a more conservative approach than trying to obtain such a prior based on data from each dSph individually.}

\afterpage{
	\thispagestyle{empty}
	\begin{figure}
		\vspace*{-2.25cm}
		\centering
		\captionsetup{width=1.2\linewidth}
		\setlength\tabcolsep{0.07\textwidth}
		\renewcommand{\arraystretch}{0.01}
		\begin{tabular}{cc}
			\includegraphics[height=4cm]{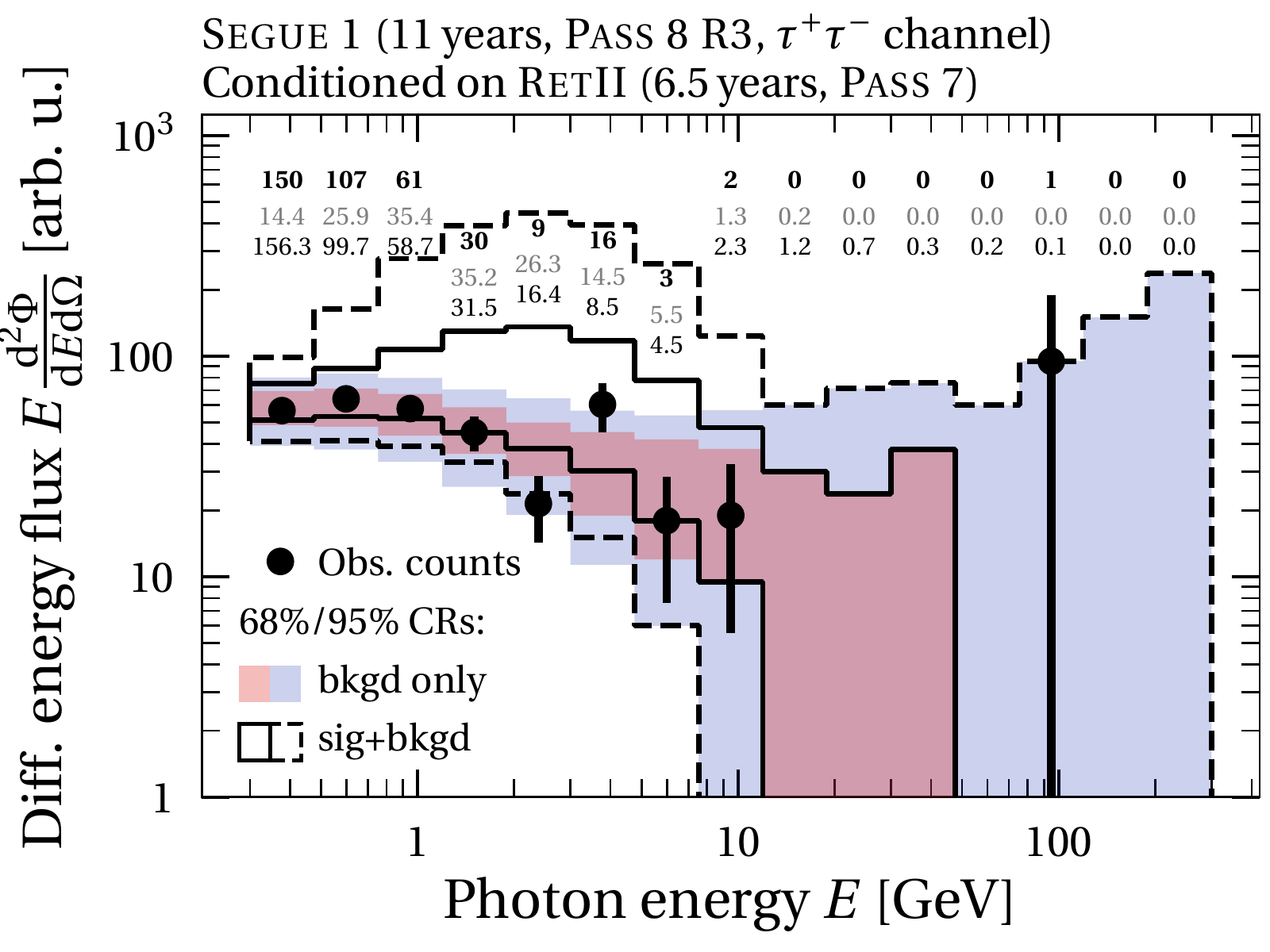} &
			\includegraphics[height=4cm]{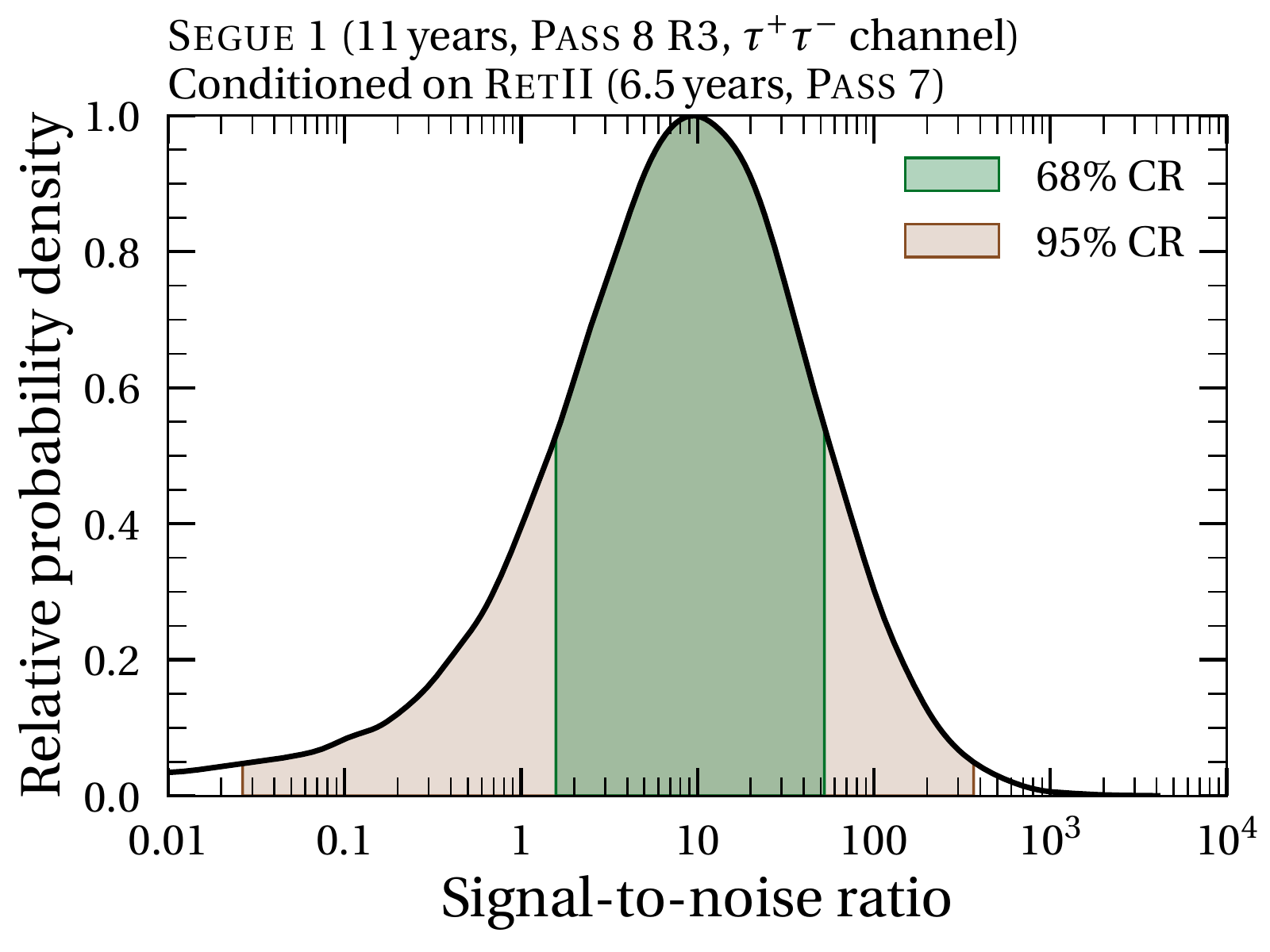} \\
			\includegraphics[height=4cm]{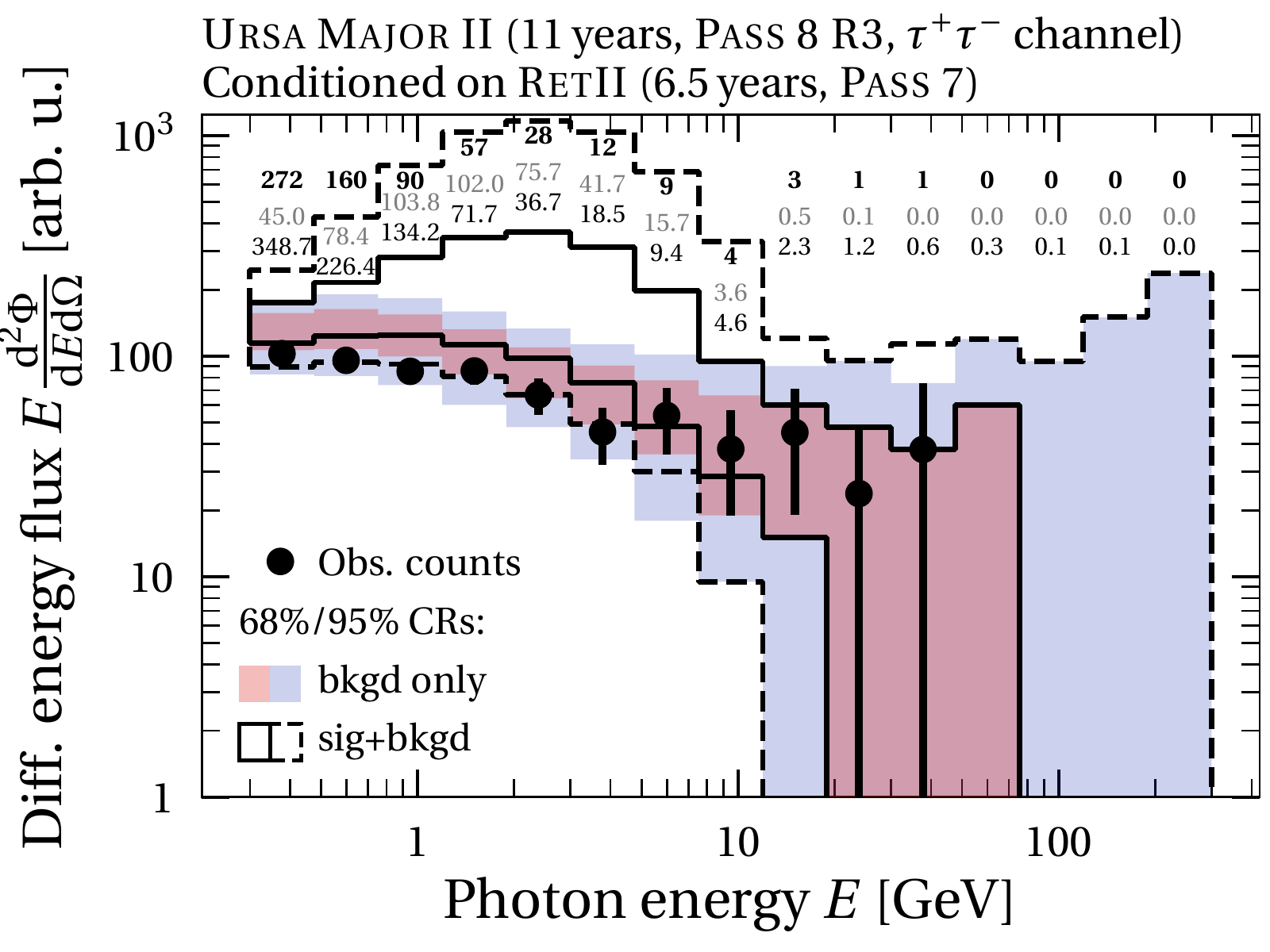} &
			\includegraphics[height=4cm]{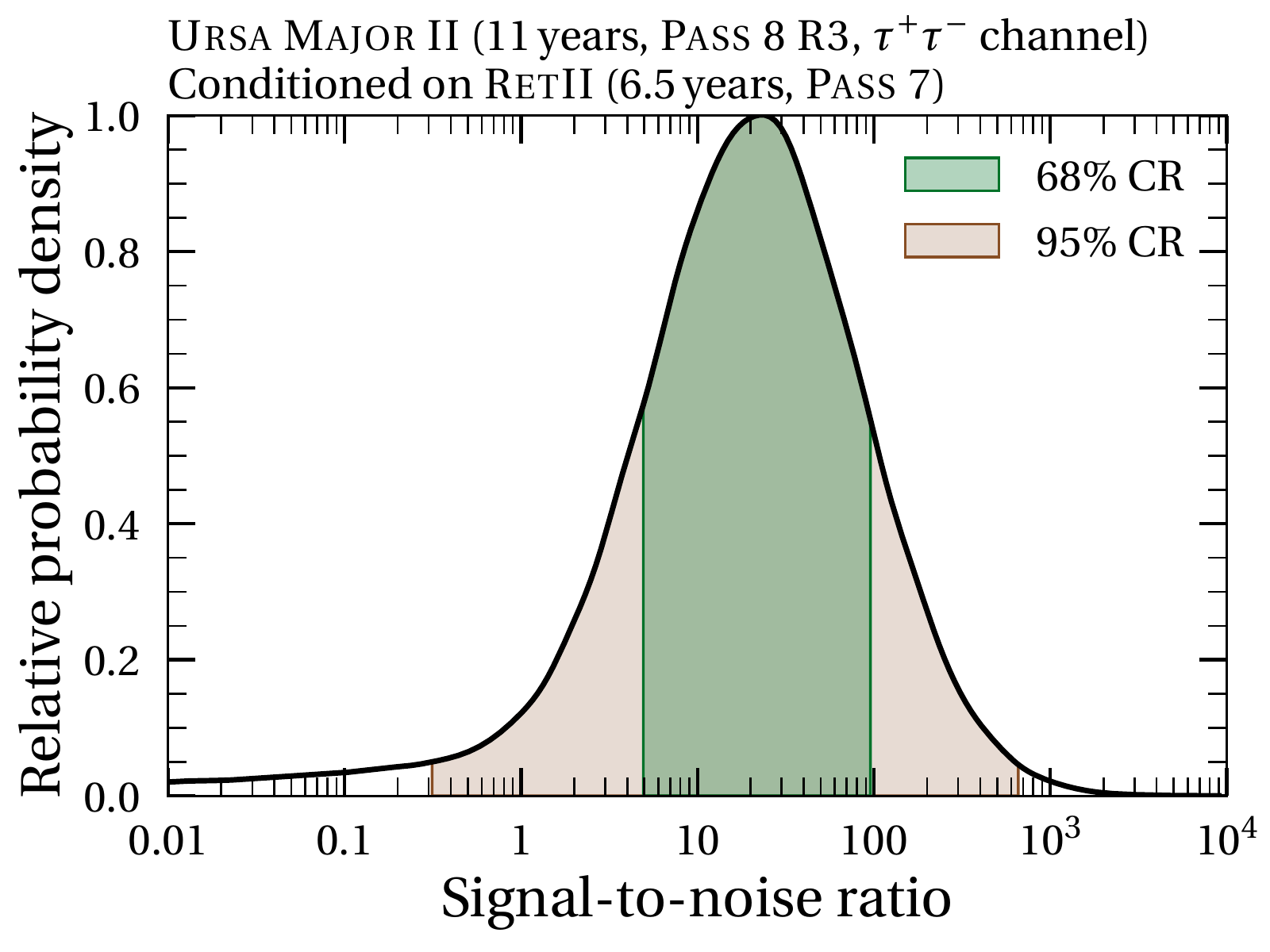} \\
			\includegraphics[height=4cm]{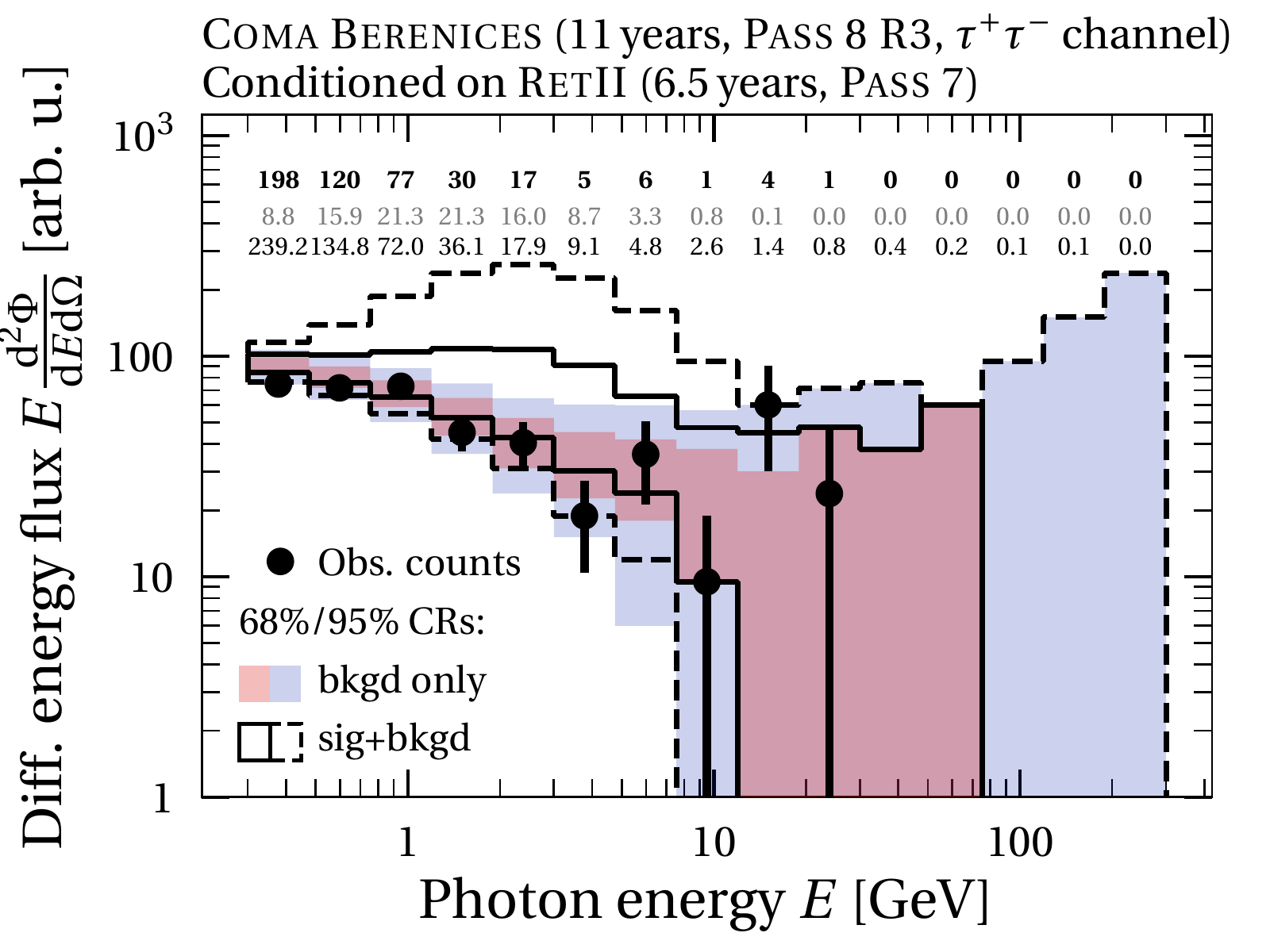} &
			\includegraphics[height=4cm]{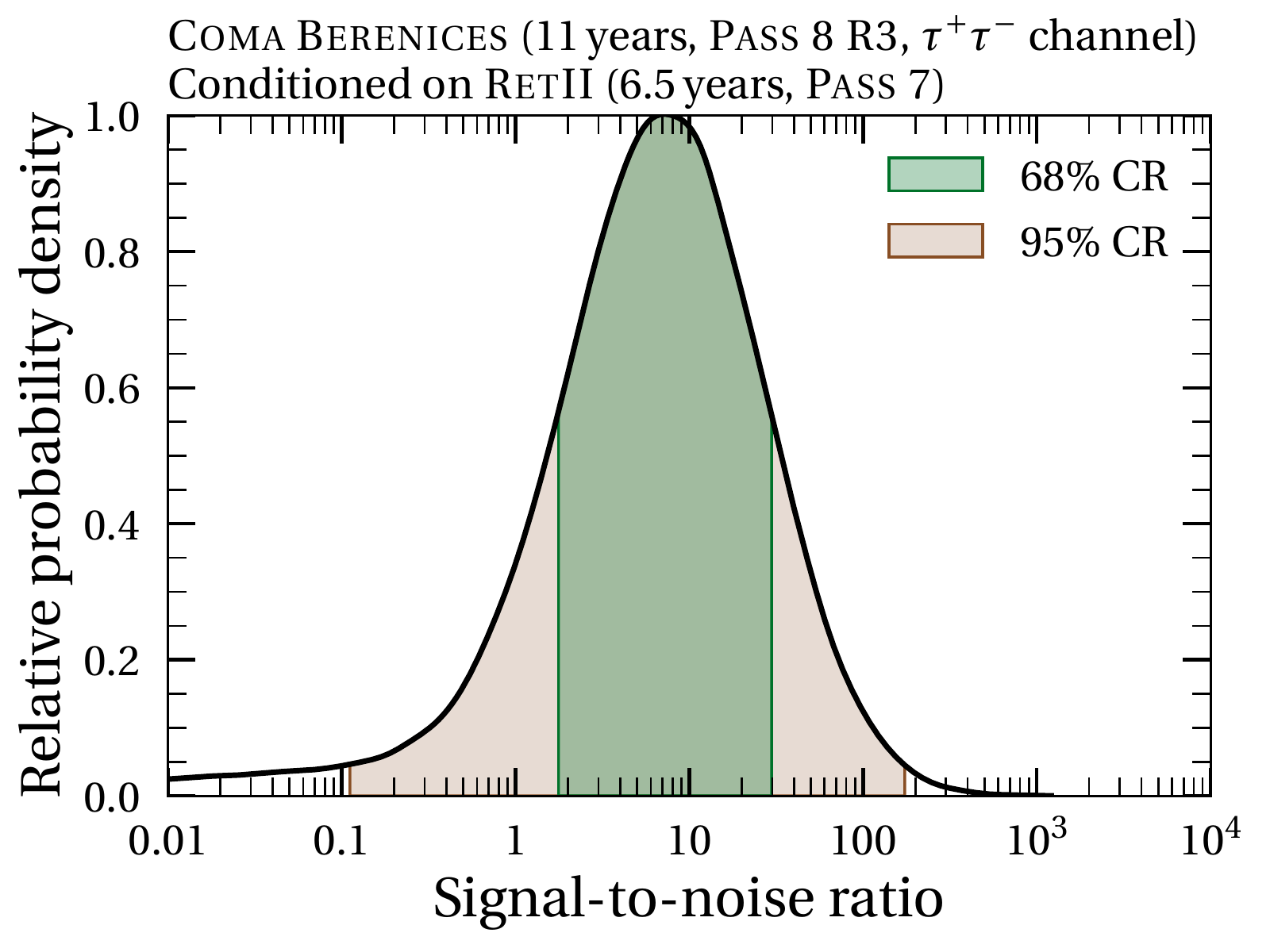} \\
			\includegraphics[height=4cm]{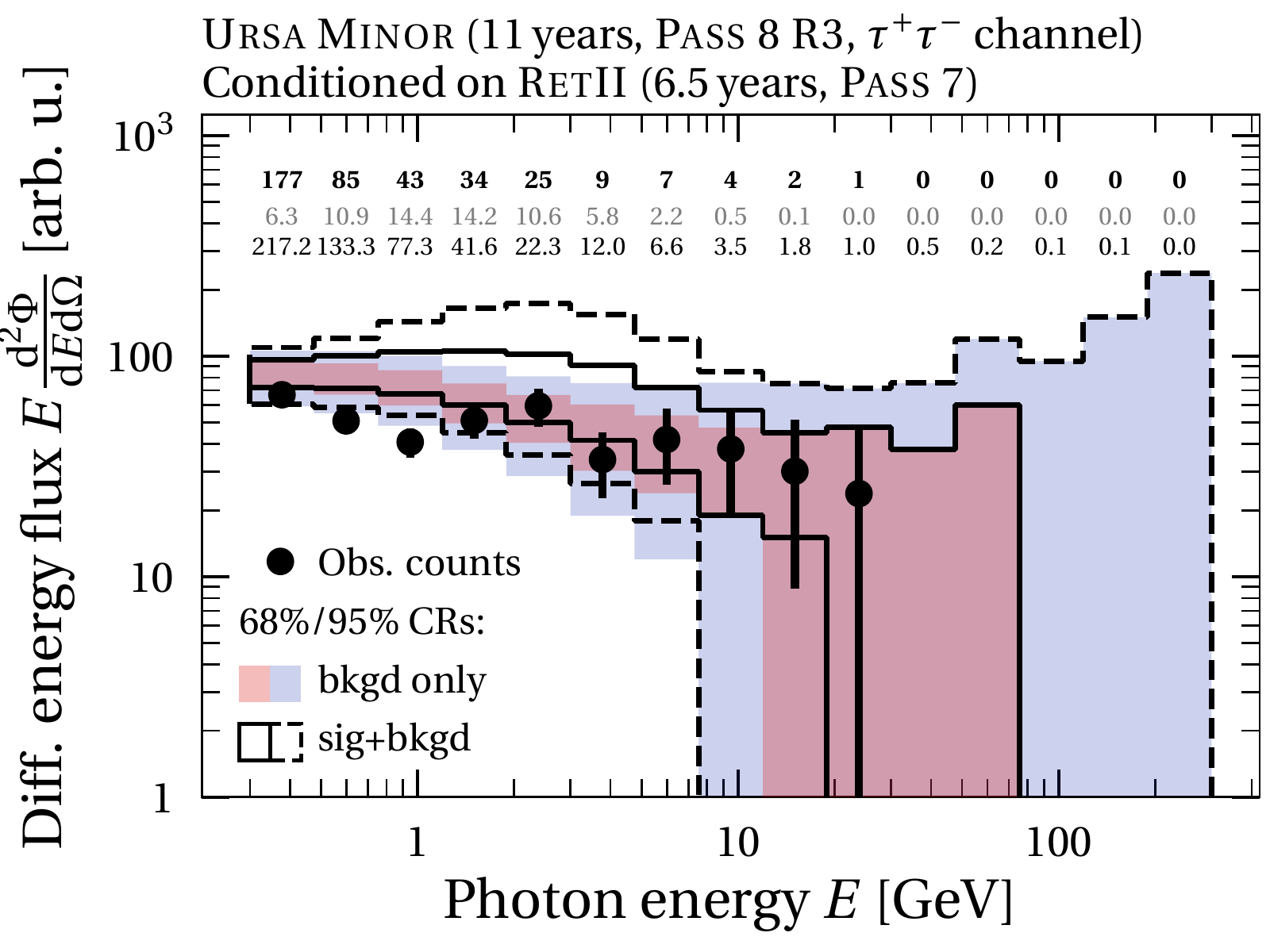} &
			\includegraphics[height=4cm]{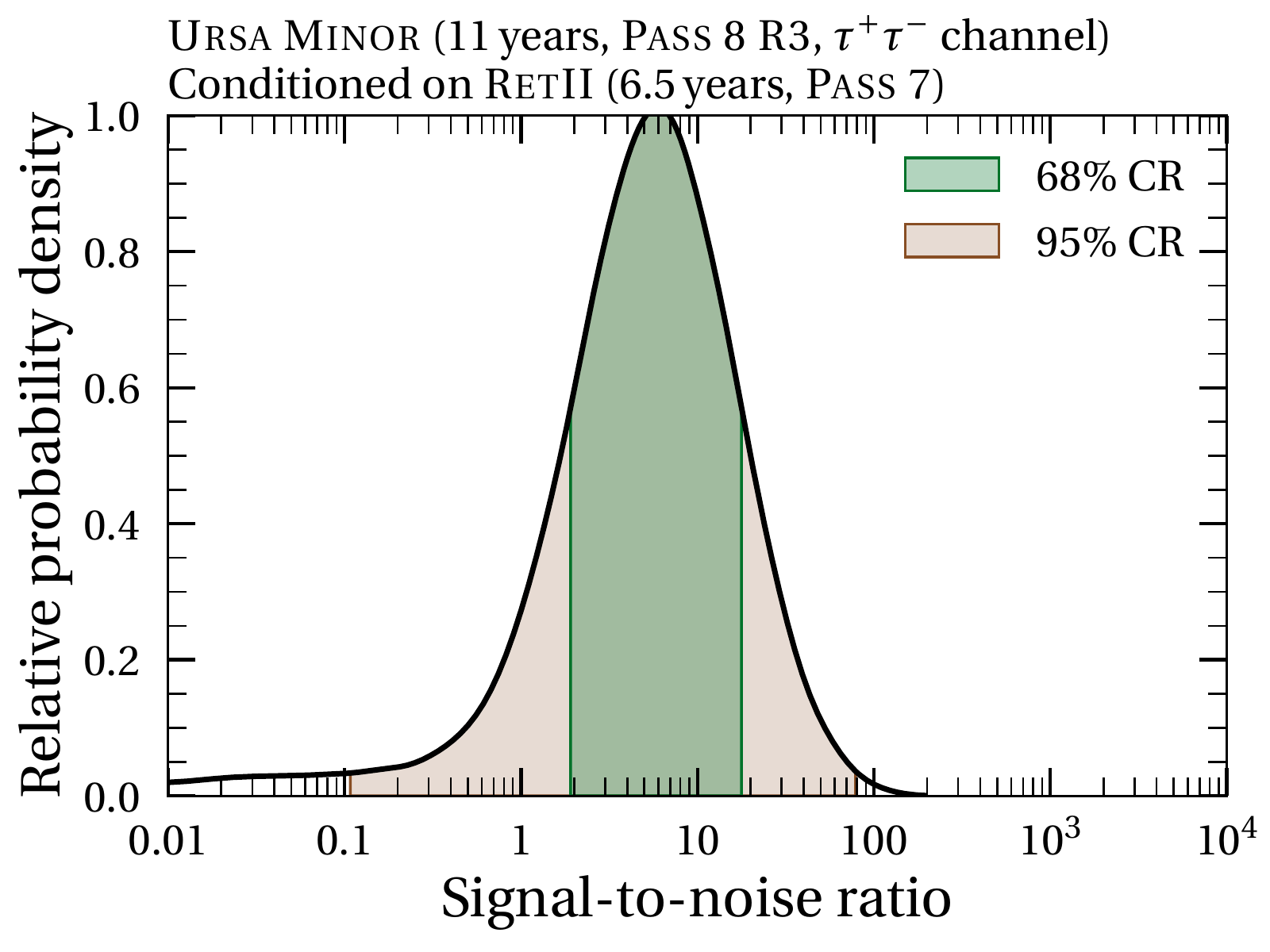} \\
			\includegraphics[height=4cm]{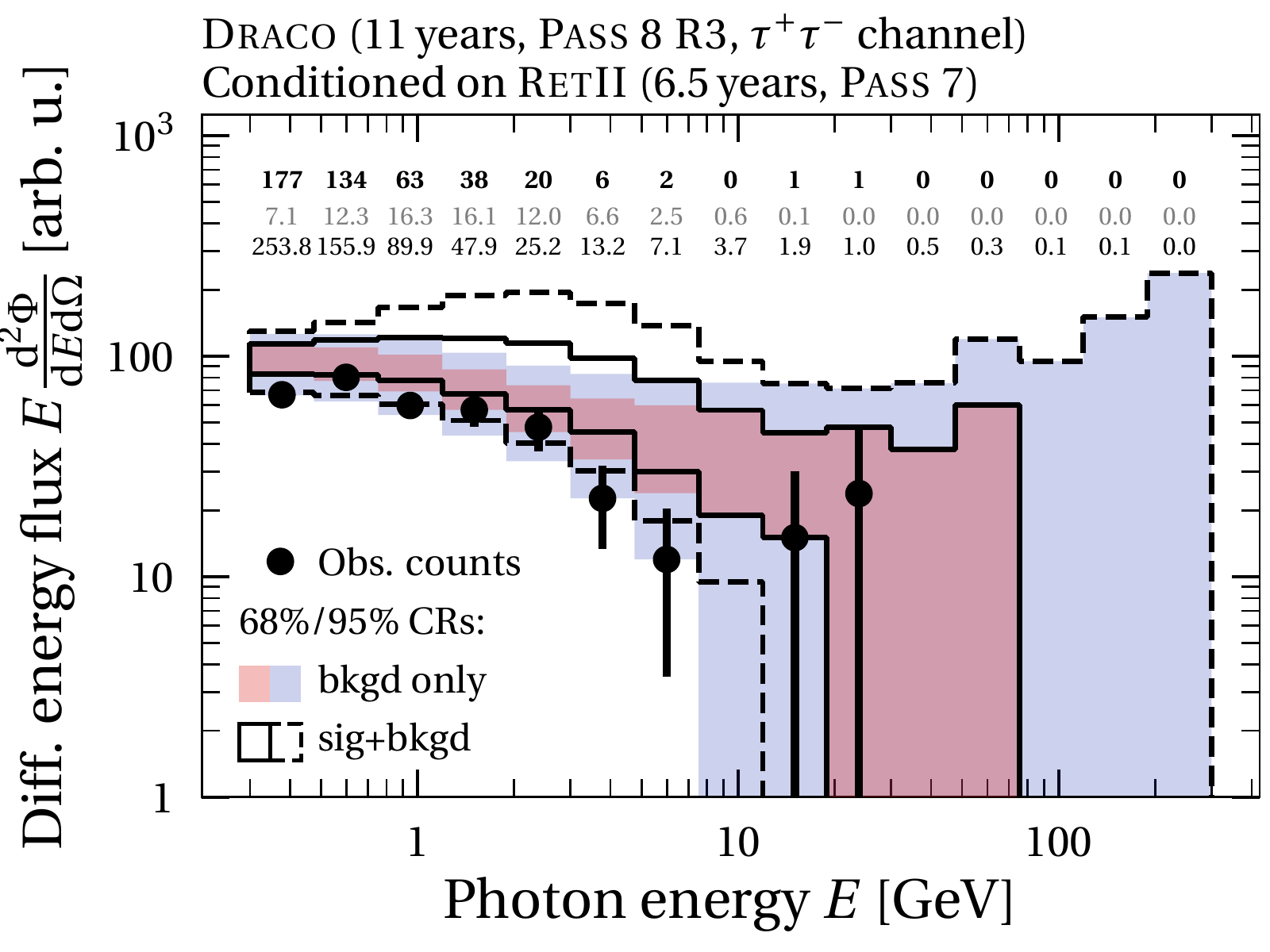} &
			\includegraphics[height=4cm]{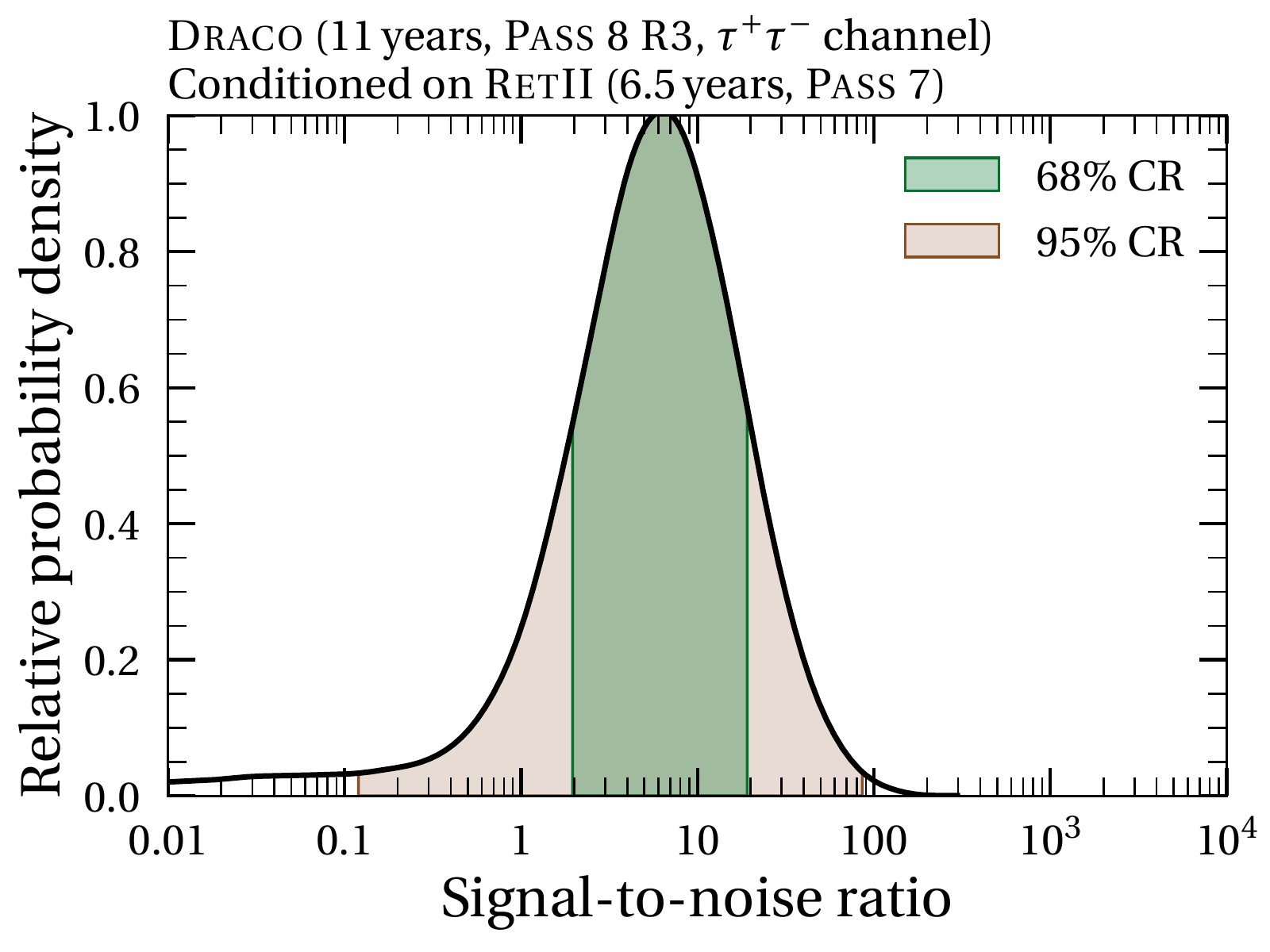} \\
			\includegraphics[height=4cm]{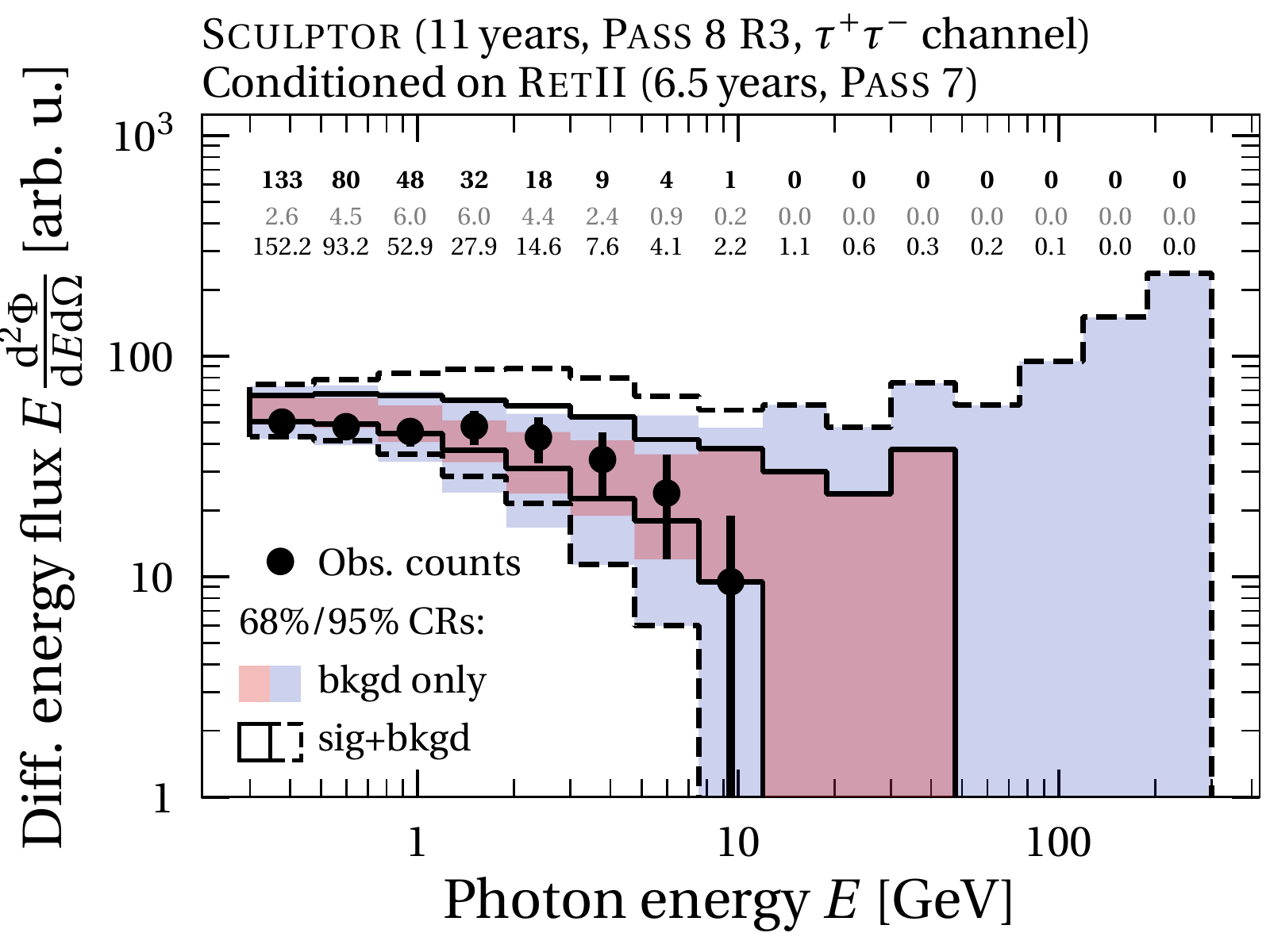} &
			\includegraphics[height=4cm]{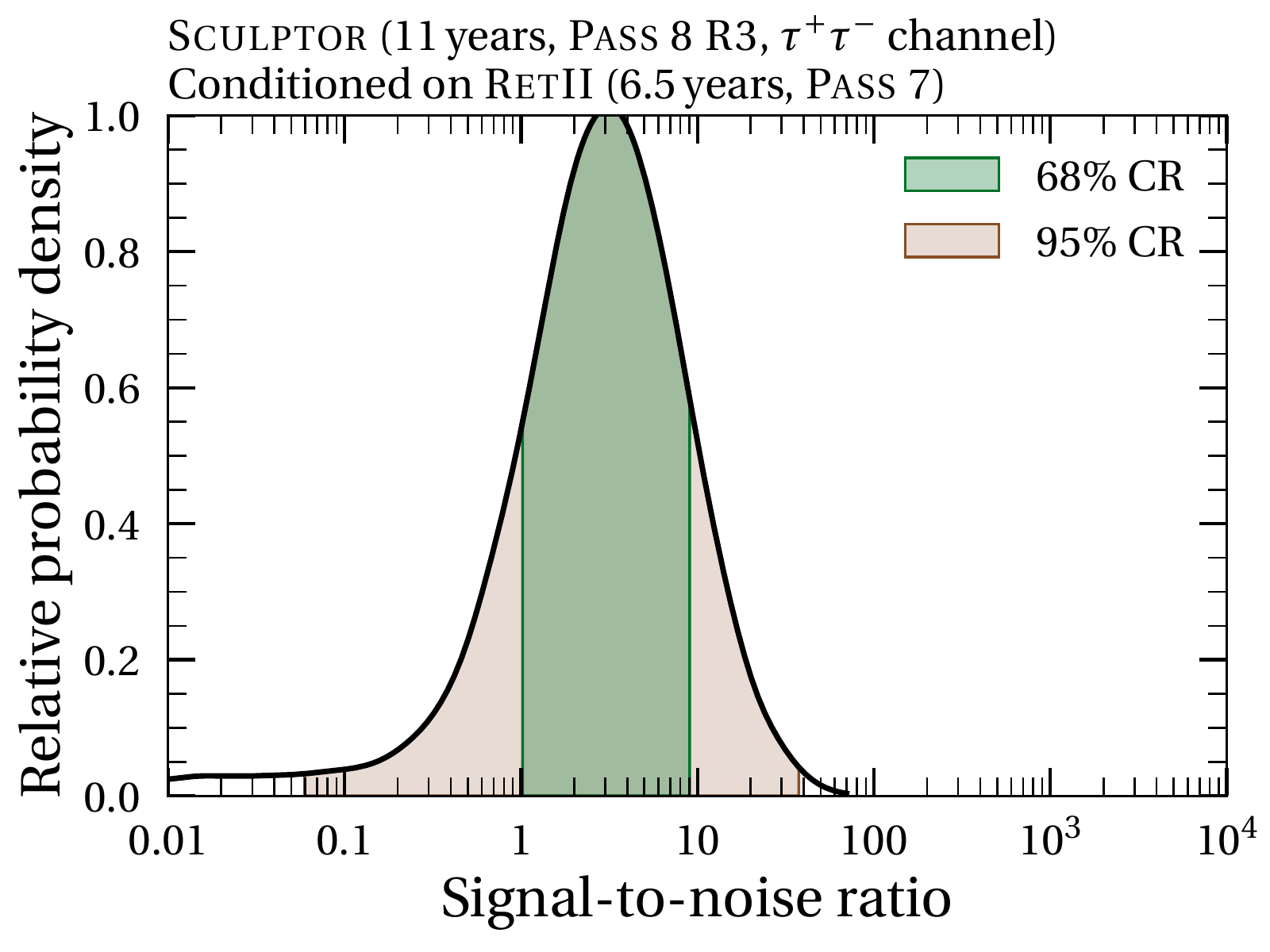}
		\end{tabular}
		\vspace*{-0.25cm}
		\caption{Posterior predictive distributions for photon flux~(\textit{left}) and the SNR~(\textit{right}) for 11\,years of \peight data, conditional on 6.5\,years of \pseven data (\rettwo only, log-uniform prior on $\sigv$, $\tptm$~channel). \textit{Left:} We show the observed counts~(black circles with Poisson error bars; bold numbers give the number of photons in each bin), the 68\%/95\%~CRs for the background-only model (blue/red shaded regions; black numbers give the mean predicted number of counts) and for the signal-plus-background model (solid/dashed empty regions; grey numbers give mean predicted signal counts), respectively. \textit{Right:} The green/brown shaded regions delimit the 68\%/95\% predicted probability interval for the SNR.\label{fig:predpost}}
	\end{figure}
	\clearpage
}
Figure~\ref{fig:predpost} demonstrates how the PPDs conditional on data from one dSph (\rettwo) can be used as a diagnostic tool to identify discrepancies between the predicted vs the observed data in the other dSphs and with a larger amount of data, thus identifying the most promising targets for analysis. The figure shows predictions for the spectra~(left panel) of the background-only model~(filled regions) and for the background-plus-signal model~(empty regions) in six different dSphs after 11\,years of \peight observations, conditional on the posterior samples from the \pseven analysis of \rettwo only, adopting a log-uniform prior on~$\sigv$. We show the PPDs for the three classical and three ultrafaint dSphs with the highest median $J$-factors as examples for the discussion. From a Bayesian model averaging perspective, for each dwarf one could obtain an averaged prediction by summing the two models' PPDs with weight given by each model's posterior probability. Given that the outcome of the Bayesian model selection in the previous section was essentially inconclusive, for non-informative priors for each model the models' posterior probabilities are almost equal. Therefore, a model-averaged prediction would be an approximately equal mixture of the PPDs for each separate model.

We observe in general that, after accounting for all uncertainties, the 68\% regions of the PPD predictions for the background-only and for the signal-plus-background model overlap for all energy bins and all dwarfs shown here. This means that a detection of a dark matter signal in these dwarfs would rely on a fortuitous upwards fluctuation of the signal (or downward fluctuation of the background). This is true even for the dSphs with the largest median $J$-factors, \textsc{Ursa~Major~II} and \textsc{Segue~1}, which exhibit the more prominent ``bump'' in the background-plus-signal model in the region of the putative excess between~\SI{1}{\GeV} and~\SI{10}{\GeV}. \updated{We also see some downward fluctuations of the observed counts in some energy bins when compared to the background-only predictions, in case of \textsc{Ursa Minor} even outside the 95\% predicted probability~band. This indicates that the prior on the background parameter disagrees with the data for this particular dSph. In fact, the global analysis revealed that e.g.\ the best-fit parameter is $\hat{\beta}_\textsc{UrMi} = 0.47$, i.e.\ noticeably lower than the reference value of~1.}

A more quantitative way of assessing how promising a dSph is in terms of detecting a dark matter signal is the predicted distribution for the signal-to-noise ratio~(SNR) in the background-plus-signal model. The SNR for the new dwarf, conditioned on \rettwo data, is computed according to the prescription in Ref.~\cite[][Eq.~18]{1410.2242}. That study introduces a test statistic~$T$ which is shown to have more power than any other at distinguishing a dark matter signal from background. The SNR is a measure of expected detection significance when using this test statistic to search for a signal. It is defined as
\begin{equation}
	\text{SNR} \equiv \frac{\mathrm{E}[T\mid H_1]-\mathrm{E}[T\mid H_0]}{\mathrm{Std}[T\mid H_0]} \, ,
\end{equation}
where $\mathrm{E}[T\mid H_1]$ is the expected value of the test statistic under the signal-plus-background hypothesis, while $\mathrm{E}[T\mid H_0]$ and $\mathrm{Std}[T\mid H_0]$ are the expected value and standard deviation of the test statistic under the background-only hypothesis. Following Ref.~\cite[][Eq.~18]{1410.2242}, the SNR for a dSph is computed using $\text{SNR}^2 = \sum_{i,j}(s^\text{DM}_{i,j})^2 \left/ b_{i,j} \right.$, where the sum is over spatial and energy bins for the dSph and $b_{i,j} \equiv \lambda_{i,j}-s^\text{DM}_{i,j}$~(see Sec.~\ref{sec:methods}).

The resulting distributions and highest-posterior density CRs for the SNRs are shown in the right panel of \reffig{fig:predpost} in order of decreasing predicted median SNR (from top to bottom). Notice that given our definition of SNR, a value of $\text{SNR} > 5$ would correspond to a $5\sigma$~detection (assuming a Gaussian distribution of $T$ under the background-only hypothesis). While the maximum of the predictive distribution is in all cases above unity (and often above a value of 5, corresponding to a $5\sigma$~detection), the predictive distribution has very long tails as a consequence of the uncertainties in the $J$-factor and background normalisation, which are fully accounted for in the prediction. There is hence a large fractional probability that the SNR will be smaller than unity and that any signal will be undetectable. We can thus classify each dwarf in terms of the probability to obtain a detection (defined as an SNR value larger than~5), which we call the ``discovery probability''. This is simply the integrated posterior predictive probability density for $\text{SNR} > 5$. Another measure of the chance of a dark matter detection is the 95\% lower limit from the PPD for the SNR. This gives the value of the SNR above which lies 95\% of the predicted probability density.

\begin{figure}[t]
	\hfill
	\includegraphics[width=0.47\textwidth]{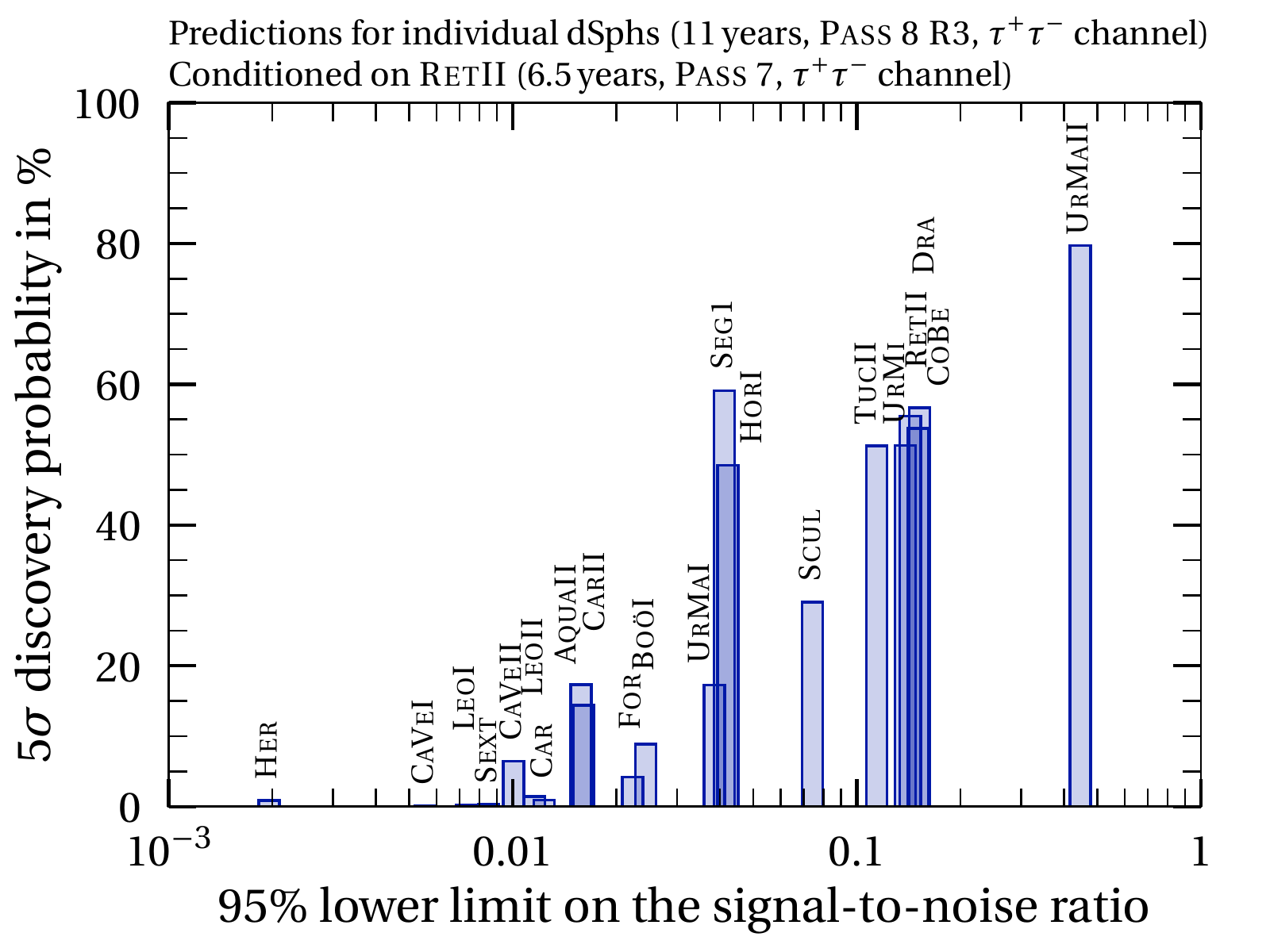}\hfill
	\includegraphics[width=0.47\textwidth]{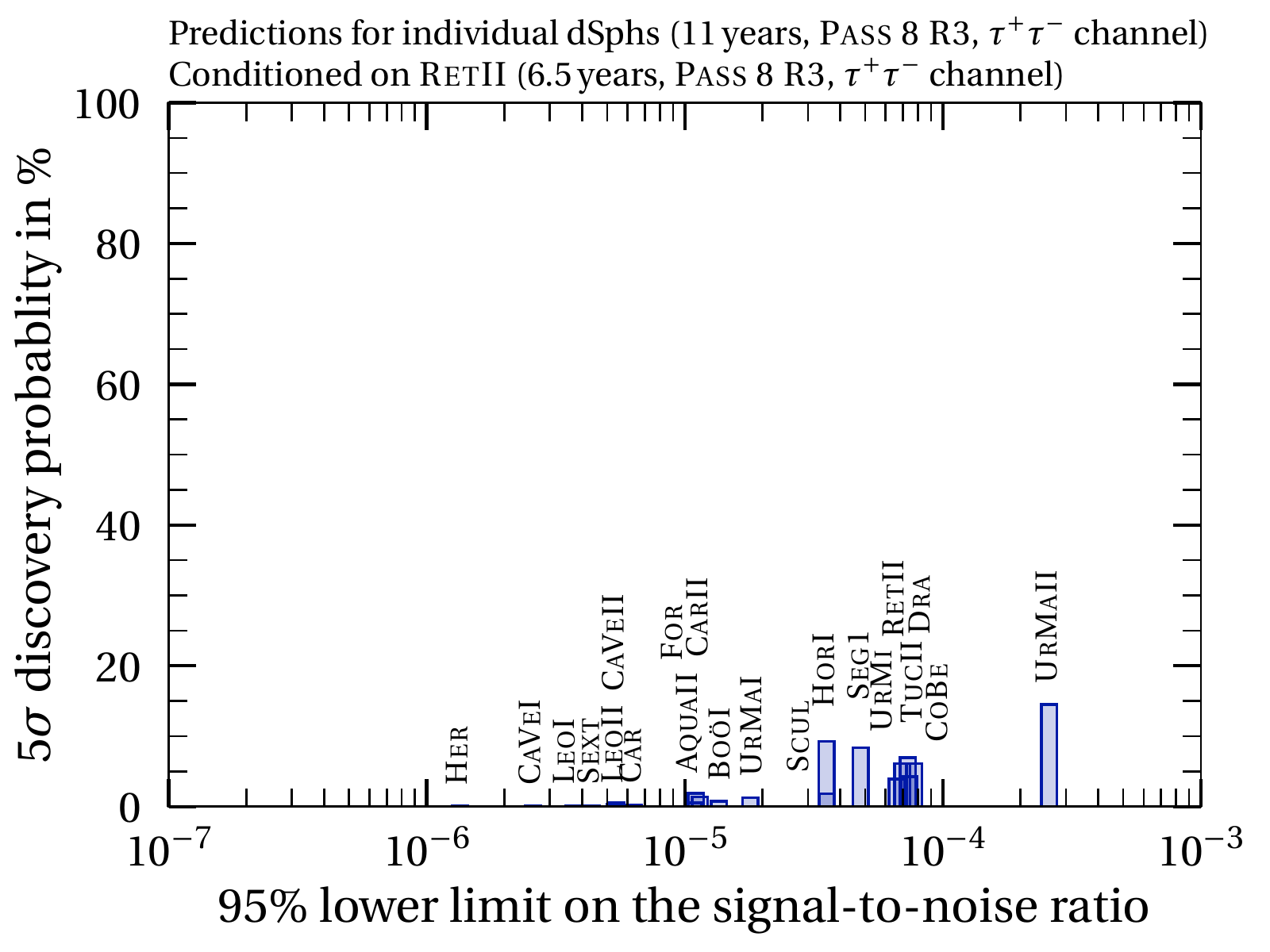}
	\hfill
	\caption{Posterior predictive probability of making a detection of a DM signal at more than~$5\sigma$ significance in another dwarf, including all relevant sources of uncertainty, conditional on \rettwo data. We show the results, sorted by their lower SNR limit, conditioned on 6.5\,years of \pseven~(\textit{left}) and \peight~(\textit{right}) \rettwo data ($\tptm$ channel, log-uniform prior on~$\sigv$). The probability of detection drops dramatically from \pseven to \peight data. \label{fig:predpostsummary}}
\end{figure}
We show these values in \reffig{fig:predpostsummary}, conditioned on 6.5\,years of \rettwo data in \pseven~(left panel) and \peight~(right panel) for the $\tptm$ channel and a log-uniform prior on $\sigv$. The individual dSphs are sorted by the 95\%~lower limit on the SNR. Conditional on \pseven data, \updated{the probability of a discovery in an individual dSph is larger than 50\% in seven~of them. The highest individual discovery~probability occurs for \textsc{Ursa Major~II} with a value of about 80\%.} Note that the six~ Sphs with long tails (cf.\ Sec.~\ref{sec:jfactors}) do not appear in the plot because they have much lower limits with vanishingly low $\text{SNR} < 10^{-12}$. However, \textsc{Pegasus~III} and \textsc{Draco~II} still show a fairly reasonable discovery probability of~17\% and~18\%, respectively.

We can also evaluate the probability that at least one of the dSphs yields a $5\sigma$ or higher detection.\footnote{Technically, each dSph detection probability is a local probability that ignores the so-called ``look-elsewhere effect'', arising from multiple testing when looking at many dSphs. Even if there is no DM signal, statistical fluctuations in the background would be expected to yield a $5\sigma$~detection if one tests a sufficiently high number of dSphs. Estimating the look-elsewhere effect would require evaluating the probability of a false detection from any one of the tested dSphs. However, this cannot be done in our framework since the SNR we use is undefined in the absence of a signal.} If each dwarf were independent, the probability of a detection in any one of the $N = \ndwarfs$~ dwarfs (ignoring the ``look-elsewhere effect'') would simply be given by
\begin{equation}
	P_\text{det} = 1 - \prod_{k=1}^{N} \, \Big(1 - \int_5^\infty \! \prob{\text{SNR}_k}{d} \, \dd (\text{SNR}_k) \Big) \, ,
\end{equation}
where $d$ are the data that the PPD is conditioned upon, and $p$~is the PPD for the SNR in dwarf~$k$. However, the signal is of course fully correlated in all dSphs since, in the presence of dark matter, the WIMP mass and cross section are exactly the same for all dSphs. Therefore, we must instead estimate the probability of making a detection in at least one dwarf numerically, as the fraction of posterior samples for which $\text{SNR} > 5$ in at least one dwarf, i.e. the SNR surpasses the detection threshold. Doing so using the \pseven posterior samples results in a probability of~90\%. This means that, conditional on \pseven \rettwo data, there is a 90\%~probability that at least one of the other dwarfs yields a $5\sigma$ detection.

Conditional on 6.5 years of \peight data from \rettwo, on the other hand, there is no strong preference for a DM~signal and it is therefore not surprising that the predicted SNR in the other dSphs will generally not yield a detection (right panel of \reffig{fig:predpostsummary}). However, a few of the dwarfs, such as \textsc{Ursa Major~II}, \textsc{Horologium~I} or \textsc{Segue~1}, still have a sizeable probability for an individual detection. Still, \updated{the probability to make a detection in at least one dSph, conditional on 6.5\,years of \rettwo data in~\peight, drops to~24\%}. The six~dwarfs with long tail have again a much lower limit on the SNR than the others, with \textsc{Pegasus~III} and \textsc{Draco~II} showing comparable discovery probabilities to other dSphs (2.4\% and 5.3\%, respectively).

PPDs can therefore be used as a tool to determine which dSphs are likely to result in a detection (or rule out a model, depending on the statistical question being asked). While a full global analysis is always desirable, it can become computationally very expensive as more dSphs are added to the likelihood, perhaps with additional nuisance parameters. In this case, a potential solution is to include only the ``most promising'' dSphs such that the outcome of the analysis is ideally approximately as strong as the result of a global, complete analysis. While the relevance of a dSph might be determined by, e.g., the highest $J$-factor or their likelihood contribution, PPDs can help identify these systems in a more statistically principled way, while at the same time accounting for all relevant uncertainties in the prediction. We use the findings of this section to inform our choices for the Bayesian model comparison in the global analysis.

\subsection{Global analysis of \ndwarfs~dwarfs}\label{sec:globalfit}
We now turn to our findings from a global analysis, using 11\,years of \peightrthree data for \ndwarfs~dwarfs in the likelihood function. Our analysis has a total of \nparams~parameters: the background normalisations and $J$-factors for each of the dSphs, plus $m_\chi$ and $\sigv$. For parameter estimation, we increase the number of parallel chains for \tw (\texttt{sqrtR}: \num{1.01} with 2016~\textsf{MPI} processes), resulting in around 200~million equally weighted samples for each channel, and use more demanding settings for \diver (\texttt{NP}: \num{e5}, \texttt{convthresh}: \num{e-7}, \texttt{jDE}: true, \texttt{lambdajDE}: false).

\begin{figure}[t!]
	\hfill
		\includegraphics[width=0.49\textwidth]{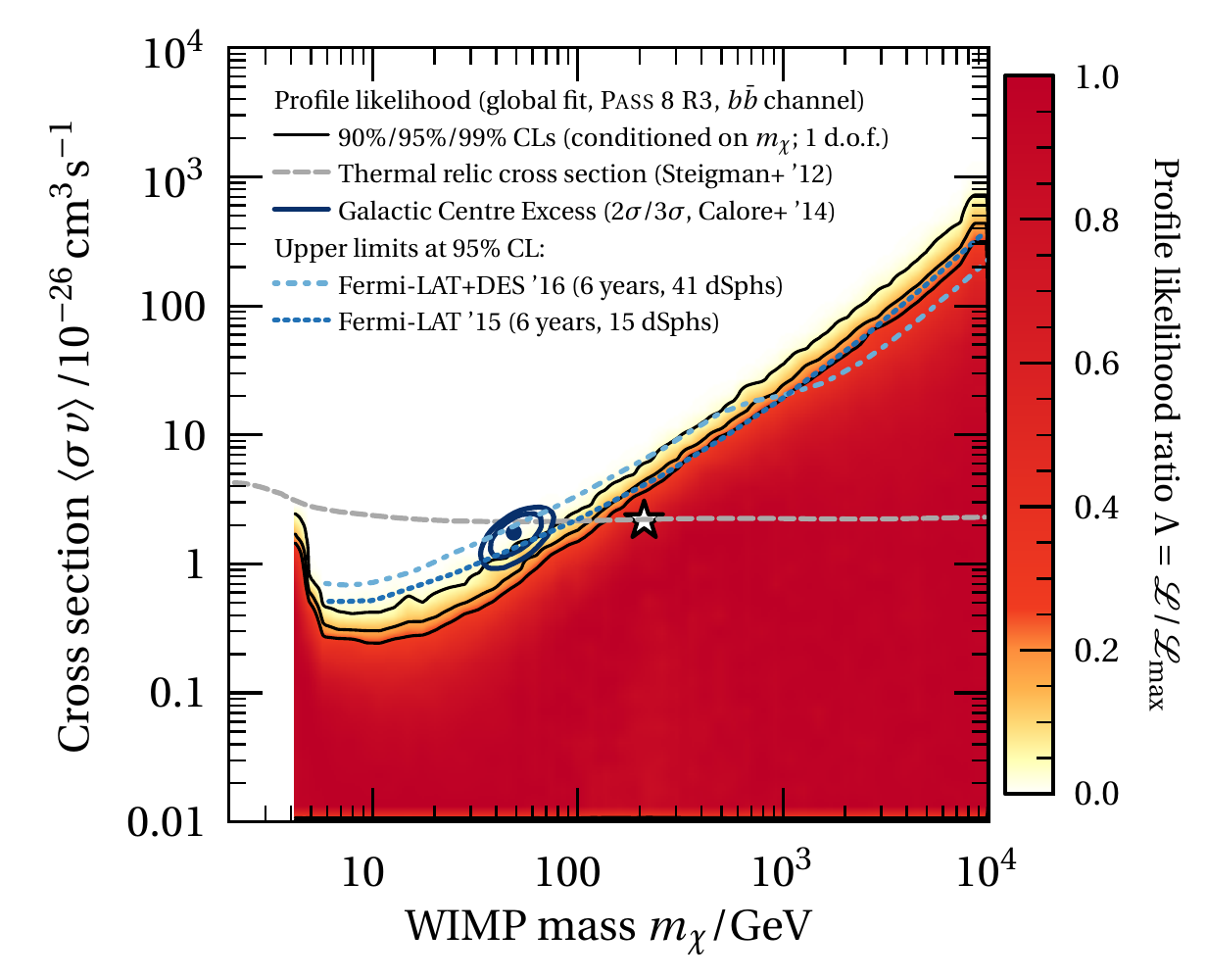}\hfill
		\includegraphics[width=0.49\textwidth]{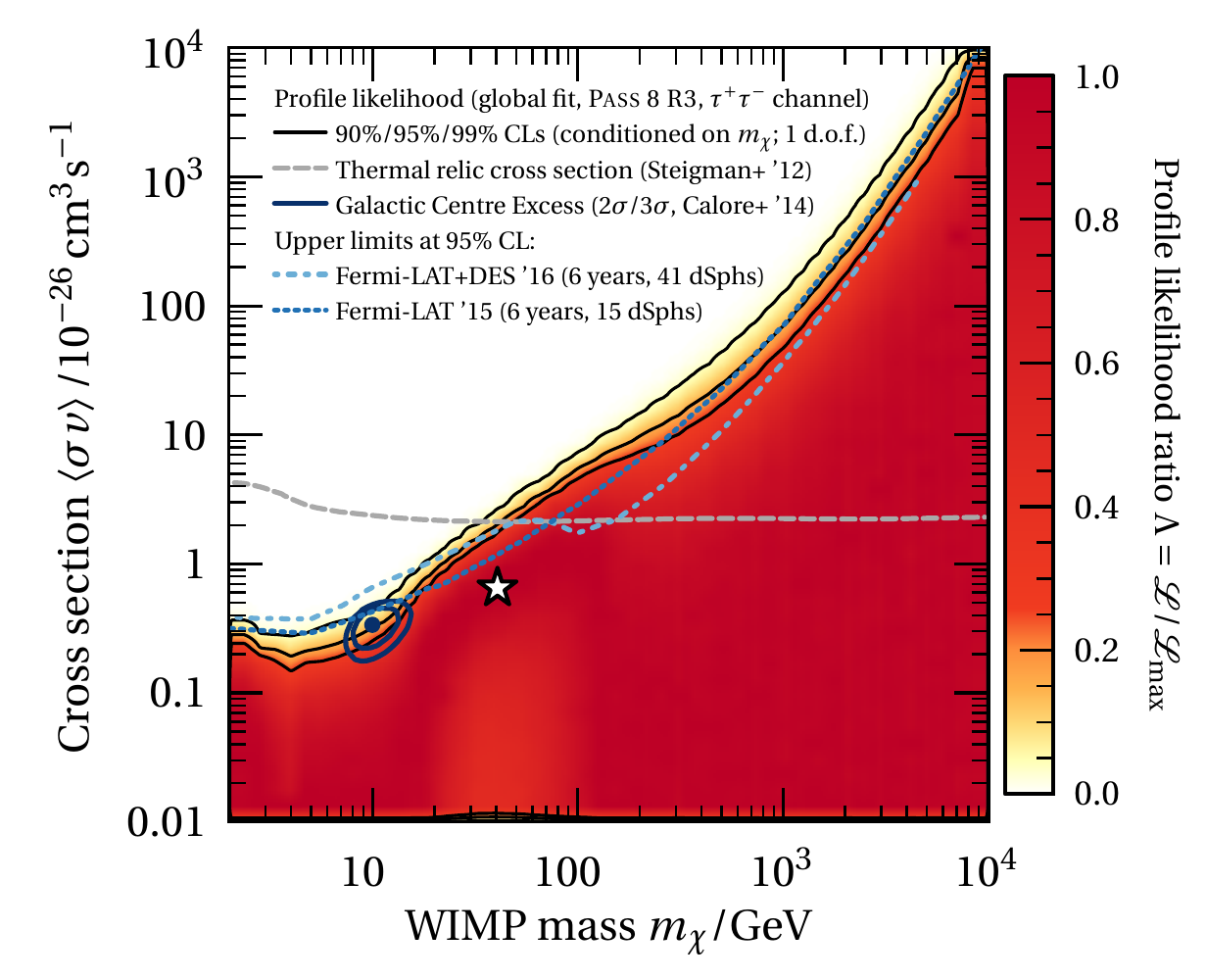}
	\hfill\\
	\hfill
		\includegraphics[width=0.49\textwidth]{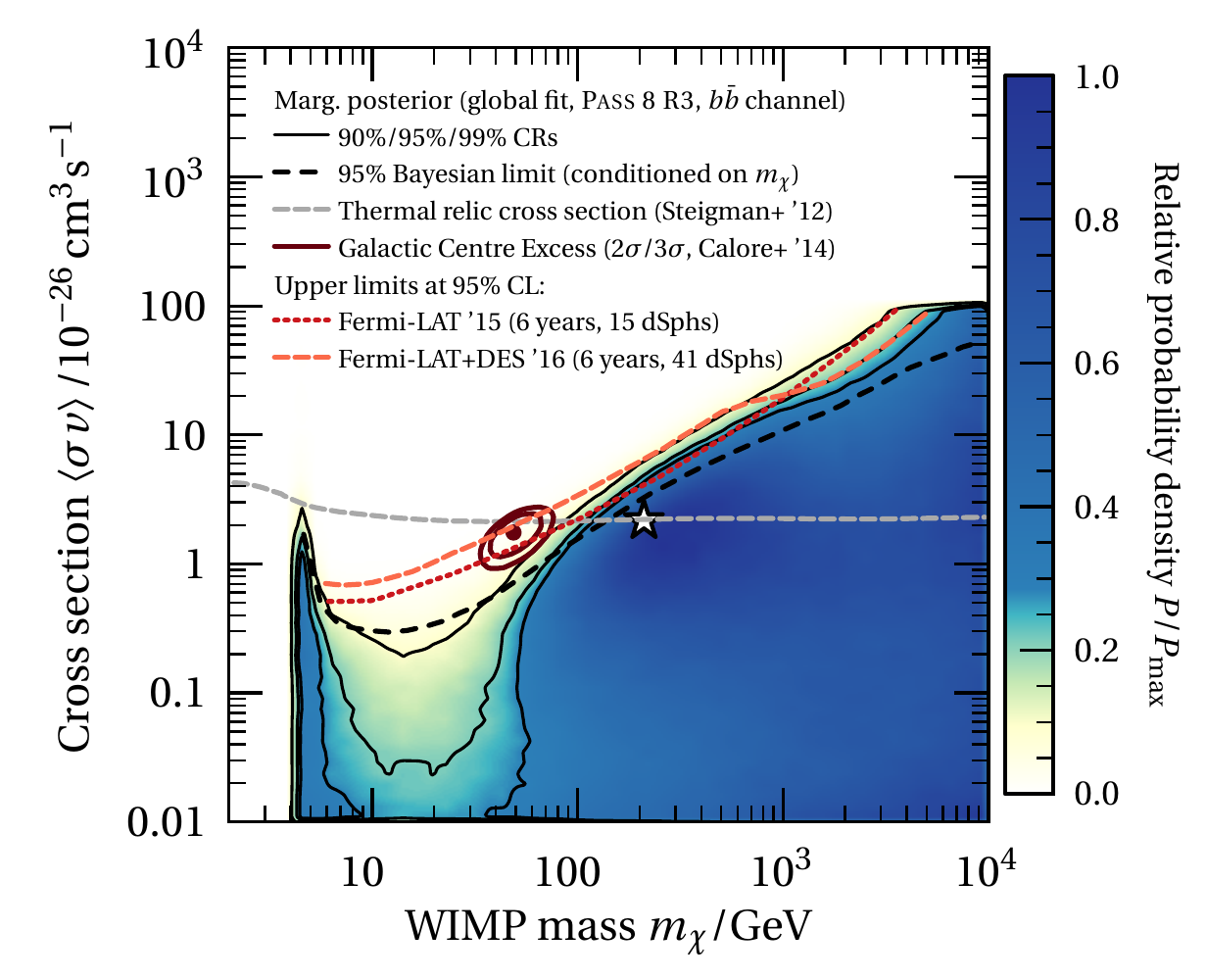}\hfill
		\includegraphics[width=0.49\textwidth]{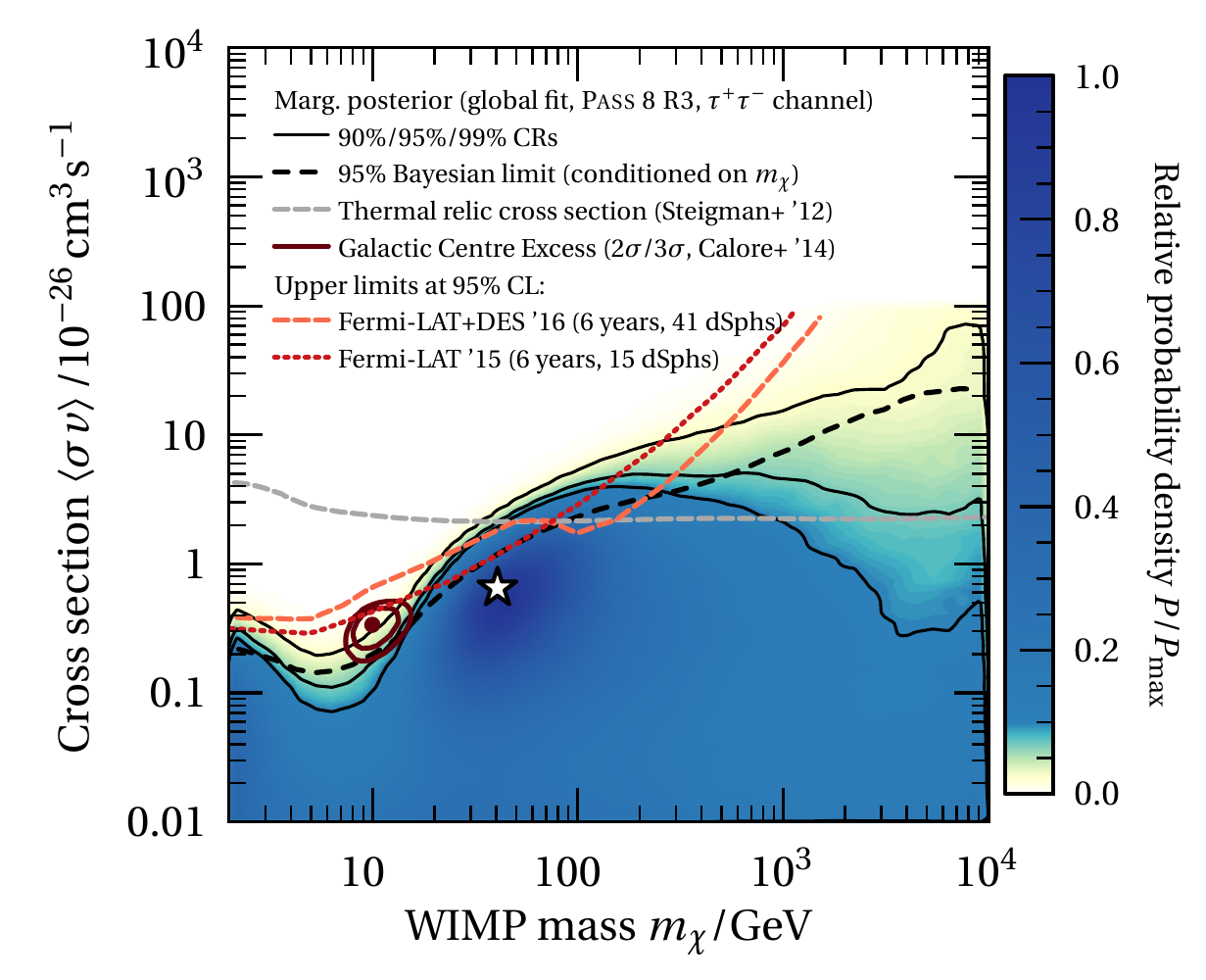}
	\hfill
	\caption{Global analyses using 11\,years of \peight data for \ndwarfs~dwarfs. We show profile likelihoods~(\textit{top}) and marginalised posteriors~(\textit{bottom}) for the $\tptm$ (\textit{left}) and $\bbb$ (\textit{right}) channels. The star denotes the best-fit point for a given channel. We compare our results with the thermal relic cross~section~\cite{1204.3622}, previous limits~\cite{1503.02641,1611.03184}, and parameters associated with the purported Galactic~Centre excess~\cite{1411.4647}. Note that the posteriors are restricted to the region $\sigv_{-26} < 100$ due to the prior range on~$\sigv$ for the Bayesian analysis.\label{fig:globalresults}}
\end{figure}
The DM parameter constraints from the global analysis, separately considering the $\tptm$ and $\bbb$~channels,\footnote{We also performed an analysis using the branching fraction~$b_{\tptm}$ into $\tptm$ as an additional free parameter with a uniform prior in the range \prrange{1}{0}. We found no strong preference for either channel: annihilation into mostly~$\bbb$ ($b_{\tptm} < 0.25$) has a 23\% posterior probability, while annihilation into mostly~$\tptm$ ($b_{\tptm} > 0.75$) has a 31\% posterior probability. Since these values are close to 25\% (the value under the uniform prior), this means that the data cannot constrain this additional parameter.} are shown in \reffig{fig:globalresults}. The top row shows results from the profile likelihood~(where the likelihood has been maximised over the nuisance parameters), while the bottom row shows the Bayesian posterior~(where nuisance parameters have been marginalised over). We also display for reference the thermal relic cross section~\cite{1204.3622} and limits from previous analyses at 95\% CL. In the top panels, the confidence limits have been obtained conditional on the value of the mass, in order to make them exactly comparable to those obtained by the Fermi-LAT Collaboration~\cite{1503.02641,1611.03184}. \updated{The best-fit WIMP parameters in the global fit for the $\tptm$~($\bbb$) channel are $\widehat{m}_\chi = \SI{40.5}{\GeV}$ and $\widehat{\sigv}_{-26} = \num{0.67}$~($\widehat{m}_\chi = \SI{210}{\GeV}$ and $\widehat{\sigv}_{-26} = \num{2.23}$).}

For WIMP~masses below about \SI{100}{\GeV}, we obtain stronger limits compared to previous frequentist analyses. As a consequence, we can exclude the best-fit DM~parameter values for the $\bbb$ channel DM~interpretation of the Galactic~Centre~(GC) excess~\cite{0910.2998,1306.5725,1402.4090,1402.6703,1411.4647} at more than 99\%~CL.\footnote{We only consider the best-fit point and associated confidence regions from the analysis of Ref.~\cite{1411.4647}, since this gives the smallest (and therefore least constrained) value for the DM cross~section, when compared with other GC studies.} For the $\tptm$ channel, the GC excess best fit point can be excluded at the 95\% CL. This improves the exclusion strength of dSphs compared to previous studies~\cite[e.g.][]{1510.06424,1611.03184,1710.03215}.

The Bayesian analysis (bottom panels of \reffig{fig:globalresults}) disfavours the GC~excess DM~interpretation even more strongly. Indeed, we notice that for low WIMP mass the Bayesian credible region is noticeably more constraining than the profile likelihood. The best-fit point and part of the associated confidence regions for the DM interpretation of the GC excess from Ref.~\cite{1411.4647} is outside the 99\% credible regions in the Bayesian analyses for both channels~(bottom panel of \reffig{fig:globalresults}). Recall that our choice of prior lower boundary for the Baysian analysis is conservative, i.e. yields looser upper limits than would be obtained by increasing the prior range at the low end. Notice that the Bayesian contours in the bottom panels are two-dimensional credible regions, which cannot be directly compared with the one-dimensional profile likelihood limits in the upper panels (which instead condition on the value of the mass). To facilitate a direct comparison, we also compute a ``Bayesian 95\% limit'' on the cross~section~(shown as black dashed line in the bottom panel of \reffig{fig:globalresults}). This is obtained by integrating the posterior \emph{conditional} on the given value of~$m_\chi$, in order to mimic the procedure used for the frequentist conditional~CL. This Bayesian limit is somewhat closer to the frequentist 95\%~CL than the corresponding CR for lower WIMP~masses, but still overall stronger than the frequentist result.

\updated{In our analysis of \rettwo, we observed that the best-fit value for the background normalisation, $\mrm{\hat{\beta}}{\rettwo} = 0.58$ (for \peight data), was noticeably lower than the reference value of $\mrm{\beta}{\rettwo} = 1$ in both \pseven and \peight. In the global analysis, we find again that $\mrm{\hat{\beta}}{\rettwo} \approx 0.67 < 1$. There seems to be a slight re-absorption of the excess photons in \rettwo by the higher background component as non-detections in other dwarfs force the value of~$\sigv$~($m_\chi$) to lower (higher)~values than in the \rettwo-only analysis. These WIMP~parameters require a slightly higher background normalisation to provide a good fit the shape of the \rettwo spectrum. Regarding the other dSphs, the 95\% HPD region 11~of them contains~$\beta_{k} = 1$, for 15~of them it lies below~$\beta_{k} = 1$, and only for \textsc{Leo~V} is lies above~$\beta_{k} = 1$.}

Combining the posteriors for all~$\beta_{k}$ into one set results in a distribution that allows us to estimate the mean and spread of the $\beta_{k}$. We find that \updated{the samples of all dwarfs combined have a posterior mean and standard deviation of~0.76 and~0.27, respectively}. Similar to how we used this information in Sec.~\ref{sec:ppds}, the mean, standard deviation, or both may be used to inform prior choices in future studies with comparable background modelling.

\updated{Tabulated exclusion limits, profile likelihood and marginal posterior density maps for \reffig{fig:globalresults} and the limits presented in Appendix~\ref{app:other_limits} are freely available on \textsf{Zenodo}~\cite{zenodo_dsphs}.}

\begin{table}
	\renewcommand{\arraystretch}{1.05}
	\caption{Bayes factors for comparing a background-only model with a model including an additional DM signal (11\,years of \peight data; we use the most promising 10~dSphs, selected using the PPD of their~SNR). A positive (negative) value of $\ln\left(\mathcal{B}_{10}\right)$ indicates evidence in favour of (against) the model with an additional DM~signal. The value of $\mathcal{B}_{10}$ gives the posterior odds between the DM model and the background only model if each model has equal prior probability.\label{tab:global:modelcomp}}
	\vspace{0.25em}
	\centering
	\begin{tabular}{lcccc}
		\toprule
		Prior on $\sigv$ & \multicolumn{2}{c}{$\tptm$ channel} & \multicolumn{2}{c}{$\bbb$ channel} \\
		& uniform & log-uniform & uniform & log-uniform \\
		\midrule
		Bayes factor~$\ln\left(\mathcal{B}_{10}\right)$ & \updated{$-0.88 \pm 0.05$} & \updated{\phantom{$-$}$0.01 \pm 0.05$} & \updated{$-1.09 \pm 0.05$} & \updated{\phantom{$-$}$0.02 \pm 0.05$} \\
\phantom{Bayes factor~$\ln($}$\mathcal{B}_{10}$ & \phantom{$-$}1:2 & \phantom{$-$}1:1 & \phantom{$-$}1:3 & \phantom{$-$}1:1\\
		\bottomrule
	\end{tabular}
\end{table}
Finally, we also performed a model comparison by calculating Bayes factors for the two hypotheses with~\mn. Obtaining reliable estimates for the Bayesian evidence proved difficult due to the relatively large dimensionality of the parameter space (the efficiency of \mn drops quickly above about 30~dimensions~\cite{1502.01856}). We therefore reduced the dimensionality of the parameter space by making a smaller selection of dSphs, choosing those with a median predicted SNR greater than unity when conditioned on \pseven data\footnote{These also correspond to the dSphs with the highest lower limit on the SNR as well as the highest discovery probabilities, except \textsc{Ursa Major~I}, whose detection probability of~17.3\% is smaller but similar to that of \textsc{Aquarius~II}~(17.3\%) and \textsc{Draco~II}~(17.8\%).} using the PPD approach in the previous section. These are the following ten~dSphs: \textsc{Coma Berenices}, \textsc{Draco}, \textsc{Horologium~I}, \textsc{Reticulum~II}, \textsc{Sculptor}, \textsc{Segue~1}, \textsc{Tucana~II}, \updated{\textsc{Ursa Major~I}}, \textsc{Ursa Major~II}, and \textsc{Ursa Minor}. We also use slightly less demanding settings for \mn (\texttt{nlive}: \num{2e4}, \texttt{tol}: \num{e-3}). The resulting Bayes factors from this analysis can be found in Table~\ref{tab:global:modelcomp}. Since we have adopted the most constraining (in terms of predicted SNR) dwarfs in this analysis, we expect it to be close to what would have been obtained by including all of the \ndwarfs~dwarfs.

Since we found no preference for a signal in the parameter estimation part of the global analysis, it is not surprising that the model comparison finds no evidence for an additional signal either. On the other hand, we obtain only \textit{weak evidence} \emph{against} the signal-plus-background model in the $\bbb$ channel (using a uniform prior on~$\sigv$).

The outcome of the model comparison is inconclusive, as there is only \textit{weak evidence} against the signal-plus-background model for the $\bbb$ channel and a uniform prior. Compared the \rettwo section, the trend that a uniform prior on~$\sigv$ gives lower Bayes~factors continues also in the global analysis. Indeed, since the WIMP~parameter space always allows for the \emph{possibility} of having an essentially irrelevant gamma-ray signal, the signal-plus-background model can only be disfavoured compared to the background-only model due to the Occam's razor effect. This happens when the region of WIMP~parameter~space resulting in a negligible gamma-ray signal from DM is only a small fraction of the prior volume -- the Occam's razor effect penalising models with ``wasted'' parameter space. The size of the prior volume of the WIMP~parameters is therefore the most important ingredient that would allow the odds to swing in favour of the background-only model. This also explains the trend we observe with the two adopted priors since this fraction is smaller for a prior uniform in~$\sigv$ than for a prior uniform in the log of~$\sigv$.

\section{Conclusions}\label{sec:conclusions}
We have revisited dark matter searches in dSphs in a systematic way, comparing Bayesian and frequentist methods in the largest dSph sample and highest-exposure search performed to date. When looking for a signal while only having imperfect knowledge of background, relevant sources of uncertainties should be accounted for in the analysis in order to obtain robust results. We therefore included scaling factors for the Galactic~diffuse background component. Since $J$-factors can only be approximated via fitting {formul\ae} or determined from stellar data, it is also important to account for their theoretical and statistical uncertainties. For this analysis, we relied on dwarf~spheroidal galaxies with posteriors for their $J$-factors as determined by spectroscopic data. To properly account for uncertainties we adopted these posteriors as priors for the gamma-ray analysis, without resorting to the usual log-normal approximation that can be inaccurate.

Using \rettwo as an example, we illustrated our methodology and investigated the differences between Bayesian and frequentist analyses as well as between \pseven and \peightrthree data. We showed that the putative excess is not significant in \peight data in neither Bayesian nor frequentist analyses, while there is evidence in \pseven data for a non-zero signal contribution. In line with previous literature, we conclude that the differences in significance of the excess are due to the data sets rather than details of the analysis, since we applied the identical methodology and equivalent data selection criteria. We also introduced the posterior predictive distribution into gamma-ray signal searches as a useful tool to determine the consistency of a potential signal amongst a sample of dwarf~spheroidal galaxies. The posterior predictive distribution combines the posterior uncertainties in model parameters with the Poisson fluctuations expected in observations to create predicted data set distributions, clearly highlighting the difference between the background-only and signal-plus-background hypotheses in different dSphs. These can be used to easily and robustly quantify the probability of a future dark~matter detection. Our predictive formalism has wide applicability in dark~matter searches beyond gamma-ray and dwarf~spheroidal galaxy analyses.

In the global analysis of \peight data, which includes \ndwarfs~dwarfs with measured $J$-factors and 11\,years of observations, we did not find any indication for an excess and hence derived upper limits on the dark matter cross~section as a function of the WIMP~mass. The best-fit parameters associated with a dark matter interpretation of the Galactic Centre excess that remained previously viable are ruled out by the frequentist analysis at 95\% confidence level for the $\tptm$ and $\bbb$~channels. The Bayesian analysis excludes the entirety of the $3\sigma$ confidence region for the Galactic~Centre excess at more than 99\% probability.

The global analysis comprises a total of \nparams~parameters, which is a fairly high number of dimensions for sampling algorithms. Thanks to using \diver and \tw, two dedicated algorithms for profile likelihood mapping and posterior sampling, we could perform parameter estimation without major problems. However, we also saw that calculating the Bayesian evidence for the global analysis with many dwarf spheroidal galaxies can present a challenge that can make a full global analysis prohibitive. We overcame this problem by using information from posterior predictive distributions of the signal-to-noise ratio to select a subset of the most relevant dwarf~spheroidal galaxies.

\section*{Acknowledgements}
\renewcommand\baselinestretch{1.25}\footnotesize
SH gratefully acknowledges funding by the Imperial College President's Scholarship, the Alexander von Humboldt Foundation, and the German Federal Ministry of Education and Research. AGS and RT are supported by Grant ST/N000838/1 from the Science and Technology Facilities Council~(UK). AGS was supported in part by Fermi Guest Investigator grant NNX16AR33G. RT was partially supported by a Marie {Sk\l{}odowska-Curie} RISE Grant (H2020-MSCA-RISE-2015-691164) provided by the European Commission. The authors would like to thank David van~Dyk, Germ\'{a}n~A. G\'{o}mez Vargas, Roberto Ruiz~de~Austri, Pat Scott and an anonymous referee for helpful suggestions and Andrew~B. Pace and Louis~E. Strigari for making their $J$-factor samples available. For computational resources, we gratefully acknowledge the UK Materials and Molecular Modelling Hub, which is partially funded by EPSRC~(EP/P020194/1), and the Imperial College Research Computing Service~(\doi{10.14469/hpc/2232}).

This work underwent an additional review by the \textit{Fermi} LAT Collaboration, for which we acknowledge input and suggestions by Michael Gustafsson, Gudlaugur {J\'ohannesson}, Mattia Di~Mauro, and Gabrijela Zaharijas. The \textit{Fermi} LAT Collaboration acknowledges generous ongoing support from a number of agencies and institutes that have supported both the development and the operation of the LAT as well as scientific data analysis. These include the National Aeronautics and Space Administration and the Department of Energy in the United States, the Commissariat \`a l'Energie Atomique and the Centre National de la Recherche Scientifique / Institut National de Physique Nucl\'eaire et de Physique des Particules in France, the Agenzia Spaziale Italiana and the Istituto Nazionale di Fisica Nucleare in Italy, the Ministry of Education, Culture, Sports, Science and Technology (MEXT), High Energy Accelerator Research Organization (KEK) and Japan Aerospace Exploration Agency (JAXA) in Japan, and the K.~A.~Wallenberg Foundation, the Swedish Research Council and the Swedish National Space Board in Sweden. Additional support for science analysis during the operations phase is gratefully acknowledged from the Istituto Nazionale di Astrofisica in Italy and the Centre National d'\'Etudes Spatiales in France. This work performed in part under DOE Contract DE-AC02-76SF00515.

\appendix
\section{Limits on other annihilation channels}\label{app:other_limits}
\begin{figure}[t]
	\hfill
	\includegraphics[width=0.49\textwidth]{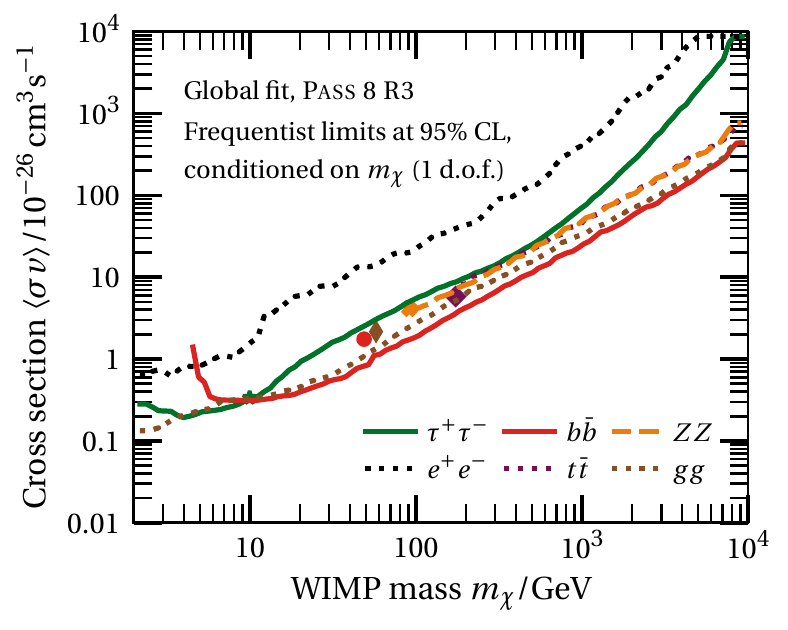}\hfill
	\includegraphics[width=0.49\textwidth]{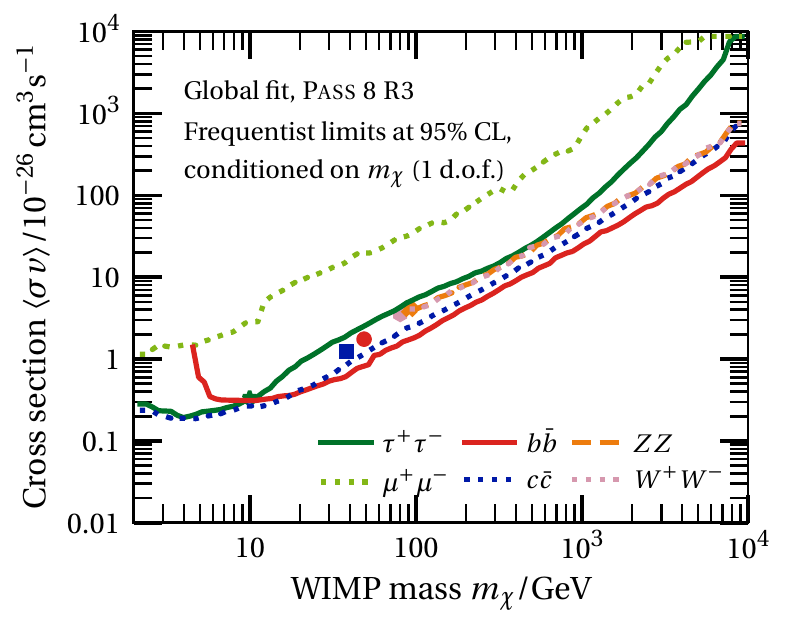}\\
	\hfill
	\includegraphics[width=0.49\textwidth]{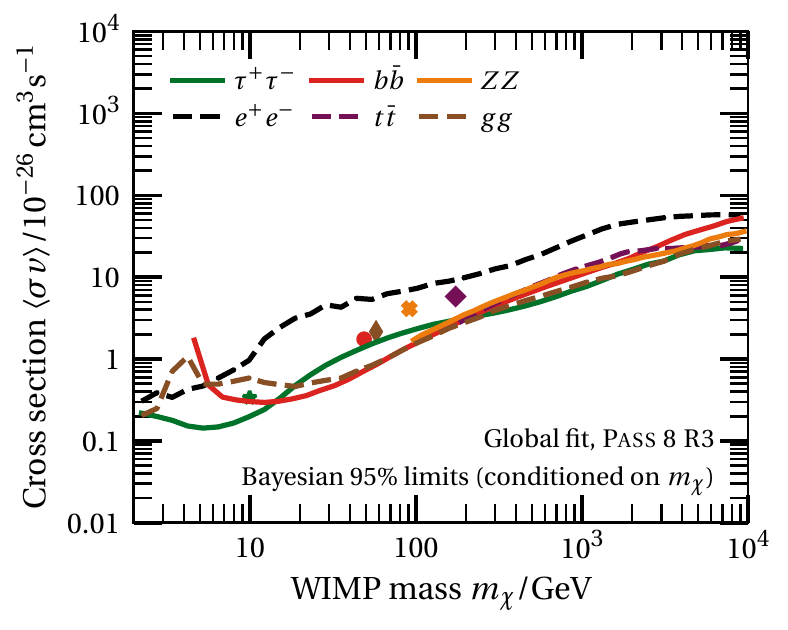}\hfill
	\includegraphics[width=0.49\textwidth]{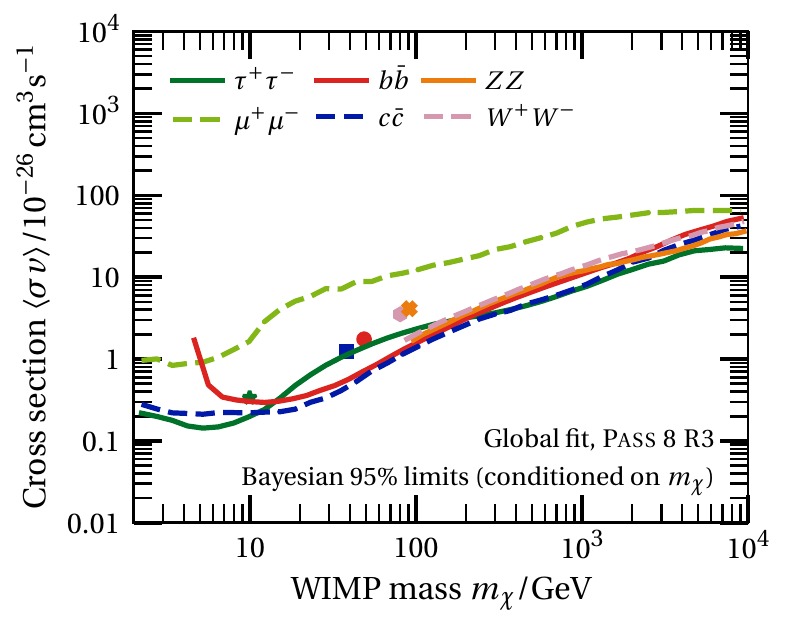}
	\hfill
	\caption{Limits on the cross~section~$\sigv$ for nine different annihilation channels. \textit{Top:}~Frequentist limits at 95\% CL~(conditioned on~$m_\chi$; 1 d.o.f.). \textit{Bottom:}~Bayesian 95\% limits~(conditioned on~$m_\chi$). Where available, markers with matching colours indicate the best-fit points for the corresponding dark matter interpretation of the Galactic~Centre excess in Ref.~\cite{1411.4647}.\label{fig:other_limits}}
\end{figure}

\renewcommand\baselinestretch{1.25}\normalsize
In addition to the $\bbb$ and $\tptm$ channels considered in the main text, we provide limits for seven other channels in \reffig{fig:other_limits}. Calculation and implementation of each model's signal prediction is equivalent to Sec.~\ref{sec:data}~(using \texttt{sqrtR} < \num{1.05} in \tw). We choose $\SI{2}{\GeV}$ as the lower edge of the prior on the WIMP mass for final states with masses lower than that value~($e$, $\mu$, $\tau$, $c$, $g$), or at the mass of the particle for the others~($W$, $Z$). Similar to the $\bbb$ and $\tptm$~channels investigated in the main text, both Bayesian and frequentist limits rule out the best-fit points of the DM interpretations of the GC~excess.

All corresponding Bayesian and frequentist limits, profile likelihood maps and marginal posterior density maps corresponding to \reffig{fig:other_limits} are freely available on \textsf{Zenodo}~\cite{zenodo_dsphs}, together with a description of the data set and format.

\newpage

\renewcommand{\baselinestretch}{0.5}\footnotesize
\setlength{\bibsep}{0.75em plus 0.3ex}
\bibliographystyle{apsrev_mod}
\bibliography{dwarfs}

\end{document}